\documentclass{article}
\usepackage{amsmath}
\usepackage{sw20bams}
\usepackage{latexsym}
\usepackage{amsfonts}
\usepackage{amssymb}
\usepackage{graphicx}
\newtheorem{theorem}{Theorem}

\newtheorem{definition}[theorem]{Definition}

\newtheorem{lemma}[theorem]{Lemma}

\newtheorem{proposition}[theorem]{Proposition}
\newtheorem{remark}[theorem]{Remark}

\begin{document}
\thispagestyle{empty}
%\hfill{{\protect\footnotesize CBPF-NF-019/01}} 

\author{Bert Schroer\\Institut f\"{u}r Theoretische Physik\\FU-Berlin, Arnimallee 14, 14195 Berlin, Germany\\presently: CBPF, Rua Dr. Xavier Sigaud, 22290-180 Rio de Janeiro, Brazil\\email schroer@cbpf.br}
\title{Lectures on Algebraic Quantum Field Theory and Operator Algebras}
\date{based on lectures presented at the 2001 Summer School in Florianopolis and at
the IMPA in Rio de Janeiro, April 2001 }
\maketitle
\begin{abstract}
In this series of lectures directed towards a mainly mathematically oriented
audience I try to motivate the use of operator algebra methods in quantum
field theory. Therefore a title as ``why mathematicians are/should be
interested in algebraic quantum field theory'' would be equally fitting.

Besides a presentation of the framework and the main results of local quantum
physics these notes may serve as a guide to frontier research problems in
mathematical physics with applications in\ particle and condensed matter
physics for whose solution operator algebraic methods seem indispensable. The
ultraviolet problems of the standard approach and the recent holographic
aspects belong to this kind of problems.
\end{abstract}
\tableofcontents

\thispagestyle{empty}
%\hfill{{\protect\footnotesize CBPF-NF-019/01}} 
%
%
%
%
%
%
%

%\date{arXiv:math-ph/0102018 v3 1 Mar 2001}
%\date{arXiv:math-ph/0102018 v3 1 Mar 2001}
%
%
%
%
%
%
%

\thispagestyle{empty}
%\hfill{{\protect\footnotesize CBPF-NF-019/01}} 
%
%
%
%
%
%
%
%
%
%

\newpage\setcounter{page}{1}

\section{Quantum Theoretical and Mathematical Background}

The fact that quantum field theory came into being almost at the same time as
quantum mechanics often lead people to believe that it is ``just a
relativistic version of quantum mechanics''. Whereas it is true that both
theories incorporated the general principles of quantum theory, the additional
underlying structures, concepts and mathematical methods are remarkably
different and this contrast manifests itself most visibly in the operator
algebra formulation of local quantum physics (LQP) \cite{Haag} whereas their
use in quantum mechanics would be an unbalanced formalistic exaggeration. This
distinction is less evident if one employs the standard quantization
formulation which has close links with differential geometry.

Mathematicians who were exposed to the mathematical aspects of some of the
more speculative ideas in contemporary high-energy/particle theory
(supersymmetry, string theory, QFT on noncommutative spacetime), which despite
their mathematical attraction were unable to make contact with physical
reality (in some cases this worrisome situation already prevails for a very
long time), often are not aware that quantum field theory (QFT) stands on
extremely solid rocks of experimental agreements. To give one showroom example
of quantum electrodynamics i. e. the quantum field theory of
electrons/positrons and photons, the experimental and theoretical values of
the anomalous magnetic moment of the electron relative to the Bohr magneton (a
natural constant) $\mu_{0}$ are
\begin{align}
\left(  \frac{\mu}{\mu_{0}}\right)  _{exper}  &  =1,001159652200(10)\\
\left(  \frac{\mu}{\mu_{0}}\right)  _{theor}  &
=1,001159652460(127)(75)\nonumber
\end{align}
where the larger theoretical error refers to the uncertainty in the knowledge
of the value for the fine-structure constant and only the second uncertainty
is related to calculational errors in higher order perturbative calculations.
The precision list can be continued to quantum field theoretic effects in
atomic physics as the Lamb shift, and with somewhat lesser accuracy in the
agreement with experiments may be extended to the electroweak generalization
of quantum electrodynamics and remains qualitatively acceptable even upon the
inclusions of the strong interaction of quantum chromodynamics.

Mathematician may even be less aware of the fact that only a few quantum field
theoretician who have had their experience with the mathematical intricacies
and conceptual shortcomings of the standard approach still believe that the
present quantization approach (which uses classical Lagrangian and formal
functional integrals as the definition of quantum electrodynamics (QED) or the
standard electroweak model) has a mathematical existence outside perturbation
theory\footnote{Perturbative QFT does not have the mathematical meaning of a
well-defined object which is being perturbatively expanded, but is rather a
formal deformation theory whose consistency does not imply anything about the
existence of a possible associated QFT.}, inspite of the mentioned amazing
experimental agreement with perturbation theory. In fact there is hardly any
theoretician who would be willing to take a bet about the mathematical
existence of these Lagrangian models. Arguments to that extend are often
presented in the physics literature by stating that ``QED does not exist''. Of
course there is a theory involving electrons and photons (even if we presently
do not know its correct mathematical description) and the critical arguments
only go against the Lagrangian/functional quantization definition of the
theory and not against the underlying principles of LQP which in fact
developed to a large degree from ideas about QED.

The cause of this critical attitude inspite of the overwhelming numerical
success is twofold. On the one hand it is known that renormalized perturbation
theory does not lead to convergent series in the coupling strength; rather the
series is at best asmptotically convergent i.e. the agreement with experiments
would worsen if one goes to sufficiently high perturbative orders (assuming
that one could seperate out the contributions coming from interactions outside
of QED). But there is also another more theoretical reason. To introduce
interactions via polynomial pointlike coupling of free fields is pretty much
ad hoc, i.e. if this recipe would not have worked, hardly anybody would have
been surprised. In fact a sufficient intrinsic understanding of what
constitutes interaction is the still missing cornerstone, even after 70 years
of QFT. Only in recent years there have been serious attempts and partial
successes on which we will comment in sections 3 and 4 of these notes.

Whereas in low spacetime dimensions (d=1+1, d=1 chiral models) the
mathematical existence of interacting quantum field theories has been
demonstrated by the presentation of certain controllable models, 4-dimensional
local quantum physics beyond free systems has largely resisted attempts at
demonstrating existence via construction of nontrivial models or otherwise.

This situation of having a perturbatively extremely successful description of
particle physics whose existence as a bona fide QFT on the other hand has
remained outside mathematical control is quite unique and in fact without
parallel in the history of physics. But it should not be viewed as something
embarrassing to be suppressed or covered by excuses (``there has to be a
cutoff at the Plank length anyhow'') because this situation is also the source
of fascination and a great challenge; its enigmatic power should not be
squandered. In the history of physics each conceptual framework (classical
mechanics, classical \ field theory, statistical mechanics, quantum mechanics)
was eventually shown to be mathematically consistent (usually by finding
nontrivial models), i.e. the necessity of finding an incorporation into a more
general framework was almost\ never coming from mathematical inconsistencies,
but either from new experimental facts or from the theoretical merger of two
different frameworks (example: relativity). If LQP build on Einstein causality
and quantum principles should really turn out to be mathematically
inconsistent, this would constitute a remarkable and enigmatic piece of
insight which should be made visible and not covered behind
cut-offs\footnote{Of course the phenomenological parametrzation in terms of
cut-off integrals is not effected by these remarks of how to deal with physcal
principles.}. There is hardly any contrast in fundamental physics comparable
to that between the verbal ease with which the word ``nonlocal'' is injected
into discussions and on the other hand the conceptual problems faced in
implementing nonlocality without destroying the whole fabric of an intrinsic
interpretation of the formalism including the derivation of the all important
scattering theory. For example string theory in its present formulation does
not permit an intrinsic derivation of time-dependent scattering theory (rather
the S-matrix is imposed by the Veneziano prescription).

Despite 50 years of attempts to render short distance properties more mild by
ad hoc nonlocal/noncausal modific]ations each proposal has proven to cause
more problems then it set out to solve \cite{cau} and it remains to be seen
whether the proposals of achieving short distance improvements via
noncommutativity of the spacetime localization pass the acid test of a
complete physical interpretation which includes in particular the derivation
of time dependent scattering theory.

The general message in the many failed attempts is that principles as
causality and locality can not be overcome by pedestrian ad hoc mathematical
modifications, but rather require the discovery of more general physical
principles of which they are limiting cases. The suggestion of the present LQP
approach with respect to short distance problems is conservative or
revolutionary depending on where one wants to put the emphasis; conservative
in that it does not temper with the causality principles and revolutionary in
that it views the short distance problems as an aspect of the limitation of
the quantization method using ``pointlike field coordinatizations'' (akin to
singular coordinates) and not part of the intrinsic frontiers set by the
principles but only of their implementation.

In pursuit of this challenge there have been new and deep conceptual and
mathematical inroads and investments over the last two decades; some of the
older ones were described in \cite{Haag}. The characteristic feature of those
achievements obtained with operator algebra methods is, as already previously
indicated, that they combine a revolutionary approach with respect to concepts
and mathematical formalism with a conservative attitude concerning physical
principles. In view of the fact that the very difficult and expensive high
energy experiments did not reveal any indication of incompatibility with the
general principles, this is a very reasonable procedure indeed.

Of course physicists need sometimes to move into the (following Feynman's
saying) ``blue yonder''. But at times of poor experimental guidance, taking
off without solid theoretical grounds under one's feet, such a trip may like
that of the legendary flying dutchman end without finding a physical landing
place over many generations.

I think that the framework of algebraic quantum field theory on which these
lectures are based offers such a firm soil. In particular it provides a
profound mathematical anchor to the concept of Einstein causality (and the
closely related Haag duality) in the form of the Tomita-Takesaki modular
theory of operator algebras. This is of course welcome because it maintains
the radical nature of such important future projects as the approach to
quantum gravity by elevating it to its deserved conceptual distinguished position.

In my contacts with mathematicians I often encountered a strong curiosity
about the motivations and history of the physical concepts behind the various
formalisms. In these lectures I will try to pay attention to this legitimate desire.

In the following I will give an exposition of some particle physics aspects of
the operator algebra approach, but before I start to emphasize the differences
to quantum mechanics, it is useful to present some concepts which actually
originated there. We will freely use such acronyms as QM, QT, QFT, AQFT
(algebraic QFT), LQP (local quantum physics) and OA (operator algebras).

\subsection{Restrictions of the Superposition Principle: Superselection}

It is helpful to recall briefly von Neumann world of QM and the changes it
suffered subsequently. Von Neumann form of QM was basically the quantum
mechanics of a single particle and its mathematical formulation which was that
of Weyl's reformulation of Heisenberg's commutation relation in the unitary
exponential form (for one degree of freedom)
\begin{align}
\left[  q,p\right]   &  =i\hbar\\
W(\alpha)W(\beta)  &  =e^{-i(\alpha,\beta)}W(\alpha+\beta)\nonumber\\
W(\alpha)  &  =e^{i(\alpha_{1}q+\alpha_{2}p)},\,(\alpha,\beta)=\alpha_{1}%
\beta_{2}-\alpha_{2}\beta_{1}\nonumber
\end{align}
This algebraic structure defines a unique $C^{\ast}$-algebra and the Stone-von
Neumann theorem says that there is only one regular irreducible
representation. As a consequence there is no loss of generality in the use of
the Schroedinger representation where pure states are represented by vectors
(modulo constant phase factors) in the Hilbert space of $L^{2}$-integrable
wave functions\footnote{More precisely states in QM are identified with unit
rays since the mutiplication if a vector with a phase factor does not change
the physical expectation values.} and mixed state with density matrices
(positive trace-class operators). This simple situation leads to the standard
Hilbert space setting of QT. As von Neumann pointed out, the irreducibility of
the representation of the Heisenberg-Weyl algebra in which the observables
correspond to Hermitian operators leads to the unrestricted superposition
principle: with two vectors describing physically realizable states also their
linear combinations are physically realizable (although in most cases,
different from classical wave theory, one does not know from what source such
a superposed state is produced). It was von Neumann who emphasized this
pivotal conceptual difference of quantum mechanics from classical wave optics,
a difference which even in modern textbooks is often squandered for some
superficial calculational gains which this conceptually incorrect analogy offers.

Here it we should also recall another of von Neumann's contributions to
quantum theory, namely his famous relation of commuting operators and
commutant algebras to commensurability of measurements. The totality of
operators which commute with a given set of observables forms a \textit{weakly
closed operator algebra} which in von Neumann's honor carries his name. They
belong to the larger class of operator algebras which are closed with respect
to the hermitian conjugation operation and the operator norm, whose abstract
version (forgetting the Hilbert space) is called $C^{\ast}$-algebra. Although
von Neumann algebras are special $C^{\ast}$-algebras it would not be
appropriate to subsume them under the heading of $C^{\ast}$-algebras.

The von Neumann algebra generated by an irreducible representation of the Weyl
algebra (the $C^{\ast}$-algebra of quantum mechanics) is $B(H)$, the algebra
of all bounded operators in the Hilbert space $H$.

It gradually became clear that von Neumann's mathematical framework of quantum
mechanics, which admitted unrestricted coherent superposition of state vectors
(or equivalently gave the status of an observable to each selfadjoint
operator), had to be amended in the presence of particles with different spin
and of multiparticle states. There were certain superpositions of state
vectors which cannot be physically realized for geometric reasons.

A historically famous example is that of Wick Wightman and Wigner \cite{Haag}.
They pointed out that if $\psi_{1}$ and $\psi_{2}$ are wave function of a
particle with halfinteger respectively integer spin, their coherent
superposition
\begin{equation}
\psi=\alpha_{1}\psi_{1}+\alpha_{2}\psi_{2}%
\end{equation}
which under the action of a $2\pi$-rotation changes to
\begin{equation}
\psi^{\prime}=-\alpha_{1}\psi_{1}+\alpha_{2}\psi_{2}%
\end{equation}
cannot carry a direct physical meaning since the expectation values of
unrestricted observables from the algebra of all bounded operators in Hilbert
space $B(H)$ are different in these two wave function. The observability of
the relative phase in the change $\psi_{1}\rightarrow e^{i\varphi}\psi_{1}$ of
$\psi$, which is one of the most characteristic aspects of quantum theory, is
prevented by the existence of a superselection rule: there is no observable
$A$ which can connect halfinteger and integer spins
\begin{align}
\left(  \psi_{1},A\psi_{2}\right)   &  =0\,\forall\,observables\,A\\
\Longleftrightarrow E_{\psi}(A)  &  \equiv\left(  \psi,A\psi\right)  =\left(
\psi^{\prime},A\psi^{\prime}\right)  \equiv E_{\psi^{\prime}}(A)\nonumber
\end{align}
In contrast to \textit{selection rules} for e.g. electromagnetic transitions
in atoms (relating final with initial spin of the atomic state) which suffer
corrections in higher orders, these superselection rules are exact and
therefore the prefix \textit{super} has a sound physical meaning. In their
presence the Hilbert space $\mathcal{H}$ becomes the direct sum of Hilbert
$\mathcal{H}_{i}$ spaces inside which the unrestricted superposition principle
holds
\begin{equation}
\mathcal{H}=\sum_{i}\mathcal{H}_{i}%
\end{equation}
The modern terminology is to call the labels $i$ summarily (eigenvalues of)
superselection charges (in the above example they are the $\pm$ univalence of
spin). The algebra generated by the observables consists of a direct sum of
algebras with no connecting operators between the different subspaces. The sum
could in principle also be a continuous integral in which case the similarity
with von Neumann's central decomposition into factors is not accidental.

Another related important example comes from the notion of identical particles
in multiparticle state vectors $\psi_{N}$ containing $N$ particles.
Permutations of particle labels in multiparticle states of identical particles
can only change the representing state vector but not the associated physical
state (the expectation values of observables) since they commute with the
algebra $\mathcal{A}$ generated by the observables
\begin{align}
&  \left[  U(\sigma),\mathcal{A}\right]  =0,\,\sigma\in S_{N}\\
&  U(\sigma)\psi_{N}=\psi_{N}^{\sigma},\,E_{\psi_{N}}(A)=E_{\psi_{N}^{\sigma}%
}(A)\nonumber
\end{align}
The different irreducible representations of the permutation group $S_{N}$ are
conveniently depicted in terms of \ Young tableaus. The above commutation
relations of permutations with observables which express the
indistinguishability of identical particles impose a superselection rule
between inequivalent representations belonging to different Young tableaus.
The standard argument in the QM literature to explain that \textit{only
abelian permutation group representations} are realized in nature is based on
a fallacious tautology: it uses tacitly Schur's lemma which is of course
synonymous with the triviality of the commutant and hence with the claimed
abelianess of the representations of permutations. A well-known mathematical
counterexample is obtained by imagining a quantum mechanics in which the spin
is not accessible to measurements\footnote{Spin is of course a spacetime
symmetry directly accessible to (Stern-Gerlach) experiments, whereas inner
symmetries (isospin and generalizations) are not.}. In such a world of hidden
spin degrees of freedom the spatial wave functions belong to different
symmetry-types with orbital Young tableaus which are conjugate to the hidden
N-particle spin Young tableaus (so that the antisymmetry refers to the tensor
product space: orbitals$\otimes$spin). Since $s=\frac{1}{2}$ states N-particle
belong to hight 2 tableaus, it is easy to see that the possible spin tableaus
(and therefore also the conjugate orbital ones) are uniquely determined by the
various values of the total spin. The energy eigenvalues of the different
symmetry types are generally different even without a spin dependent
interaction. In this example the nonabelian ``parastatics'' of the different
contributing permutation group representations can of course be reprocessed
back into ordinary Fermion statistics by reintroducing the spin multiplicity
(which was lost by assuming nonobservability i.e. averaging over spin degrees
of freedom). But the general statement, that it is always possible to convert
parastatistics into Fermi/Bose statistics (plus multiplicities for an internal
symmetry group to act on), is one of the most nontrivial theorems in particle
physics. For its proof one needs the full power of the superselection theory
in local quantum physics as well as some more recent group theoretical tools
\cite{DR}.

In local quantum physics\footnote{It should be clear by now that we prefer to
use the terminology LQP instead of QFT whenever we want the reader not to
think primarily about the standard text-book formalism (Lagrangian
quantization, functional integrals) but rather about the underlying physical
principles and alternative conceptually more satisfactory implementations
\cite{Haag}.} (LQP) superselected charges, despite their global aspects have a
local origin which complies with the (Einstein-) causality and spectral
stability principles; as in the classical theory of Maxwell and Einstein the
Global originates from the Local. Therefore ``Topological Field Theories'' are
not directly physical since they originate from LQP by ignoring localization
aspects. The only way to interpret them physically is to remember the physical
``flesh'' of localization and transportability which was separated from these
topological ``bones''. A good example is provided by the reference
endomorphisms and the intertwiner formalism of the next section (and the
ensuing Markov traces on the infinite braid group which have a natural
extension to the mapping class group and 3-manifold invariants). In fact
topological field theories are only topological from a differential geometric
viewpoint, whereas in LQP the terminology ``combinatorial'' would be more appropriate.

In fact one could define LQP as being the theory of spacetime dynamics of
local densities of superselected charges. It turns out that the localized
version of these charges constitute the backbone of the observable algebra
which, as we will see in more detail in the next section, is described by a
map (a net) of spacetime regions into $C^{\ast}$- algebras. Nothing turns out
to be lost if we define this algebra in terms of its vacuum representation
i.e. as a map into concrete operator algebras in a common Hilbert space which
contains a distinguished vacuum vector. On the present level of understanding
of observable nets there is also no loss if we assume that the individual
spacetime-indexed operator algebras are weakly closed i.e. are von Neumann
algebras. In fact the physically admissable representations of this algebra
are just the localized representations of the observable vacuum net.

Apart from the above quantum mechanical superselection rules of multiparticle
statistics and the before mentioned Wick-Wightman-Wigner univalent spin
superselection rules, the only other mechanism for encountering interesting
superselection sectors in QM is by leaving the setting of Schroedinger quantum
mechanics and admitting topologically nontrivial configuration spaces. Take
for example quantum mechanics on a circle (instead of a line) or in more
physical terms the Aharonov-Bohm effect i.e. the quantum physics in the vector
potential associated with a stringlike idealized solenoid generating a
$\delta$-function-like magnetic flux of strength $\theta$ through the x-y
plane in the z-direction of the solenoid (a situation which shares the same
nontrivial topology with the circle). In this case the quantum mechanics is
not unique but rather depends on an angle $\theta$ which in the A-B case has
the physical interpretation of the magnetic strength. The intuitive reason
(which can be made mathematically rigorous) is that if one realizes the
maximal abelian subalgebra generated by the multiplication operator
$x\,mod\,2\pi$ (the angular coordinate) on the space of periodic wave
functions $L^{2}(S^{1})$ then the canonical conjugate p is equal to
$-i\partial_{x}$ plus an operator which commutes with an irreducible system x
and, p and hence is central-valued and therefore a numerical constant say
$\theta$ in each irreducible representation, so that $p=-i\partial_{x}%
+\theta.$ If the x-space would be the real axis, the $\theta\,$ can be
transformed away by a nonsingular unitary ``gauge transformation'' but the
topology prevents its elimination in this case. An equivalent formulation
would be to keep the $x$ and $p$ as in the Schroedinger representation and to
encode the $\theta$ as a quasiperiodicity angle into wave functions (which
hereby turn into trivializing wave sections in a complex line bundle on the
circle). But the way which would suit our present purpose (which was to find
illustrations of superselection rules in QM) best would be to encode the
quantum mechanics on the circle into an abstract ($\theta$-independent)
$C^{\ast}$-algebra and have the $\theta$ appear as a representation label for
the various irreducible Hilbert space representations of this algebra. This
can be done, but we will not pursue his matter, since our main interest is QFT
which offers a quite different mechanism for obtaining superselection-sectors
as equivalence classes of irreducible representations of $C^{\ast}$algebras,
namely the mechanism of infinite degrees of freedoms which will be illustrated
in the sequel.

Let us look at a standard example (following \cite{Fred}) which requires only
a modest amount of concepts from physics namely spins on a linear lattice, or
mathematically a tensor product of an arbitrary large number $N$ of
two-dimensional matrix-algebras $Mat_{2}(\mathbb{C})=alg\left\{  \sigma
_{i},\mathbf{1}|i=1,2,3\right\}  ;$ here the right hand side is the
physicists's notation for this complex algebra in terms of the three hermitian
Pauli matrices which together with the identity form a linear basis of the
space $Mat_{2}(\mathbb{C})$. The pure states on this algebra are described in
terms of the unit rays associated with Hilbert space vectors in a
two-dimensional Hilbert space $H_{2}.$ In this simple case one even has an
explicit parametrization of all density matrices (mixed states) in terms of a
3-dim. unit ball $\vec{n}^{2}\leq1$
\begin{equation}
\omega(A)=\frac{1}{2}Tr(\mathbf{1}+\vec{n}\vec{\sigma})A,\,\,A\in
Mat_{2}(\mathbb{C})
\end{equation}
where the pure states reside on the surface of the unit ball. The object of
our interest is the tensor product algebra
\begin{equation}
\mathcal{A}_{N}=\otimes^{N}Mat_{2}(\mathbb{C})=Mat_{2^{N}}(\mathbb{C})
\end{equation}
acting irreducibly on the tensor Hilbert space $H_{2^{N}}=\otimes^{N}H_{2}.$
Since we are interested in the ``thermodynamic'' (inductive) limit, we first
define the infinite dimensional Hilbert space in which the limiting algebra
can act as an operator algebra
\begin{equation}
H=\left\{  \sum_{s}c(s)\left|  s\right\rangle \,|\,\sum_{s}\left|
c(s)\right|  ^{2}<\infty,\,s:\mathbb{Z}\rightarrow\left\{  \pm1\right\}
\right\}
\end{equation}
i.e. we choose a (nonseparable) Hilbert space spanned by a basis of binary
sequences with the generators of the algebra being Pauli-matrices labeled by
the points of the linear chain $x\in\mathbb{Z}$%

\begin{align}
\sigma_{3}(x)\left|  s\right\rangle  &  =s(x)\left|  s\right\rangle
\label{act}\\
\sigma_{1}(x)\left|  s\right\rangle  &  =\left|  s^{\prime}\right\rangle
\nonumber\\
\sigma_{2}(x)\left|  s\right\rangle  &  =is(x)\left|  s^{\prime}\right\rangle
\nonumber
\end{align}
with $s^{\prime}(y)=s(y)\forall y$ except for $y=x$ where $s^{\prime
}(x)=-s(x).$ Although the limiting Hilbert space is nonseparable, the
inductive limit algebra $\mathcal{A}$ remains separable and simple (no ideals)
even after its uniform closure (this is a rather general property of
inductively defined limiting $C^{\ast}$-algebras); a basis is given by
\begin{equation}
\sigma_{k_{1}}(x_{1})...\sigma_{k_{N}}(x_{j}),\,x_{1}<...<x_{k}%
\end{equation}
i.e. a product of Pauli-matrices at finitely many chain pointswhich acts on
$H$ according to (\ref{act}). The big nonseparable Hilbert space decomposes
into a noncountable sum of separable Hilbert spaces that are invariant under
the $\mathcal{A}$-action
\begin{align}
H  &  =\oplus_{\left[  s\right]  }H_{\left[  s\right]  }\\
H_{\left[  s\right]  }  &  =\overline{\mathcal{A}\left|  s\right\rangle
}\nonumber
\end{align}
where one equivalence class evidently consists of binary sequences which have
common two-sided ``tails'' $s\sim s^{\prime}$ if $s(x)=s^{\prime}(x)$ for
sufficiently large $x\in\mathbb{Z}.$ By applying $\mathcal{A}$ to a vector in
$H_{\left[  s\right]  }$ one cannot change the two-sided ``tail'' i.e.
\begin{equation}
\left\langle s^{\prime}\left|  A\right|  s\right\rangle =0,\,\,\forall
A\in\mathcal{A},\,\left[  s\right]  \neq\left[  s^{\prime}\right]
\end{equation}
A physicist would use the magnetization to show the presence of superselection
rules, e.g. he would consider the sequence of averaged magnetization in a
piece of the chain of size 2n
\begin{equation}
M_{n}=\frac{1}{2n+1}\sum_{x=-n}^{n}\sigma_{3}(x)
\end{equation}
As a result of the decreasing pre-factor in the magnetization $M_{n}$ commutes
for n$\rightarrow\infty$ with any basis element of $\mathcal{A}.$ Since
vectors $\left|  \psi_{\pm}\right\rangle \in H_{\left[  s_{\pm}\right]  }$
$s_{\pm}(x)=\pm1,\,\forall x$ are eigenvectors of $M_{n}$ with eigenvalues
$\pm1$ we have
\begin{align}
&  \left\langle \psi_{+}\left|  A\right|  \psi_{-}\right\rangle
=lim_{n\rightarrow\infty}\left\langle \psi_{+}\left|  M_{n}A\right|  \psi
_{-}\right\rangle =\\
&  lim_{n\rightarrow\infty}\left\langle \psi_{+}\left|  AM_{n}\right|
\psi_{-}\right\rangle =-\left\langle \psi_{+}\left|  A\right|  \psi
_{-}\right\rangle \nonumber\\
&  \curvearrowright\left\langle \psi_{+}\left|  A\right|  \psi_{-}%
\right\rangle =0\nonumber
\end{align}
This illustration shows clearly the mechanism by which infinitely many
degree's of freedom generate superselection rules: there a too many different
configurations at infinity which cannot be connected by operators from the
quasilocal $\mathcal{A}$ (the uniform limit of local operations); whereas the
$\pm$ magnetization (and certain others as alternating anti-ferromagnetic
states) sectors with respect to a chosen spin quantization direction (in our
case the 3-direction) have a clearcut physical meaning and can take on the
role of ground state vectors of suitably chosen dynamical systems with
(anti)ferromagnetic interactions, this is not the case for most of the myriads
of other sectors.

This also makes clear that infinite degrees of freedom in quantum systems do
not only lead to inequivalent representations, but also that there are far too
many of them in order to be physically relevant. One needs a selection
principle as to what states are of physical interest. In particle physics
Einstein causality of the observable algebra and a suitable definition of
localization of states relative to a reference state (the vacuum) constitute
the cornerstone of what is often appropriately referred to as ``Local Quantum
Physics'' (LQP). How to prove theorems and derive significant results in such
a framework will be explained in the second lecture. Whereas the topological
superselections in QM which we illustrated by the Aharonov-Bohm model
(mathematically the QM on a circle) are the result of an over-idealization
(infinitely thin stringlike solenoids, in order to reach mathematical
simplicity\footnote{The problem with a ``physical'' solenoid of finite
extension (which creates a less than perfect homogeneous magnetic field with a
small contribution outside the solenoid) is a complicated boundary value
problem \textit{within the Schroedinger theory} and hence subject to the
Stone- von Neumann uniqueness. In the idealized limit the field strength
becomes encoded into the topological $\theta$-angle.}), the so-called vacuum
polarization nature of the all-pervading vacuum reference state makes the
infinite degree of freedom aspect of LQP an immutable physical reality. In the
scaling limit of LQP which leads to conformal quantum field theory, there is
also a topological aspect which enters through the compactification of
Minkowski spacetime and the ensuing structure of the algebra (see the third
section). But this mechanism is quite different and more fundamental than the
over-idealizations of the Aharonov-Bohm solenoid. The typical situation is
that the $C^{\ast}$-algebra $\mathcal{A}$ describing the observables of a
system in LQP has a denumerable set of superselection sectors and that the
full Hilbert space which unites all representations is a direct sum
\begin{equation}
H=\bigoplus H_{i}%
\end{equation}
This decomposition is reminiscent of the decomposition theory of group
algebras of compact groups apart from the fact that for the superselection
sectors in LQP there exist ``natural'' intertwining operators which transfer
superselection charges and connect the component spaces $H_{i}$ (this has no
analog in group theoretic representation theory). They are analogous to
creation and annihilation operators in Fock space which intertwine the
different N-particle subspaces.

Even though they themselves are not observables, they are nevertheless
extremely useful. The best situation one can hope for is to have a full set of
such operators which create a field algebra $\mathcal{F}$ in with no further
inequivalent representations i.e. in which all representation labels
(``charges'') have become inner (charges within the field algebra). Note that
the word ``field'' in this context does not necessarily refer to pointlike
operator-valued distributions \cite{St-Wi} but rather to the charge-transfer
aspect of charge carrying operators which intertwine between different
superselection sectors wheras the observables by definition stay in one sector.

Doplicher and Roberts \cite{DR} proved that observable algebras in four
spacetime dimensions indeed allow a unique construction (after imposing the
conventions of ``normal commutation'' relations between operators carrying
different superselected charges \cite{St-Wi}) of such a field algebra
$\mathcal{F}$ from its observable ``shadow'' $\mathcal{A}$. This
(re)construction of $\mathcal{F}$ from $\mathcal{A}$ is therefore reminiscent
of Marc Kac's famous aphorism about a mathematical inversion problem which
goes back to Hermann Weyl namely: ``how to hear the shape of a drum?'' The
most startling aspect of their result is that the inclusion\footnote{Both
algebras consist of spacetime indexed subalgebras (nets) and for each there is
an inclusion $\mathcal{A(O)}\subset\mathcal{F(O)}.$} of the two nets
$\mathcal{A\subset F}$ (apart from low-dimensional QFT) is completely
characterized by the category of compact groups and that for each compact
topological group there is a pair ($\mathcal{A}\subset\mathcal{F}$) within the
setting of LQP which has the given group $G$ as the fixed point group of
$\mathcal{F}$%
\begin{equation}
\mathcal{A}=\mathcal{F}^{G}%
\end{equation}
This equality also holds for each algebra in the net i.e. $\mathcal{A(O)}%
=\mathcal{F}^{G}(\mathcal{O}).$

This observation goes a long way to \textit{de-mystify the concept of inner
symmetries} (``inner'' in the physicists sense refers to symmetries related to
superselected charges which commute with spacetime symmetries, whereas
operator algebraist use inner/outer for unitarily
implementable/nonimplementable auto- or endo- morphisms) which originated from
Heisenberg's phenomenological introduction of isospin in nuclear physics and
played a pivotal role through its group action on multiplicity indices of
multicomponent Lagrangian fields. What results is an (presumably even for
mathematicians) unexpected new road to group theory and group representations
\cite{DR} via a ``group dual'' which is very different from that of the
well-known Tanaka-Krein theory. The DR form of the group dual in turn emerges
from the DHR superselection sector analysis, i.e. from an input which consists
only of Einstein causality (known to mathematicians through its classical
manifestation from relativistic wave propagation in the context of partial
differential equation) and spectral stability (energy positivity).

In LQP of lower spacetime dimensions the rigid separation between spacetime
and inner symmetries looses its meaning\footnote{Whereas in higher dimensions
($d\geq1+3$) with B/F statistics inner (compact group) and outer (spacetime)
symmetries cannot be nontrivially ``married'' \cite{Haag}, in low dimensions
they cannot be unequivocally ``divorced''.} and one is entering the realm of
subfactor theory of V. Jones (which in a certain sense constitutes an
extension of group theory), with braid group statistics being the main
physical manifestation. Again a short interlude concerning the physicists use
of the word ``statistics'' may be helpful. Historically the main physical
manifestation of the difference between spacelike commuting Boson and
anticommuting Fermi fields (or between (anti)symmetrized multiparticle tensor
product spaces) was observed in the thermodynamical behavior resulting from
statistical ensembles of such particles i.e. from their statistical mechanics.
Even in situations in which there is no statistical mechanics involved,
physicist continue to use the word ``statistics'' for the characterization of
commutation relations of the fields which describe those particles.

Here it is helpful to remember a bit of history on the mathematical side.
Group theory originated from Galois studies of inclusion of a (commutative)
number field into an extended field (extended by the roots of a polynomial
equation). The new subfactor theory in some way generalizes this idea to
inclusions of particular families of nonabelian algebras. At the threefold
junction between abstract quantum principles, the geometry of spacetime and
inner symmetries stands one of the most startling and impressive mathematical
theories: the Tomita-Takesaki modular theory, which in the more limited
context of thermal aspects of open quantum systems was independently
discovered by physicists \cite{Bo}. It is, according to the best of my
knowledge, the only theory capable to convert abstract domains of quantum
operators and ranges of operator algebras within the algebra of all bounded
operators  $B(H)$ of the underlying Hilbert space into geometry and spacetime
localization; and although this is not always obvious, this is also behind the
geometrical aspects of subfactor theory.

The standard Hilbert space setting which one learns in a course of QM is only
a sufficient tool if the algebraic structure of observables allows for only
one (regular) representation as the p's and q's (encoded into the Weyl
C$^{\ast}$-algebra of QM). Whereas it is possible to present the standard
particle theory in terms of computational recipes in this restricted setting
(as it is in fact done in most textbooks), the extension of LQP into yet
unexplored directions of particle physics requires a somewhat broader basis,
including additional mathematical concepts.

In passing we mention that superselection rules also play a role in an
apparently quite different area of fundamental quantum physics. It is commonly
accepted that the Schr\"{o}dinger cat paradox of QM (which is a dramatic
setting of the von Neumann ``reduction of the wave packet'' dictum) becomes
more palatable through the idea of a decoherence process in time which is
driven by an environment i.e. there is a transition process in time from a
pure quantum state of the object to a mixed (in a classical sense) state, and
that for all practical purposes the coherence in the superposition
$\psi_{live}+\psi_{dead}$ will have been lost in the limit of infinite time as
a result of interactions with the infinite degrees of freedom of the
environment. Ignoring the environment and the decoherence time, this takes on
the form of the von Neumann wave packet collapse
\begin{equation}
\psi\overset{collapse}{\longrightarrow}\,\left|  P\psi\right\rangle
\left\langle P\psi\right|  +\left|  \left(  1-P\right)  \psi\right\rangle
\left\langle \left(  1-P\right)  \psi\right|  \overset{pointer}{\underset
{reading}{\longrightarrow}}\left\{
\begin{array}
[c]{c}%
P\psi\,with\,\;prob.\,\left\|  P\psi\right\|  ^{2}\,\,or\\
\left(  1-P\right)  \psi\,with\,\,prob.\,\left\|  \left(  1-P\right)
\psi\right\|  ^{2}%
\end{array}
\right.
\end{equation}
where for simplicity we assumed that the observable is a projection operator
$P.$ The mixed state on the right hand side has been represented as a mixture
of two orthogonal components, but unlike a pure state a mixture has myriads of
other (nonorthogonal) representations and it is not very plausible that in
addition to the collapse into a mixed state, the measurement also selects that
orthogonal representation of the mixture among the myriads of other
possibilities. Rather one believes that the measurement causes an interaction
with the infinite degrees of freedom of the environment (the finite degrees of
freedom of a quantum mechanical object may itself be the result of an
idealization) \ and that what is observed at the end is a mixture of
superselected states of the big system. Since the central decomposition in
contrast to the aforementioned orthogonal mixture has an unambiguous classical
meaning, this is the more satisfactory explanation of the decoherence process
at infinite times \cite{Lands}. Whereas it is quite easy to write down unitary
time propagators which become isometric in the limit (or positive maps which
generate mixtures already in finite time), it is a much more difficult task to
create a realistic infinite degree of freedom model which describes the above
features of a measurement process in a mathematically controllable way.

There is as yet no agreement about whether the full system (including the
spacetime of the part of the universe which is spacelike separated from the
laboratory during the finite duration of the measurement) remains in a pure
state or undergoes a ``reduction'' into a mixture; it is not even clear
whether this question belongs still to physics or is of a more philosophical
nature. One idea which however resisted up to now a good conceptual and
mathematical understanding is the possible existence of a more complete form
of quantum theory which in addition to the standard dynamics also contains a
process of ``factualization'' of events i.e. an interface between the
potentiality of the standard (Copenhagen) interpretation and the
\textit{factuality of observed spacetime localized events} on a very basic
level \cite{Haag}. It seems that some of the oldest problems in quantum theory
related to the quantum mechanical measurement process is still very much alive
and that as a result of the importance of an infinite degree of freedom
environment and locality, the theory of local quantum physics (which
automatically generates the omnipresent environment of the spacelike separated
infinite degree of freedom via the vacuum polarization property) may yet play
an important role in future investigations.

It should be clear from this birds eye view of motivation and content of LQP
that the properly adapted Einstein causality concept, which in its classical
version originated at the beginning of the last century, still remains the
pillar of the present approach to particle physics. With its inexorably
related vacuum polarization structure and the associated infinite number of
degrees of freedom it has given LQP its distinct fundamental character which
separates it from QM. Despite its startling experimentally verified
predictions and despite theoretical failures of attempts at its nonlocal
modifications, its conceptual foundations for 4-dimensional interacting
particles are presently still outside complete mathematical control.

\subsection{Appendix A: The Superselection Sectors of \textbf{C}G}

As a mathematical illustration of superselection rules we are going to explain
the representation theory of \ (finite) group algebras using the setting of
superselection sectors. In this way the reader becomes acquainted with the
present notation and mode of thinking for a situation he may have already
encountered in a different way.

Let $G$ be a (not necessarily commutative) finite group. We affiliate a
natural $\mathbf{C}^{\ast}$-algebra, the group-algebra $\mathbf{C}G$ with $G $
in the following way:

\begin{itemize}
\item (i)\thinspace\thinspace\thinspace\thinspace\thinspace\thinspace The
group elements g$\in G$ including the unit e form the basis of a linear
vectorspace over $\mathbf{C}$:
\begin{equation}
x\in\mathbf{C}G,\,\,\,\,\,x=\sum_{g}x(g)g\,,\,\,\,\,with\hbox{ }%
x(g)\in\mathbf{C}%
\end{equation}

\item (ii)\thinspace\thinspace\thinspace\thinspace\thinspace This finite
dimensional vector space $\mathbf{C}G$ inherits a natural convolution product
structure from G:
\begin{equation}
\left(  \sum_{g\in G}x(g)g\right)  \cdot\left(  \sum_{h\in G}y(h)h\right)
=\sum_{g,h\in G}x(g)y(h)g\cdot h=\sum_{k\in G}z(k)k
\end{equation}%
\[
with\,\,z(k)=\sum_{h\in G}x(kh^{-1})y(h)=\sum_{g\in G}x(g)y(g^{-1}k)
\]

\item (iii)\thinspace\thinspace\thinspace\thinspace A *-structure, i.e. an
antilinear involution:
\begin{equation}
x\rightarrow x^{*}=\sum_{g\in G}x(g)^{*}g^{-1}\,\,,\,\,\,\,i.e.x^{*}%
(g)=x(g^{-1})^{*}%
\end{equation}
Since\thinspace\thinspace:
\begin{equation}
\left(  x^{*}x\right)  \left(  e\right)  =\sum_{g\in G}\left|  x(g)\right|
^{2}\geq0,\,\,\,\,(=iff\,\,\,\,\,x=0)
\end{equation}
\thinspace\thinspace\thinspace\thinspace this *- structure is nondegenerate
and defines a positive definite inner product:
\[
\left(  y,x\right)  \equiv(y^{*}x)(e)
\]

\item (iv)\thinspace\thinspace\thinspace\thinspace The last formula converts
$\mathbf{C}G$ into a Hilbert space and hence, as a result of its natural
action on itself, it also gives a $C^{\ast}$ norm (as any operator algebra):
\begin{equation}
\left|  \left|  x\right|  \right|  =\sup_{\left\|  y\right\|  =1}\left\|
xy\right\|  ,\,\,\,\,\,\,\,\,C^{\ast}-condition:\,\left|  \left|  x^{\ast
}x\right|  \right|  =\left|  \left|  x^{\ast}\right|  \right|  \left|  \left|
x\right|  \right|
\end{equation}
A $\mathbf{C}^{\ast}-$norm on a *-algebra is necessarily unique (if it exists
at all). It can be introduced through the notion of spectrum.
\end{itemize}

It is worthwhile to note that (iii) also serves to introduce a tracial state
on $\mathbf{C}G$ i.e. a positive linear functional $\varphi\,$with the trace
property:
\begin{equation}
\varphi\left(  x\right)  :=x(e),\,\,\,\,\varphi(x^{\ast}x)\geq0,\,\,\,\varphi
(xy)=\varphi(yx)
\end{equation}
This state (again as a result of (iii)) is even faithful, i.e. the scalar
product defined by:
\begin{equation}
\left(  \hat{x},\hat{y}\right)  :=\varphi(x^{\ast}y)
\end{equation}
is nondegenerate. On the left hand side the elements of $\mathbf{C}G$ are
considered as members of a vector space. The nondegeneracy and the
completeness of the algebra with respect to this inner product (a result of
the finite dimensionality of $\mathbf{C}G$) give a natural representation (the
regular representation of $\mathbf{C}G$) on this Hilbert space:
\begin{equation}
x\hat{y}:=\widehat{xy}%
\end{equation}
The norm of these operators is identical to the previous one.

This construction of this ``regular'' representation $\lambda_{reg}$ from the
tracial state on the $\mathbf{C}^{\ast}$-group-algebra is a special case of
the general Gelfand-Neumark-Segal (GNS-)construction presented in a later section.

Returning to the group theoretical structure, we define the conjugacy classes
$K_{g}\,$and study their composition properties.
\begin{equation}
K_{g}:=\left\{  hgh^{-1},h\in G\right\}
\end{equation}
In particular we have $K_{e}$=$\left\{  e\right\}  $. These sets form disjoint
classes and hence:
\begin{equation}
G=\cup_{i}K_{i},\,\,\,\left|  G\right|  =\sum_{i=0}^{r-1}\left|  K_{i}\right|
,\,\,\,\,K_{e}=K_{0},\,\,\,K_{1,}....K_{r-1},\,\,\,\,r=\#classes
\end{equation}
We now define central ``charges'':
\begin{equation}
Q_{i}:=\sum_{g\in K_{i}}g\,\in\mathcal{Z}(\mathbf{C}G):=\left\{  z,\,\left[
z,x\right]  =0\,\,\,\,\,\forall x\in CG\right\}  \,\, \label{QB}%
\end{equation}
It is easy to see that the center $\mathcal{Z}(\mathbf{C}G)$ consists
precisely of those elements whose coefficient functions $z(g)$ are constant on
conjugacy classes i.e. $z(g)=z(hgh^{-1})$ for all h. The coefficient functions
of $Q_{i}$:
\begin{equation}
Q_{i}(g)=\left\{
\begin{array}
[c]{c}%
1\,\,\,\,if\,\,\,g\in K_{i}\\
0\,\,\,otherwise
\end{array}
\right.
\end{equation}
evidently form a complete set of central functions. The composition of two
such charges is therefore a linear combination of the r independent
$Q_{i}^{\prime}s$ with positive integer-valued coefficients (as a result of
the previous formula (\ref{QB})):
\begin{equation}
Q_{i}Q_{j}=\sum_{l}N_{ij}^{l}Q_{l}%
\end{equation}
The fusion coefficients $N$ can be arranged in terms of $r$ commuting
matrices
\begin{equation}
\mathbf{N}_{j},\,\,\,with\,\,\left(  \mathbf{N}_{j}\right)  _{i}^{l}%
=N_{ij}^{l}%
\end{equation}
The associativity of the 3-fold product $QQQ$ is the reason for this
commutativity, whereas the the $N_{j\text{ }}$ would by symmetric matrices iff
the group itself is abelian.

Functions on conjugacy classes also arise naturally from characters $\chi$ of
representations $\pi$
\begin{equation}
\chi^{\pi}(g)=Tr\pi(g)\,,\,\,\,\,\,\,\chi^{\pi}(g)=\chi^{\pi}(hgh^{-1})
\end{equation}
This applies in particular to the previously defined left regular
representation $\lambda$ with $\left(  \lambda_{g}x\right)  \left(  h\right)
=x(g^{-1}h).$ Its decomposition in terms of irreducible characters goes hand
in hand with the central decomposition of\textbf{\ }$\mathbf{C}G$:
\begin{equation}
\mathbf{C}G=\sum_{l}P_{l}\mathbf{C}G,\,\,\,\,\,\,\,\,\,Q_{i}=\sum_{l}Q_{i}%
^{l}P_{l}%
\end{equation}
The central projectors $P_{l}$ are obtained from the algebraic spectral
decomposition theory of the $Q_{i}^{\prime}s\,$ by inverting the above
formula. The ``physical'' interpretation of the coefficients is: $Q_{i}%
^{l}=\pi_{l}(Q_{i})$ i.e. the value of the $i^{th}$ charge in the $l^{th}$
irreducible representation. The $P_{l}$ are simply the projectors on the
irreducible components contained in the left regular representation. Since any
representation of $G$ is also a representation of the group algebra, every
irreducible representation must occur in $\lambda_{reg}(\mathbf{C}G)$. One
therefore is supplied with a complete set of irreducible representations, or
in more intrinsic terms, with a complete set of $r$ equivalence classes of
irreducible representations. As we met the intrinsic (independent of any basis
choices) fusion rules of the charges, we now encounter the intrinsic fusion
laws for equivalence classes of irreducible representations.
\begin{equation}
\pi_{k}\otimes\pi_{l}\simeq\sum_{m}\tilde{N}_{kl}^{m}\pi_{m}%
\end{equation}
Whereas the matrix indices of the $N^{\prime}s$ label conjugacy classes, those
of $\tilde{N}$ refer to irreducible representation equivalence classes. The
difference of these two fusions is typical for nonabelian groups and
corresponds to the unsymmetry of the character table: although the number of
irreducible representations equals the number of central charges
(=\thinspace\# conjugacy classes), the two indices in $\pi_{l}(Q_{j})$ have a
different meaning. With an appropriate renormalization this mixed matrix which
measures the value of the $j^{th}$ charge in the $l^{th}$ representation we
obtain the unitary \textit{character matrix} $S_{lj}\equiv\sqrt{\frac{\left|
K_{j}\right|  }{\left|  G\right|  }}Tr\pi_{l}(g_{j})$ (Tr is the normalized
trace) which diagonalizes the commuting system of $N^{\prime}s$ as well as
$\tilde{N}^{\prime}s$:
\begin{align}
\frac{S_{kj}}{S_{0j}}\frac{S_{lj}}{S_{0j}}  &  =\sum_{m}\tilde{N}_{kl}%
^{m}\frac{S_{mj}}{S_{0j}},\quad\label{S}\\
\frac{\sqrt{\left|  K_{j}\right|  }S_{ki}}{S_{k0}}\frac{\sqrt{\left|
K_{b}\right|  }S_{kj}}{S_{k0}}  &  =\sum_{c}N_{ij}^{c}\frac{\sqrt{\left|
K_{c}\right|  }S_{kc}}{S_{k0}}\quad\nonumber
\end{align}
The surprise is that $S$ shows up in two guises, once as the unitary which
diagonalizes this $\tilde{N}_{l}(N_{b})$-system, and then also as the system
of eigenvalues $\frac{S_{la}}{S_{0a}}\,(\frac{S_{kb}}{S_{k0}})$ which can be
arranged in matrix form. We will not elaborate on this point. In section 3.2
we will meet an analogous situation outside of group theory which is symmetric
in charge- and representation labels i.e. $N=\tilde{N}$ and $S$ is a symmetric
matrix,yet the composition of representations is not commutative.

In passing we mention that closely related to the group algebra $\mathbf{C}G$
is the so-called ``double'' of the group (Drinfeld):
\begin{equation}
D(G)=C(G)\Join_{ad}G
\end{equation}
In this crossed product designated by $\bowtie$ , the group acts on the
functions on the group $C(G)$ via the adjoint action:
\begin{equation}
\alpha_{h}(f)(g)=f(h^{-1}gh)
\end{equation}
The dimension of this algebra is $\left|  G\right|  ^{2}$ as compared to
dim$\mathbf{C}G=\left|  G\right|  $. Its irreducible representations are
labeled by pairs $(\left[  \pi_{irr}\right]  ,K)$ of irreducible
representation and conjugacy class and therefore their matrices $N$ and $S$
are \textit{selfdual}. In this sense group doubles are ``more symmetric'' than
groups. In chapter 7 we will meet selfdual matrices $S$ which cannot be
interpreted as a double of a group and which resemble the $S$ of abelian groups.

Returning to the regular representation we notice that the equivalence classes
of irreducible representations appear with the natural multiplicity:
\begin{equation}
mult(\pi_{l}\,\,in\,\,\lambda_{reg})=\,dim\pi_{l}%
\end{equation}
The results may easily be generalized to compact groups where they are known
under the name of Peter-Weyl theory.

Since group algebras are very special, some remarks on general finite
dimensional algebras are in order.

Any finite dimensional $C$*-algebra $\mathcal{R}$ may be decomposed into
irreducible components, and any finite dimensional irreducible C*-algebra is
isomorphic to a matrix algebra $Mat_{n}(\mathbf{C})$ . If the irreducible
component $Mat_{n_{i}}(\mathbf{C})$ occurs with the multiplicity $m_{i}$ , the
algebra $\mathcal{R}$ has the form is isomorphic to the following matrix
algebra:
\begin{equation}
\mathcal{R}=\bigoplus_{i}Mat_{n_{i}}(\mathbf{C})\otimes1_{m_{i}}%
\,\,\,\,\,\,in\,\,\,\mathcal{H}=\oplus_{i}\mathcal{H}_{n_{i}}\otimes
\mathcal{H}_{m_{i}}%
\end{equation}
and the multiplicities are unrelated to the dimensionalities of the
components. The commutant of $\mathcal{R}$ in $\mathcal{H\,\,}$ is:
\begin{equation}
\mathcal{R}^{\prime}=\oplus_{i}1_{n_{i}}\otimes Mat_{m_{i}}(\mathbf{C}%
)\,,\,\,\,\,\,\,\,\,Z:=R\cap R^{\prime}=\oplus_{i}\mathbf{C\cdot}1_{n_{i}%
}\otimes1_{m_{i}}%
\end{equation}
The last formula defines the center.

\subsection{Appendix B: Some Operator Algebra Concepts}

Since operator algebras still do not quite belong to mainstream mathematics,
it would be unrealistic for me to assume that a mainly mathematical audience
is familiar with allthe mathematical concepts which I will use in these
lectures. Whereas mathematicians usually have a stock of very profound
knowledge which covers a rather small specialized region, the mathematical
physicists knowledge tends to be better described by the Fouriertransform of
the former. This is because a theoretical physicist cannot indulge in the
luxury of being highly mathematically selective. Contrary to mathematicians he
has to live at least part of his life with half-truths (however without ever
loosing the urge to convert them into full truths). If one wants to understand
the physical nature one has to be prepared for the unexpected and to create a
large supply of mathematical knowledge for all potential future physical
applications would be totally unrealistic. Unsolved problems in mathematical
physics often cannot be that clearly formulated as e.g. Hilbert formulated the
important mathematical problems at the beginning of last century. Only if a
promising new theoretical Ansatz has been found, a physicist is willing to
learn and invest in depth into the appropriate mathematics. For quantum theory
this has occurred a long time ago and therefore the majority of mathematical
physicist know Hilbert space theory and even a bit about operator algebras.

The fact that in recent decades Fields medals have been twice awarded to
operator-algebra related work (Alain Connes, Vaughn Jones) shows however that
this area is receiving an increased attention and recognition within mathematics.

The following collection of definitions and theorems are not to be confused
with a mini-course on operator algebras. Their only purpose is a reminder of
the kinds of objects I will use and to urge the uninitiated reader to consult
some of the existent literature on the subject.

The objects to be represented are C$^{\ast}$-algebras $\mathcal{A}$ i.e.
normed (Banach) algebras with an antilinear involutive $^{\ast}$-operation
with the following consistency relations between them
\begin{align}
\left\|  A^{\ast}\right\|   &  =\left\|  A\right\|  ,\,\,A\in\mathcal{A}\\
\left\|  A^{\ast}A\right\|   &  =\left\|  A\right\|  ^{2}\nonumber
\end{align}
It is a remarkable fact that the norm of a $^{\ast}$-algebra (which is already
complete in that norm) is unique and solely determined by its algebraic
structure namely through the formula
\begin{equation}
\left\|  A\right\|  =inf\left\{  \rho\in\mathbb{R}_{+},\,A^{\ast}%
A-r^{2}\mathbf{1}\text{ is invertible in }\mathcal{A}\text{ }\forall
r<\rho\right\}
\end{equation}
Since only the subalgebra generated by $A^{\ast}A$ and 1 is used in this
definition, embedding of $C^{\ast}$-algebras into larger ones are
automatically isometric and homomorphisms $\phi$ are automatically contracting
i.e. $\left\|  \phi(A)\right\|  \leqslant\left\|  A\right\|  .$

A mathematical physicist often prefers the more concrete illustration of a
C$^{\ast}$-algebra as a norm-closed operator algebra in a Hilbert space. The
following two theorems show that this is no restriction of generality

\begin{theorem}
Every commutative unital C$^{\ast}$-algebra is isomorphic to the
multiplication algebra of continuous complex-valued functions on an
$L^{2}(\mathcal{M},\mu)$ where $\mathcal{M}$ is an appropriately chosen
measure space with measure $\mu$.
\end{theorem}

\begin{theorem}
Every C$^{\ast}$-algebra is isomorphic to a norm-closed $^{\ast}$-algebra of
operators in a Hilbert space
\end{theorem}

The space $\mathcal{M}$ in the first theorem is the space of maximal ideals
(closely related to the notion of spectrum which is related to the complement
of the above notion of invertibility) of $\mathcal{A}$.

For the second theorem the concept of a state $\omega$ on $\mathcal{A}$ is important.

\begin{definition}
A state $\omega$ on $\mathcal{A}$ is a linear functional on $\mathcal{A}$ with
the following properties

(1) \ $\omega(A^{\ast}A)\geqslant0\,\,\,\forall A\in\mathcal{A}\,\,\,(positivity)$

(2) \ $\omega(\mathbf{1})=1\,\,\,\,(normalization)$
\end{definition}

The relation $\left|  \omega(A)\right|  \leq\left|  A\right|  $ and hence
continuity is an easy consequence of this definition; this inequality also
allows to introduce a norm $\left\|  \omega\right\|  $ on the state space of a
$C^{\ast}$-algebra.

Each state has a representation associated with it and the canonical
construction which establishes this relation is called the GNS construction
(after Gelfand, Neumark and Segal). It basically consists in using the
positivity for defining an inner product in a linear space (defined by the
algebra)
\begin{equation}
\left(  A,B\right)  _{\omega}\equiv\omega(A^{\ast}B)
\end{equation}
It is obvious that for a faithful state (no null ideal) this would be a
positive definite inner product and on its Hilbert space closure one can
define a faithful representation with \textbf{1} a cyclic vector $\left|
\mathbf{1}\right\rangle $ for the representation
\begin{equation}
\pi(A)\left|  B\right\rangle =\left|  AB\right\rangle
\end{equation}
The presence of a null ideal $\mathcal{N}$ requires to construct the Hilbert
space from equivalence classes $A$ mod $\mathcal{N}$ which forces the positive
semidefinite inner product to become positive definite. Vice versa a cyclic
representation with a distinguished cyclic vector (in physics usually the
vacuum, a ground- or a KMS thermal- state vector) $\psi_{0}$ defines via
$\omega_{\psi_{0}}(A)=\left\langle \psi_{0}\left|  A\right|  \psi
_{0}\right\rangle $ a state.

The convex set $S(\mathcal{A})$ of states on $\mathcal{A}$ is a subset of the
C$^{\ast}$ dual $\mathcal{A}^{\ast}$ and
\begin{align}
&  S_{\pi}(\mathcal{A})\equiv\left\{  \omega\in S(\mathcal{A})|\,\exists
\rho\in B(H),\omega(A)=Tr\rho\pi(A)\right\} \\
&  \rho\,\,density\,\,matrix\,\ i.e.\,\rho\geq0,\,\,Tr\rho=1\nonumber
\end{align}
is a norm-closed subset of $S(\mathcal{A})$ called the folium associated with
$\pi.\,$\ Evidently $S$($\mathcal{A}$) decomposes into disjoint folia.

Whereas the notions of irreducibility/factoriality of a representation $\pi$
as well as unitary equivalence $\pi_{1}\simeq\pi_{2},$ and
disjointedness$\,\,\pi_{1}\overset{|}{\circ}\pi_{2}$ of two representations in
terms of spaces of intertwiners ($\pi_{1},\pi_{2}$) from $H_{1}\rightarrow
H_{2}$ are mostly familiar to mathematicians and mathematical physicist,
\begin{align*}
irred.  &  :\left(  \pi,\pi\right)  =\mathbb{C}\mathbf{1}%
\,,\,\,i.e.\,Schur^{\prime}s\,\,Lemma\\
factorial  &  :center\,\,of\,\left(  \pi,\pi\right)  =\mathbb{C}\mathbf{1}\\
unit.\,equiv.  &  :\exists\,unitary\,U\in\left(  \pi_{1},\pi_{2}\right) \\
disjoint  &  :\left(  \pi_{1},\pi_{2}\right)  =\left\{  \mathbf{0}\right\}
\end{align*}
quasiequivalence $\approx$ is less well known
\begin{equation}
\pi_{1}\approx\pi_{2}\,\Longleftrightarrow S_{\pi_{1}}(\mathcal{A})=S_{\pi
_{2}}(\mathcal{A})
\end{equation}

For a factorial representations $\pi$ is equivalent to all its
subrepresentation (think of an allegory to Greek mythology: Laokoon and the
multi-headed snake with heads growing immediately again after their beheading)
i.e. $S_{\pi}(\mathcal{A})$ does not contain a closed subfolium. Two factorial
representations are either quasiequivalent or disjoint (an extension of the
situation presented by two irreducible representations).

Von Neumann algebras originate in QFT typically through the representation
theory of $C^{\ast}$-algebras. Their Hilbert space representation
$\pi(\mathcal{A})$ in $H$ allows to take the weak closure which according to
von Neumann's famous commutant theorem is equal to the double commutant
$\pi(\mathcal{A})^{\prime\prime}$ of $\pi(\mathcal{A})$ in $H.$ The so
obtained von Neumann algebras are special weakly closed $C^{\ast}$-algebras
which have no interesting representation theory since all representations
which maintain a natural continuity property (normal representations) turn out
to be quasiequivalent and have only one ``folium of states''. The physically
relevant spacetime-indexed local algebras $\mathcal{A(O)}$ are in typical
(presumably even in all) cases hyperfinite type III$_{1}$ von Neumann factors
and hence are even unitarily equivalent, in other words there is up to unitary
equivalence only one such algebra. The importance of factors i.e. von Neumann
algebras with trivial center results from the fact that any von Neumann
algebra allows a (generally continuous) central decomposition into factors and
the latter can be classified in terms of equivalences between their projectors
and will be explained in the sequel.

The system of all projectors $\mathcal{P}(M)$ in a von Neumann algebra $M$
obeys the mathematical structure of a lattice. It is clear that unitarily
equivalent projectors should be considered as part of an equivalence class and
the first aim would be to understand the class structure. In order to have
coherence of this equivalence notion with additivity of orthogonal projectors,
one need to follow Murray and von Neumann and enlarge the class of equivalent
projectors in the following way \cite{Sunder}

\begin{definition}
Let $e,f\in\mathcal{P}(M),$ then

\begin{enumerate}
\item  the two projectors are equivalent $e\sim f$ if there exists an partial
isometry such that $e$ and $f$ are the source and range projectors:
$u^{*}u=e,\,\,uu^{*}=f$

\item $e$ is subequivalent to $f,$ denoted as $e\preceq f$ if $\exists
g\in\mathcal{P}(M)$ such that $g$ is dominated by $f$ and equivalent to
$e:e\sim g\leq f$
\end{enumerate}
\end{definition}

One easily checks that this definition indeed gives a bona fide equivalence
relation in $\mathcal{P}(M).$ Via the relation between projectors and
subspaces, these definitions and the theorems of the Murray von Neumann
classification theory can be translated into relations between subspaces. The
main advantage to restrict to factors is the recognition that \textit{any two
projectors are then subequivalent}. One calls a von Neumann algebra finite if
a projector is never equivalent to a proper subprojector. Example: in $B(H)$
infinite dimensional spaces allow a partially isometric mapping on infinite
dimensional subspaces and therefore this factor is infinite. $Mat_{n}%
(\mathbb{C})$ is of course a finite factor. It was a great discovery of Murray
and von Neumann, that there exist infinite dimensional finite factors. In fact
they defined:

\begin{definition}
A factor $M$ is said to be one of the following three types:
\end{definition}

\begin{enumerate}
\item $I,$ if it possesses pure normal states (or minimal projectors).

\item $II$, if it not of type $I$ and has nontrivial finite projectors.

\item $III,$ if there are no nontrivial finite projectors.
\end{enumerate}

Murray and von Neumann were able to refine their classification with the help
of the trace. In more recent terminology a trace without an additional
specification is a weight $Tr$ with $Trxx^{*}=Trx^{*}x$ $\forall x\in M.$ A
tracial state is a special case of a tracial weight.

The use of tracial weights gives the following refinement:

\begin{definition}
Using normal tracial weight one defines the following refinement for factors:
\end{definition}

\begin{enumerate}
\item  type$I_{n}$ if $ranTr\mathcal{P}(M)=\left\{  0,1,...,n\right\}  ,$ the
only infinite type I factor is type$I_{\infty}$. Here the tracial weight has
been normalized in the minimal projectors (for finite n this weight is in fact
a tracial state).

\item  type$II_{1}$ if the $Tr$ is a tracial state with $ranTr\mathcal{P}%
(M)=\left[  0,1\right]  $; type $II_{\infty}$ if $ranTr\mathcal{P}(M)=\left[
0,\infty\right]  .$

\item  type III if no tracial weight exists i.e. if $ranTr\mathcal{P}%
(M)=\left\{  0,\infty\right\}  .$
\end{enumerate}

In particular all nontrivial projectors (including \textbf{1}) are Murray-von
Neumann equivalent.

The classification matter rested there, up to the path-breaking work of Connes
in the 70ies which in particular led to an important gain in understanding and
complete classification of all hyperfinite type III factors.

\begin{remark}
In LQP only the type I factor and the hyperfinite type III$_{1}$ factor are
directly used. Besides the representation of the global algebra the type I
factor features in the local split property (see next section). Local (wedge-,
double cone-) algebras are of hyperfinite type III$_{1}.$ For the formulation
of the intertwiner formalism (topological field theory) of the superselection
theory one also employs tracial hyperfinite type II$_{1}$ factors as an
auxiliary tool.
\end{remark}

Although von Neumann algebras have no interesting representation theory, they
are the ideal objects for the study of properties related to the relative
positions of several of them in one common Hilbert space in particular
inclusions of one into another. A baby version of an inclusion is as follows.
Suppose that $Mat_{2}(\mathbb{C})$ acts not on its natural irreducible space
$\mathbb{C}^{2}$ but by left action on the 4-dim Hilbert space $\mathcal{H}%
(Mat_{2}(\mathbb{C}),\frac{1}{2}Tr)$ where the inner product is defined in
terms of the usual trace. In that space the commutant is of equal size and
consists of $Mat_{2}(\mathbb{C})$ acting in the opposite order from the right
which will be shortly denoted as $Mat_{2}(\mathbb{C})^{opp}$. Explicitly the
realization of $\mathcal{H}$ as $\mathbf{C}^{4}$ may be defined as
\begin{equation}
\left(
\begin{array}
[c]{ll}%
\xi_{11} & \xi_{12}\\
\xi_{21} & \xi_{22}%
\end{array}
\right)  \rightarrow\left(
\begin{array}
[c]{l}%
\xi_{11}\\
\xi_{21}\\
\xi_{12}\\
\xi_{22}%
\end{array}
\right)
\end{equation}
and the action of $\mathcal{A}=Mat_{2}(\mathbb{C})$ takes the following form:
\begin{equation}
a=\left(
\begin{array}
[c]{llll}%
a_{11} & a_{12} & 0 & 0\\
a_{21} & a_{22} & 0 & 0\\
0 & 0 & a_{11} & a_{12}\\
0 & 0 & a_{21} & a_{22}%
\end{array}
\right)  \simeq\left(
\begin{array}
[c]{ll}%
a_{11} & a_{12}\\
a_{21} & a_{22}%
\end{array}
\right)  \otimes\underline{1}%
\end{equation}
The most general matrix in the commutant $a^{\prime}\in\mathcal{A}^{\prime}$
has evidently the form:
\[
a^{\prime}=\left(
\begin{array}
[c]{llll}%
a_{11}^{\prime} & 0 & a_{12}^{\prime} & 0\\
0 & a_{11}^{\prime} & 0 & a_{12}^{\prime}\\
a_{21}^{\prime} & 0 & a_{22}^{\prime} & 0\\
0 & a_{21}^{\prime} & 0 & a_{22}^{\prime}%
\end{array}
\right)  \simeq\underline{1}\otimes\left(
\begin{array}
[c]{ll}%
a_{11}^{\prime} & a_{12}^{\prime}\\
a_{21}^{\prime} & a_{22}^{\prime}%
\end{array}
\right)
\]
The norm $\left\|  \xi\right\|  =\left(  \frac{1}{2}Tr\xi^{\ast}\xi\right)
^{\frac{1}{2}}$ is invariant under the involution $\xi\rightarrow\xi^{\ast}$
which in the $\mathbb{C}^{4}$ representation is given by the isometry:
\begin{equation}
J=\left(
\begin{array}
[c]{llll}%
K & 0 & 0 & 0\\
0 & 0 & K & 0\\
0 & K & 0 & 0\\
0 & 0 & 0 & K
\end{array}
\right)  ,\quad K:\hbox{natural conjugation in }\mathbb{C}%
\end{equation}
We have:
\begin{equation}
j(\mathcal{A}):=J\mathcal{A}J=\mathcal{A}^{\prime},\quad\hbox{antilin. map
}\mathcal{A}\rightarrow\mathcal{A}^{\prime}%
\end{equation}
which may be rewritten in terms of a linear anti-isomorphism:
\begin{equation}
a\rightarrow Ja^{\ast}J,\quad\mathcal{A}\rightarrow\mathcal{A}^{\prime}%
\end{equation}
Consider now the trivial algebra $\mathcal{B}=\mathbb{C}\mathbf{\cdot1}_{2}$
as a subalgebra of $\mathcal{A}=Mat_{2}(\mathbb{C})$. In the $\mathbb{C}^{4}$
representation the B-algebra corresponds to the subspace:
\begin{equation}
\mathcal{H}_{B}=\left\{  \left(
\begin{array}
[c]{l}%
\xi\\
0\\
0\\
\xi
\end{array}
\right)  ,\xi\in\mathbb{C}\right\}  ,\quad\mathcal{H}_{B}=e_{B}\mathcal{H}%
,\quad e_{B}=\left(
\begin{array}
[c]{llll}%
\frac{1}{2} & 0 & 0 & \frac{1}{2}\\
0 & 0 & 0 & 0\\
0 & 0 & 0 & 0\\
\frac{1}{2} & 0 & 0 & \frac{1}{2}%
\end{array}
\right)
\end{equation}
The projector $e_{B}$ commutes clearly with $\mathcal{B}$ i.e. $e_{B}%
\in\mathcal{B}^{\prime}$ . We now define a measure for the relative size of
$\mathcal{B}\subset\mathcal{A}$ the Jones index:
\[
\left[  A:B\right]  =\tau_{B^{\prime}}(e_{B})^{-1},\quad\tau:\hbox{
normalized trace in }\mathcal{B}^{\prime}%
\]
In our example $\tau(e_{B})=\frac{1}{4}(\frac{1}{2}+0+0+\frac{1}{2})=\frac
{1}{4}$ i.e. the satisfying result that the Jones index is 4. The same method
applied to the inclusion:
\begin{equation}
Mat_{4}(\mathbf{C})\supset Mat_{2}(\mathbf{C})\otimes\mathbf{1}_{2}=\left\{
\left(
\begin{array}
[c]{ll}%
X & 0\\
0 & X
\end{array}
\right)  ,X\in Mat_{2}(\mathbf{C})\right\}
\end{equation}
also gives the expected result:
\begin{equation}
\left[  A:B\right]  =\frac{\dim Mat_{4}(\mathbf{C})}{\dim Mat_{2}(\mathbf{C}%
)}=4
\end{equation}
If, as in the previous cases $B$ is a finite dimensional subfactor (i.e. a
full matrix algebra) of $A,$ the Jones index is the square of a natural
number. For inclusions of finite dimensional semisimple algebras the index
takes on more general values. For example:
\begin{align}
Mat_{2}(\mathbf{C})\oplus\mathbf{C}  &  =\left(
\begin{array}
[c]{lll}%
X &  & \\
& X & \\
&  & x
\end{array}
\right)  \subset Mat_{2}(\mathbf{C})\oplus Mat_{3}(\mathbf{C})\\
X  &  \in Mat_{2}(\mathbf{C}),\quad x\in\mathbf{C}1\nonumber
\end{align}
Here the index is 3. It is easy to see that instead of the projector formula
one may also use the incidence matrix formula:
\begin{equation}
\left[  \mathcal{A}:\mathcal{B}\right]  =\sum_{n,m}(\Lambda_{n}^{m})^{2}%
\end{equation}
The incidence matrix $\Lambda$ is describable in terms if a bipartite graph.
The number of say white vertices correspond to the number of full matrix
component algebras for the smaller algebra and the black vertices labelled by
the size of the components to the analogously labelled irreducible components
of the bigger algebra. A connecting line between the two sets of vertices
indicates that one irreducible component of the smaller is included into one
of the bigger algebra. In our case:
\begin{equation}
\Lambda=\left(
\begin{array}
[c]{ll}%
1 & 1\\
1 & 0
\end{array}
\right)  ,\quad\left|  \left|  \Lambda\right|  \right|  ^{2}=3
\end{equation}
>From a sequence of ascending graphs one obtains important infinite graphs
(Bratteli diagrams) which are very useful in the ``subfactor theory''
\cite{GHJ} which will appear in the mathematical appendix. In the infinite
dimensional case the inclusion of full matrix algebras corresponds to the
inclusion of von Neumann factors i.e. the ``subfactor problem''. In that case
the spectrum of inclusions shows a fascinating and unexpected quantization
phenomenon, the Vaughn Jones quantization formula for index $\leq4.$ AFD
(almost finite dimensional) $C$*-algebras obtained by sequences of ascending
Bratteli diagrams equipped with tracial states enter LQP via the intertwiner
algebra of charge transporters. A special case are the combinatorial theories
which result from Markov-traces on selfintertwining transporters which contain
the braid group and mapping class group (see subsection 2.4).

The finite dimensional inclusion theory has a very interesting infinite
dimensional generalization through the subfactor theory which was initiated by
Vaughan Jones. Whereas as in the above example inclusions of finite
dimensional full matrix algebras have Jones indices which are squares of
integers, infinite dimensional subfactors have a more interesting spectrum of
Jones indices. For a presentation of subfactor theory using concepts and
techniques of AQFT as well as its use for studying superselection sectors we
refer to \cite{Lo-Re}

\section{ Superselections and Locality in Quantum Physics}

In this second section I will explain the assumptions underlying LQP, its
relation to more standard formulations of QFT and some of its important achievements.

Let us first list some assumptions which include the main properties of
(Haag-Kastler) nets of observables.

\begin{itemize}
\item (i) \thinspace\thinspace\thinspace There is an inclusion preserving map
of \ compact regions $\mathcal{O}$ in Minkowski space into von Neumann
operator algebras $\mathcal{A(O)}$ which are subalgebras of all operators
$\mathcal{B(H)}$ in some common Hilbert space $\mathcal{H}$:
\begin{align}
\mathcal{A}  &  :\mathcal{O}\rightarrow\mathcal{A(O)}\\
\mathcal{A}(\mathcal{O})  &  \subset\mathcal{A}(\hat{O})\,\,if\,\,\mathcal{O}%
\subset\mathcal{\hat{O}}\nonumber
\end{align}
It is sufficient to fix the map on the Poincar\'{e} invariant family
$\mathcal{K}$ of double cone regions ($V_{\pm}:$forward/backward lightcone)
\begin{equation}
\mathcal{O}=\left(  V_{+}+x\right)  \cap\left(  V_{-}+y\right)  ,\,\,y-x\in
V_{+}%
\end{equation}
Since the family of algebras $\left\{  \mathcal{A}(\mathcal{O})\right\}
_{\mathcal{O}\in\mathcal{K}}$ forms a net directed towards infinity of
Minkowski space (two double cones can always be encloded into a larger one),
one can naturally globalize the net by forming its inductive limit whose
$C^{\ast}$-completion defines the quasilocal $C^{\ast}$-algebra $\mathcal{A}%
_{qua}:$%
\begin{equation}
\mathcal{A}_{qua}=\overline{\bigcup_{\mathcal{O\in M}}\mathcal{A(O)}%
}^{\left\|  .\right\|  } \label{gl}%
\end{equation}
were the superscript indicates the uniform operator norm in terms of which the
closure is taken. It is called ``quasilocal'' because its operators can still
be uniformly approximated by those of the net (which excludes truly global
operators as global charges). The $C^{\ast}$-algebras for noncompact regions
are analogously defined by inner approximation with double cones $\mathcal{O}%
$\footnote{There are also outer approximations approximations by intersections
of wedges. Even if the result is geometrically identical, the associated
algebras may lead to a genuine inclusion containing interesting physical
information.}$.$ Since they are concrete operator algebras in a common Hilbert
space they have a natural von Neumann closure $\mathcal{M}=\mathcal{M}%
^{\prime\prime}$. The closely related (but independent) assumption (for double
cones $\mathcal{O}$ of arbitray size)
\begin{equation}
\left\{  \bigcup_{a}\mathcal{M}(\mathcal{O}+a)\right\}  ^{\prime\prime
}=\mathcal{M}(M),\,\,M=Minkowski\,\,spacetime
\end{equation}
is called \textit{weak additivity} and expresses the fact that the
\textit{Global} can be constructed from the \textit{Local}.

\item (ii)\thinspace\thinspace\thinspace\thinspace\textit{Einstein causality}
and its strengthened form called Haag duality
\begin{align}
Einstein\,\,causality  &  :\mathcal{A(O)\subset A(O}^{\prime}\mathcal{)}%
^{\prime}\label{cau}\\
Haag\,\,duality  &  :\mathcal{A(O)=A(O}^{\prime}\mathcal{)}^{\prime}\\
\mathcal{O}^{\prime}  &  =causal\,\,disjoint\nonumber
\end{align}

\item (iii)\thinspace\thinspace\thinspace\thinspace\textit{Covariance and
stability} (positive energy condition) with respect to the Poincar\'{e} group
$\mathcal{P}$. For observable nets:
\begin{align}
\alpha_{(a,\Lambda)}(\mathcal{A(O))}  &  =\mathcal{A}(\Lambda\mathcal{O}+a)\\
&  =AdU(a,\tilde{\Lambda})\mathcal{A(O)}\nonumber\\
U(a,1)  &  =e^{iPa},\,\,specP\in V_{+},\,\nonumber\\
\exists\,vacuum\,vector\,\left|  0\right\rangle  &  \in H,\,P\left|
0\right\rangle =0\nonumber
\end{align}
where the unitaries represent the covering group $\mathcal{\tilde{P}}$ in $H.$
A particular case is that the P-spectrum contains the vacuum state $P=0.$ We
will call the cyclically generated subspace $H_{vac}\equiv$ $\overline
{\mathcal{A}\left|  0\right\rangle }\subset H$ the space of the vacuum sector.
It is customary in LQP to assume that the common Hilbert space $H$ in which
the net is defined is the vacuum space $H_{vac}$ and that the other physical
representations are to be computed from the vacuum data by the methods of LQP
explained in the next section.

\item (iv) \ \textit{Causal time slice property} (causal shadow property): Let
$\mathcal{O}$ be the double cone like causal shadow region associated with a
subregion $\mathcal{C(O)}$ of a Cauchy surface $\mathcal{C}$ and let $U$ be a
(timeslice) neighborhood of $\mathcal{C(O)}$ in $\mathcal{O}$, then
\begin{equation}
\mathcal{A}(\mathcal{O})=\mathcal{A}(U)
\end{equation}

\item (v) \ \textit{Phase space structure of LQP}
\begin{equation}
the\,map\,\Theta:\mathcal{A}(\mathcal{O})\rightarrow e^{-\beta P_{0}%
}\mathcal{A}(\mathcal{O})\Omega\text{\thinspace\thinspace}is\,\,nuclear\text{
\thinspace\thinspace}%
\end{equation}
i.e. the range of $\Theta$ is a ``small'' set of vectors contained in the
image of a traceclass operator in $B(H).$
\end{itemize}

\textit{Some additional comments on the physical ideas and some easy consequences.}

In QM the physical interpretation of the commutant (of a collection of
observable Hermitian operators) is that of a von Neumann algebra which is
generated by all those observable operators whose measurements are
commensurable relative to the given set of observables. The general setting of
quantum theory offers no specific physical characterization of such algebras;
however in LQP the Einstein causality property (ii) tells us that if the
original collection generates all observables which can be measured in a given
spacetime region $\mathcal{O}$, then the commensurable measurements are
associated with observables in the spacelike complement $\mathcal{O}^{\prime}%
$. Whereas Einstein causality limits the region of future influence of data
contained in localized observable algebras $\mathcal{A}(\mathcal{O})$ to the
forward closed light cone subtended by the localization region $\mathcal{O}$,
the causal shadow property (iv) prevents the appearance of new degrees of
freedom in the causal shadow $\mathcal{O}^{\prime\prime}$ from outside
$\mathcal{O}$. In fact the idea that all physical properties can be extracted
from an affiliation of an observable to a (possibly multiple-connected)
spacetime region and its refinements is the most important and successful
working hypothesis of LQP, whereas the standard formulation is ill suited to
formalize this important aspect.

Haag duality is the equality in (\ref{cau}) i.e. the totality of all
measurements relatively commensurable with respect to the observables in
$\mathcal{A(O)}$ is exhausted by the spacelike disjoint localized observables
$\mathcal{A(O}^{\prime}\mathcal{)}$. One can show (if necessary by suitably
enlarging the local net within the same vacuum Hilbert space) that Haag
duality can always be achieved in an Einstein causal net. It turns out that
the inclusion of the original net in the Haag dualized net contains intrinsic
(independent of the use of particular poitlike fields) information on
``spontaneous symmetry breaking'', an issue which will not be treated in this
survey \cite{spon}. The magnitude of violation of Haag duality in other
non-vacuum sectors is related to properties of their nontrivial superselection
charge whose mathematical description is in terms of endomorphisms of the net
(the Jones index of the inclusion $\rho(\mathcal{A})\subset\mathcal{A\,}$\ is
a quantitative measure).

Neither Einstein causality nor Haag duality guaranty causal disjointness in
the form of ``statistical independence'' for spacelike seperations i.e. a
tensor product structure between two spacelike separated algebras (analogous
to the factorization for the inside/outside region associated with a quantum
mechanical quantization box). This kind of strengthening of causality is best
formulated in terms of properties of states on the local algebras. It turns
out that the nuclearity of the QFT phase space in (v) is a sufficient
condition for the derivation of statistical independence \cite{Haag}. This is
done with the help of the so-called split property, a consequence of the
nuclearity assumption which is interesting in its own right. It states the
tensor factorization can be achieved if one leaves between the inside of a
double cone $\mathcal{O}$ and its spacelike disjoint a ``collar'' region
(physically for the vacuum fluctuations to settle down) i.e. if one takes
instead the spacelike outside of a slightly bigger double $\mathcal{\hat{O}}$
cone which properly contains the original one. In that case there exists an
intermediate type $I$ factor $\mathcal{N}$ between the two double cone
algebras $\mathcal{A(O)}\subset\mathcal{N}\subset\mathcal{A}(\mathcal{\hat{O}%
})$ (there is even a canonically distinguished one) whose localization is
``fuzzy'' i.e. cannot be described in sharp geometrical terms beyond this
inclusion. A type $I$ factor is synonymous with tensor factorization%
\begin{align}
H  &  =H_{\mathcal{N}}\bar{\otimes}H_{\mathcal{N}^{\prime}}\\
B(H)  &  =\mathcal{N}\bar{\otimes}\mathcal{N}^{\prime}\nonumber
\end{align}

The positivity of energy is a specific formulation of stability adapted to
particle physics which deals with local excitations of a Poincar\'{e}
invariant vacuum. It goes back to Dirac's observation that if one does not
``fill the negative energy sea'' associated with the formal energy-momentum
spectrum of the Dirac equation, the switching on of an external
electromagnetic interaction will create a chaotic instability. In case of
thermal states it is the so called KMS condition which secures
stability\footnote{The KMS condition is a generalization of the Gibbs formula
to open systems \cite{Haag}.}.

The energy positivity leads via analytic properties of vacuum expectation
values to the cyclicity of the vacuum with respect to the action of
$\mathcal{A}(\mathcal{O})$ i.e. $\overline{\mathcal{A}(\mathcal{O})\Omega}=H$
and for $\mathcal{O}$'s with a nontrivial causal complement the use of
causality also yields the absence of local annihilators i.e. $A\Omega
=0,\,A\in\mathcal{A(O)}\curvearrowright A=0.$ This latter property is called
separability of ($\mathcal{A(O)},\Omega$) and follows from cyclicity and
causality. Both aspects together are known under the name of Reeh-Schlieder
property and in operator algebra theory such pairs are called ``standard''.
This property very different from what one is accustomed to in QM since it
permits a creation of a particle ``behind the moon'' (together with an
antipartcle in some other far remote region) by only executing local
operations of small duration in an earthly laboratory. Mathematically this is
the starting point for the Tomita-Takesaki modular theory which we will return
to below. On the physical side the attempts to make this exotic mathematical
presence of a dense set of state vectors by local operations physically more
palatable has led to insights into the profound role of the phase space
structure (v) \cite{Haag}.

\subsection{Connection with pointlike formulation}

Before we give an account of structural theorems, in particular the
superselection structure following from these assumed properties, it is
helpful to make a relation to the traditional formulation in terms of fields
to which QFT owes its name. This arose from the canonical quantization of
classical field theories which eventually found its more covariant but still
formal formulation in terms of functional integrals using classical actions.
The first successful attempt to overcome the ``artistic'' aspects\footnote{By
this we mean properties which serve to start the calculation as e.g. the
validity of a functional integral representation which the physical
correlation functions (after renormalization) obtained at the end do not obey.
The Wightman approach avoids this mathematical imbalance.} and to characterize
the conceptual and mathematical properties of what one expects to lie behind
the formal manipulations was given by Wightman \cite{St-Wi} in terms of axioms
about (not necessarily observable) pointlike fields. These axioms separate
into two groups.

\begin{itemize}
\item $\mathcal{H}$-space and $\mathcal{P}$-group
\end{itemize}

1.\thinspace\thinspace\thinspace\thinspace Unitary representation
$U(a,\alpha)$ in $H\,$of the covering group $\widetilde{\mathcal{P}}$ of
$\mathcal{P}$, $\alpha\in SL(2,C)$

2.\thinspace\thinspace\thinspace\thinspace Uniqueness of the vacuum $\Omega,$
$U(a,\alpha)\Omega=\Omega$

3.\thinspace\thinspace\thinspace\thinspace Spectrum condition: $specU\in
\bar{V}_{+},\,$the forward light cone.

\begin{itemize}
\item \thinspace Fields

1.\thinspace\thinspace\thinspace\thinspace Operator-valued distributions:
$A(f)$=$\int A(x)f(x)d^{4}x,\,\,\,f\in\mathcal{S}$ (the Schwartz space of
``tempered'' testfunctions) is an unbounded operator with a \thinspace dense
domain $\mathcal{D}$ \thinspace such that the function $\left\langle \psi
_{2}\left|  A(x)\right|  \psi_{1}\right\rangle $ exists as a sesquilinear form
for $\psi_{i}\subset\mathcal{D}$

2.\thinspace\thinspace\thinspace\thinspace\thinspace Hermiticity: with $A $,
also $A^{\ast}$ belongs to the family of fields and the affiliated
sesquilinear forms are as follows related: $\left\langle \psi_{2}\left|
A^{\ast}(x)\right|  \psi_{1}\right\rangle =\left\langle \psi_{1}\left|
A(x)\right|  \psi_{2}\right\rangle $

3.\thinspace\thinspace\thinspace\thinspace$\widetilde{\mathcal{P}}$-covariance
of fields: $U(a,\alpha)A(x)U^{\ast}(a,\alpha)=D(\alpha^{-1})A(\Lambda
(\alpha)x+a).$ For observable fields only integer spin representations ( i.e.
representations of $\mathcal{P}$) occur.

4.\ \thinspace\thinspace Locality: $\left[  A^{\#}(f),A^{\#}(g)\right]  _{\mp
}=0$ \quad for suppf$\times$suppg ($\times:$supports are spacelike separated).
\end{itemize}

Haag duality, statistical independence and primitive causality (the causal
shadow property) allow no natural formulation in terms of individual pointlike
field coordinates, they are rather relations between algebras. The process of
Haag dualization of a net affects the relation between fields and local
algebras, there is no extension of pointlike fields involved.

Formally the fields obey a dynamical law describing their causal propagation
in timelike/lightlike direction. The relation between the two approaches is
very close as far as the intuitive content of the physical principles is
concerned. In fact the best way to relate them is to think about fields as
being akin to coordinates in geometry and the local algebras as representing
the coordinate-free intrinsic approach\footnote{However the differences
between the intrinsic algebraic approach as compared to that with pointlike
fields appears greater than that on the geometric side, since in the
coordinate-free geometric approach coordinates one still uses coordinate
patches in the definition of a manifold.}. The smeared fields $A(f)$ with
$suppf\subset\mathcal{O}$ play the role of formal (affiliated) generators of
$\mathcal{A}(\mathcal{O}),$ but since the latter are unbounded operators with
a dense domain, the relation involves domain problems which are similar (but
more difficult) than the connection between Lie algebras and Lie groups in the
noncompact case. When the first monographs on this axiomatic approach were
written \cite{St-Wi}, these domain properties were thought of as
technicalities. However as a result of recent developments in ``modular
localization'' one now knows that these domains contain physical information
in particular information about the geometric localization of the operators.

In the opposite direction from nets of algebras to fields one does not expect
in general that all degrees of freedom of a local theory can be described in
terms of pointlike covariant fields at least if the theories are not scale
invariant. In such a case the Lagrangian framework is too narrow and one must
use the LQP framework. This is in particular the case if one ``takes a
holographic image, transplants or scans'' a theory which was generated by
pointlike fields as will be explained in the last section.

The pointlike Wightman approach leads, as a result of its analytic formalism,
to fairly easy proofs of the existence of an antiunitary TCP operator $\Theta$
which implements the total spacetime reflection symmetry $x\rightarrow-x$
which simultaneously involves a conjugation of the superselection charge
\cite{St-Wi}. In the algebraic setting the proof requires presently a mild
additional assumptions. \cite{Bo-Yn}.

\subsection{Localization and Superselection}

In order to study localized states and the associated representations it is
convenient to have a global algebra which contains all $\mathcal{A}%
(\mathcal{O}).$ For this purpose we use $\mathcal{A}\equiv\mathcal{A}_{qua}$
as defined in (\ref{gl})

The strongest form of localization of states is that of Doplicher Haag and
Roberts (DHR) \cite{Haag}:

\begin{definition}
A positive energy state is DHR-localized (relative to the vacuum) if the
associated GNS representation $\pi(\mathcal{A})$ is unitarily equivalent to
the vacuum representation $\pi_{0}(\mathcal{A})$ in the spacelike complement
$\mathcal{O}^{\prime}$ of any preassigned compact region (double cone)
$\mathcal{O}$%
\begin{equation}
\pi|_{\mathcal{A(O}^{\prime})}\simeq\pi_{0}|_{\mathcal{A(O}^{\prime}%
)}\,\,\forall\mathcal{O}%
\end{equation}
\end{definition}

The definition in particular implies that a state which is strictly localized
in $\mathcal{O},$ i.e. $\omega(A)=\omega_{0}(A)\,\,\forall A\in\mathcal{A(O}%
^{\prime}),$ is also localized in the DHR sense \cite{Haag}. This localization
underlies the standard formulation of QFT which is based on covariant
pointlike fields and covers in particular the fields featuring in the
Lagrangian quantization formalism. The important step which converts the
localization of states/representations into the localized superselected
charges uses the Haag duality

\begin{proposition}
Localized DHR representations can be expressed in terms of ``localized
charges'' which are described in terms of localized endomorphism $\rho$ of
$\mathcal{A}$
\begin{equation}
\pi\simeq\pi_{0}\circ\rho
\end{equation}
\end{proposition}

\begin{proof}
Pick any region $\mathcal{O}$ and use the existence of a unitary partial
intertwiner $V(\mathcal{O})$ following from the above definition
\begin{align}
V(\mathcal{O})\pi_{0}(A)  &  =\pi(A)V(\mathcal{O}),\,\,A\in\mathcal{A}%
(\mathcal{O}^{\prime})\\
\hat{\pi}(A)  &  =V(\mathcal{O})^{-1}\pi(A)V(\mathcal{O}),\,\,\forall
A\in\mathcal{A}_{qua}\,\nonumber
\end{align}
where the second line is the definition of a representation which is
equivalent to $\pi$ and is identical to $\pi_{0}$ in its restriction to
$\mathcal{A}(\mathcal{O}^{\prime}).$ Therefore \textit{for all regions}
$\mathcal{O}_{1}\supset\mathcal{O}$ the range of $\hat{\pi}$ is according to
Haag duality contained in that of $\pi_{0}:\hat{\pi}(\mathcal{A}%
(\mathcal{O}_{1}))\subset\pi_{0}(\mathcal{A}(\mathcal{O}_{1})).$ This is so
because $\left[  \pi_{0}(A^{\prime}),\hat{\pi}(A)\right]  =\pi\left(  \left[
A^{\prime},A\right]  \right)  =0$ for $A^{\prime}\in\mathcal{A}(\mathcal{O}%
_{1}^{\prime}),A\in\mathcal{A}(\mathcal{O}_{1}).$ This relation together with
Haag duality then tells us that $\hat{\pi}(\mathcal{A}(\mathcal{O}%
_{1}))\subset\pi_{0}(\mathcal{A}(\mathcal{O}_{1}^{\prime}))^{\prime}%
\overset{HD}{=}\pi_{0}(\mathcal{A}(\mathcal{O}_{1})),$ from which one
concludes that
\[
\rho:=\pi_{0}^{-1}\hat{\pi},\text{\thinspace\thinspace}\mathcal{A}%
\rightarrow\mathcal{A}%
\]
defines an endomorphism of $\mathcal{A}$.
\end{proof}

Endomorphisms are generalizations of automorphism; they are not required to be
morphisms of the algebra onto itself but may have a subalgebra as an image.
The endomorphisms in LQP are faithful. They are called localized in
$\mathcal{O}$ if $\rho(A)=A,\,\forall A\in\mathcal{A}(\mathcal{O}),$ and they
are said to be transportable if for any given $\mathcal{\tilde{O}}$ there
exists an equivalent endomorphism $\tilde{\rho}\in\left[  \rho\right]  $ with
$loc\tilde{\rho}\subset\mathcal{\tilde{O}}.$ Since the latter region contains
no limitation of a minimal size, there is no fundamental length in the DHR
setting. It is often convenient to identify the algebra $\mathcal{A}$ with its
faithful vacuum representation and write $\rho(A)\psi$ instead of $\pi
_{0}\circ\rho(A)\psi.$ A neat way to remember that we are using the action on
the vacuum Hilbert space but mediated through the endomorphism $\rho$%
\begin{equation}
A:(\rho,\psi)\mapsto(\rho,\rho(A)\psi)
\end{equation}
is to denote this different use of the vacuum Hilbert space $H_{0}$ as a
representation space for $\rho$ in form of a pair $H_{\rho}\equiv(\rho
,H_{0}).$ This notation will later be allowed to develop a life of its own; it
suggests the introduction of a $C^{\ast}$-algebra with bimodule properties
called the \textit{reduced field bundle.}

The marvelous achievement of converting localized transportable
representations $\pi$ into endomorphisms with the same properties of the
observable algebra $\mathcal{A}$ is that now one may define a product
structure of endomorphisms simply by acting successively, i.e. ($\rho_{2}%
\circ\rho_{1})(A)\equiv\rho_{2}(\rho_{1}(A))\,\ \ \forall A\in\mathcal{A}.$
With this composition we have achieved a natural definition for the product
$\ $of two representations%
\begin{equation}
\pi_{1}\circ\pi_{1}\equiv\pi_{0}\circ\rho_{2}\rho_{1}%
\end{equation}
It is now appropriate to define in more precise terms what we mean by
superselection sectors. Since unitary equivalent DHR representations
correspond precisely to inner equivalent (i.e. by unitaries in $\mathcal{A}$)
$\rho^{\prime}s,$ we call a (superselection) sector a class of inner
equivalent endomorphisms $\left[  \rho\right]  $ associated with a given
$\rho.$ It immediately follows, that whereas the individual endomorphisms
compose in a noncommutative manner, the composition of sectors is abelian
\begin{equation}
\left[  \rho_{2}\right]  \left[  \rho_{1}\right]  :=\left[  \rho_{2}\rho
_{1}\right]  =\left[  \rho_{1}\rho_{2}\right]  =\left[  \rho_{1}\right]
\left[  \rho_{2}\right]
\end{equation}
This is a consequence of causality and the localizability and transportability
of the endomorphisms which results in commutativity in case of their spacelike
separation
\begin{equation}
\rho_{2}\rho_{1}=\rho_{1}\rho_{2},\,\,if\,\,loc\rho_{1}\times loc\rho_{2}%
\end{equation}
The proof is very simple: if both sides are applied to an $A\in\mathcal{A(O}%
$)$\subset\mathcal{A}$ with $\mathcal{O}$ spacelike to both loc$\rho_{i},$ the
relation obviously holds. But this standard situation can always be achieved
by suitably transporting the $\rho_{i}$ into two ``spectators'' $\rho
_{i}^{\prime}$ with loc$\rho_{i}^{\prime}$ causally disjoint from the
$\mathcal{O}$ by using suitably localized charge transporters $U_{i}$ in such
away that the localization of these unitaries does not destroy the
commutativity in the process of changing back to the original $\rho
_{i}^{\prime}s.\,\ $In these arguments one uses the (also easily proven) fact
that the localization of the composite $\rho_{2}\rho_{1}$ is $O_{12}%
:=\mathcal{O}_{1}\mathcal{\vee O}_{2}$ i.e. the smallest double cone
containing both $\mathcal{O}_{i}.$

An important step in the development of an intertwiner calculus is the
realization that projectors $E$ which project onto subrepresentations $\pi$ on
$H_{\pi}=E_{\pi}H_{0}$ commute with $\pi_{0}\circ\rho_{2}\rho_{1}%
(\mathcal{A}(\mathcal{O}_{12}^{\prime}))=\pi_{0}(\mathcal{A}(\mathcal{O}%
_{12}^{\prime}))$ and hence belongs to the algebra $\pi_{0}(\mathcal{A}%
(\mathcal{O}_{12})).$ This permits the introduction of isometric intertwiners
$T$ which map $H_{0}$ onto the subspace $E_{\pi}H_{0}$ i.e. their source space
is $H_{0}$ and their range space $H_{\pi}.$ These isometries intertwine the
endomorphisms
\begin{align}
T\rho(A)  &  =\rho_{2}\rho_{1}(A)T,\,\,\,T^{\ast}T=1,\,\,TT^{\ast
}=E\label{inter}\\
\pi &  =\pi_{0}\circ\rho\nonumber
\end{align}
The mathematical basis of this is the ``property B'' (due to Borchers
\cite{Haag}) stating that a projection operator $E$ which is localized in
$\mathcal{O}$ allows a factorization into intertwiners $T$ with
$locT=\mathcal{\tilde{O}}$ for any $\mathcal{\tilde{O}}\supset\supset
\mathcal{O}$ (proper inclusion, i.e. no touching of boundaries so that a full
neighborhood of the trivial translation in the translation group can act on
$\mathcal{A}(\mathcal{O})$). This is a well-known property of properly
infinite von Neumann algebras of type $III$ in which case the factorization
does not need the $\tilde{O}$ enlargement. The property B follows from
additivity whereas proving that the local von Neumann algebras are hyperfinite
of type III$_{1}$ (for which the corresponding von Neumann factors are unique)
is more involved. In our special case (\ref{inter}) where $E\in\mathcal{A}%
(\mathcal{O}_{12})$ and hence according to property B $locT$ is a double cone
which properly contains $O_{12}.$

The intertwiners (\ref{inter}) form a (Banach)space in $\mathcal{A},$ for
which one sometimes uses the notation $(\rho_{2}\rho_{1},\rho)$ i.e.
$T:\rho\rightarrow\sigma$ forms the space $(\sigma|\rho)$ and their
composition in $\mathcal{A}$ again belongs to an intertwiner space $ST=S\circ
T\in(\tau|\rho)$ if $S\in(\tau|\sigma).$ Evidently for $\rho$ irreducible
(i.e. $\rho(\mathcal{A)}^{\prime}\mathcal{\cap A=}\mathbb{C}$\textbf{1),} the
$T^{\prime}$s are isometric and form a Hilbert space within the $C^{\ast}%
$-algebra $\mathcal{A}_{qua}$ i.e. $S^{\ast}T\in\mathbb{C}\mathbf{1.}$

Our main interest in the following will be the intertwiner calculus within the
set $\Delta_{0}$ of localized, transportable endomorphisms which have
conjugates and have finite statistics. We call two irreducible localized
endomorphisms $\rho$ and $\bar{\rho}$ conjugate to each other if the sector
$\left[  \rho\bar{\rho}\right]  $ contains the vacuum sector i.e. if there
exist isometries $R\in(\bar{\rho}\rho|id),\,\bar{R}\in(\rho\bar{\rho}|id);$
finiteness of statistics then leads to their uniqueness in case of irreducible
endomorphisms. Their unique existence is closely linked to the existence of a
unique ``left inverse'' $\phi$ and conditional expectation $E:\mathcal{A}%
\rightarrow\rho(\mathcal{A})$%
\begin{align}
\phi(A)  &  :=R^{\ast}\overline{\rho}(A)R,\,\,A\in\mathcal{A}\\
E(A)  &  :=\rho\circ\phi(A)\nonumber
\end{align}
This left inverse draws its name from the relation $\phi(A\rho(B))=\phi(A)B$
and $E$ has the properties of a projection of $\mathcal{A}$ onto
$\rho(\mathcal{A});$ both properties are immediately read off from the
definition of $R.$ With an additional minimality requirement the uniqueness
continues to hold in the case of general $\rho^{\prime}s$ (on which the $\phi$
and $E$ depend) \cite{Lo-Re}.

This still leaves us with explaining statistics. The statistics operator of a
pair $\rho_{1},\rho_{2}$ is a distinguished flip operator $\varepsilon
(\rho_{1},\rho_{2})\in$ $(\rho_{1}\rho_{2}|\rho_{2}\rho_{1})$ which is
explicitly defined in terms of charge transporters $U_{i}\in(\hat{\rho}%
_{i}|\rho_{i})$ which shift the localization into spacelike separated regions
$\mathcal{\hat{O}}_{1},$ $\mathcal{\hat{O}}_{2}$%
\begin{equation}
\varepsilon(\rho_{1},\rho_{2})=\rho_{2}(U_{1}^{\ast})U_{2}^{\ast}U_{1}\rho
_{1}(U_{2}),\,\,if\,\,\mathcal{\hat{O}}_{2}<\mathcal{\hat{O}}_{1} \label{ep}%
\end{equation}
If the spacelike ordering is inverted, the same formula represents
$\varepsilon(\rho_{2},\rho_{1})^{\ast}$ instead of $\varepsilon(\rho_{1}%
,\rho_{2}).$ It is easy to see that the definition (\ref{ep}) does not depend
on the choice of the ``spectator'' endomorphisms $\hat{\rho}_{i}$ as long as
one does not change their topological relation. Whenever there is no
topological separation between the two orders (which in the case of double
cones occurs for spacetime dimension $d\geq2+1)$ one has the additional
relation of their equality. It is easy to show (see the second appendix to
this section) that $\varepsilon\equiv\varepsilon(\rho,\rho)$ generates the
Artin braid group via $g_{k}\equiv\rho^{k-1}(\varepsilon)$ and that the action
of the $\rho_{i}^{\prime}s\,$\ on the $\varepsilon(\rho_{1},\rho_{2})$ define
some colored groupoid version (see appendix of \cite{FRSII}). Since for
spacetime dimension $d\geq2+1$ with compact (double cone) localization there
is only one localization class, we obtain with $\varepsilon^{2}=1$ the
permutation group statistics of standard QFT.

These intertwiner spaces are empty precisely if there are no common
subrepresentations. Similar to the use of creation and annihilation operators
for Fock spaces, these charge intertwiners can be used to intertwine between
the different charge subspaces of one ``master space''. This is done in the
following way. From the $\Delta_{0}$ one chooses one representative per
irreducible sector and defines (with $\nabla_{0}\subset\Delta_{0}$ being the
reference set)
\begin{align}
&  H=\bigoplus_{\rho_{i}\in\nabla_{0}}H_{\rho_{i}}\\
&  H_{\rho_{i}}=(\rho_{i},H_{0})\nonumber\\
&  \left(  (\rho_{i},\psi),(\rho_{j},\varphi)\right)  =\delta_{\iota j}%
(\psi,\varphi)\nonumber
\end{align}
We will call this space the reduced field bundle space and our aim is to
define on $H$ with help of the system of reference intertwiners a kind of
bimodule $C^{\ast}$-algebra$,$ the so called reduced field bundle $\mathcal{F}.$

We start with defining isometric intertwiners $T_{e}^{(i)}\in(\rho_{\alpha
}\rho|\rho_{\beta})$ associated with $\rho^{\prime}s$ from $\nabla_{0}$ where
the edge $e$ stands for the superselection channel of the three irreducible
endomorphisms with charge $c(e)=\rho,$ source $s(e)=\rho_{a}$ and range
$r(e)=\rho_{\beta}.$ The finiteness assumption insures that these algebraic
Hilbert spaces in $\mathcal{A}_{qua}$ are finite dimensional $dim(\rho
_{\alpha}\rho|\rho_{\beta})=\left(  N_{\rho}\right)  _{\alpha}^{\beta}%
<\infty.$ Assume that the $T_{e}$ have been chosen orthonormal
\begin{align}
T_{e}^{\ast}T_{e^{\prime}}  &  =\delta_{ee^{\prime}}1_{e}\\
\sum T_{e}T_{e}^{\ast}  &  =\mathbf{1}\nonumber
\end{align}
In the completeness relation the sum extends over $r(e)$ as well as over (here
suppressed) possible degeneracy index i. For the special case that $\rho$ or
$\rho_{\alpha}$ we choose $T_{e}=1$ whereas for $\rho_{\beta}=id$ (and hence
$\rho_{\alpha}=\bar{\rho}$) we take $T_{e}=R_{\rho}.$ The definition of the
reduced field bundle is now
\begin{align}
\mathcal{F}  &  =\bigoplus_{e}(e,\mathcal{A})\\
F(e,A)(\rho_{\alpha},\psi)  &  =\delta_{\rho_{\alpha}s(e)}(r(e),T_{e}^{\ast
}\rho_{\alpha}(A)\psi)\nonumber
\end{align}
The successive application of the last line leads to a product formula for the
operators in $\mathcal{F}.$
\begin{equation}
F(e_{2},A_{2})F(e_{1},A_{1})=\delta_{s(e_{2})r(e_{1})}\sum_{e,f}D_{f,e}%
^{e_{2}\circ e_{1}}F(e,A_{f})\in\mathcal{F}%
\end{equation}
with the fusion coefficients $D$ being analogs of the group theoretic
Clebsch-Gordan coefficients. With the above rules which allow to re-express
everything in terms of the $\nabla_{0}$ basis and the associated
$T_{e}^{\prime}s$ one calculates straightforwardly
\begin{align}
A_{f}  &  =T_{f}^{\ast}\rho_{1}(A_{2})A_{1}\label{comp}\\
D_{f,e}^{e_{2}\circ e_{1}}  &  =T_{e_{2}}^{\ast}T_{e_{1}}^{\ast}\rho_{\alpha
}(T_{f})T_{e}\in(r(e)|r(e))=\mathbb{C}\nonumber\\
\rho_{a}(T_{f})  &  =\sum_{e_{i},e}D_{f,e}^{e_{2}\circ e_{1}}T_{e_{1}}%
T_{e_{2}}T_{e}^{\ast}\nonumber
\end{align}
$\mathcal{F\supset\,}F(\mathbf{1},\mathcal{A})\simeq\mathcal{A}$ (where the
edge denoted by $\mathbf{1}$ corresponds to the sum over all edges with
$c(e)=0\curvearrowright s(e)=r(e)$) becomes a Banach subalgebra of $B(H).$ A
more detailed analysis \cite{FRSII} shows that there is a $^{\ast}$-operation
which renders it a $C^{\ast}$algebra such that the observable subalgebra
$\mathcal{A}\simeq F(\mathbf{1},\mathcal{A})$ acts on $\mathcal{F}$ in a
\textit{bimodule manner}
\begin{align}
F(\mathbf{1},A)F(e,B)  &  =F(e,\rho_{\alpha}(A)B)\\
F(e,B)F(\mathbf{1},A)  &  =F(e,BA),\,\,A,B\in\mathcal{A}\nonumber
\end{align}
Similarly to the composition law (\ref{comp}) $\mathcal{F}$ has a commutation
structure related to the localization of its operators. We define
$locF=\mathcal{O}$, $F\in\mathcal{F}$ to be that region $\mathcal{O}$ for
which $F$ commutes with $\mathcal{A(O}^{\prime}$) i.e.
\begin{equation}
F(\mathbf{1},\mathcal{A(O}^{\prime}\mathcal{)})F(e,A)=F(e,A)F(\mathbf{1}%
,\mathcal{A(O}^{\prime}\mathcal{)})
\end{equation}
The definition is in fact independent of the source and range projection and
can be solely expressed in terms of $\rho=c(e)$ in form of the existence of a
charge transporter $U$ with $locAdU\rho\subseteq\mathcal{O}$ and
$locUA\subseteq\mathcal{O}.$ The F's are in general nonlocal relative to each
other and obey
\begin{align}
F(e_{2},A_{2})F(e_{1},A_{1})  &  =\sum_{f_{1}\circ f_{2}}R_{f_{1}\circ f_{2}%
}^{e_{2}\circ e_{1}}(+/-)F(f_{1},A_{1})F(f_{2},A_{2})\\
locF_{1}  &  \lessgtr locF_{2}\nonumber\\
R_{f_{1}\circ f_{2}}^{e_{2}\circ e_{1}}(+/-)  &  =T_{e_{1}}^{\ast}T_{e_{2}%
}^{\ast}\rho_{a}\left(
\begin{array}
[c]{c}%
\varepsilon(\rho_{1},\rho_{2})\\
\varepsilon(\rho_{2},\rho_{1})^{\ast}%
\end{array}
\right)  T_{f_{1}}T_{f_{2}},\,\,\rho_{a}=s(e_{1})
\end{align}
where similarly to the previous $D$-matrix case the numerical R-matrices
result from expanding the flip operator $\varepsilon(\rho_{1},\rho_{2})$ in
the complete $T_{e}$ intertwiner basis
\begin{equation}
\rho_{a}\left(
\begin{array}
[c]{c}%
\varepsilon(\rho_{1},\rho_{2})\\
\varepsilon(\rho_{2},\rho_{1})^{\ast}%
\end{array}
\right)  T_{f_{2}}T_{f_{1}}=\sum_{f_{2}\circ f_{1}}R_{f_{2}\circ f_{1}}%
^{e_{2}\circ e_{1}}(+/-)T_{e_{1}}T_{e_{2}}%
\end{equation}
Up to unitary equivalence, the R-matrices are determined by the Markov traces
on the $B_{\infty}$ braid group (see appendix to this section). As already
mentioned the DHR analysis for $d\geq2+1$ gives permutation group statistics
$B_{\infty}\rightarrow S_{\infty}$. But it allows in addition to the complete
antisymmetric/symmetric Fermi/Bose permutation group representations also
``parastatistics'', i.e. irreducible $S_{n}$ representations with mixed Young
tableaus of size n. There was a suspicion since the beginning of the 70ies
that behind the reduced field bundle structure with its source and range
dependent operators, there may be a more natural description in terms of a
field algebra $\mathcal{F}$ where the basic degrees of freedom were
Fermions/Bosons but with multiplicities on which an ''inner'' symmetry group
can act. At that time it was already known that all standard QFTs (including
all models which were associated with Lagrangian quantization) with an
internal symmetry group allow a descend to an observable algebra $\mathcal{A}$
consisting of the fixed points of the action of the compact symmetry group on
the field algebra $\mathcal{A=F}^{G}$ in such a way that the latter can be
reconstructed from its ''observable shadow'' as a kind of cross product of the
observable algebra with a group dual $\hat{G},$i.e. $\,\mathcal{F}=$
$\mathcal{A}\bowtie\hat{G}$. Doplicher and Roberts finished this extensive
program of proving the (unique after imposing a natural physical convention)
existence of $\mathcal{F}$ from only using the priciples of observable
algebras around 1990 \cite{DR}. For their solution the discovery of a new
duality theory different from the Tannaka-Krein theory was essential. In this
way the mathematically efficient but conceptually somewhat mysterious internal
symmetry concept (which historically entered particle physics with Heisenbergs
introduction of isospin) was finally demystified: the inner group symmetry
resulted from the unfolding of causality and superselection rules encoded in
the observable algebras; it is part of ''how to hear (reconstruct) the shape
of a drum (the field algebra)'' using again the famous metaphor of Marc Kac.
We will not explain any of the additional concepts and theorems which led to
those amazing results and refer the interested reader to the literature
\cite{DR}.

In the case of braid group statistics such an encoding into an extended
algebra is not known and one has to be content with the field bundle (in the
context of braid group statistics also refereed to as ``exchange algebras'')
whose objects even in the pointlike limit turn out not to be ordinary fields
whose closed source and range space is the full Hilbert space but rather field
bundle (vertex) operators with a partial source and range space (in the
context of conformal quantum field theories often referred to as ``vertex operators).

For the case of chiral field theories i.e. nets indexed by the intervals on a
circle one may also develop the superselection theory by studying 2-interval
algebras. In that case one does not have to leave the vacuum representation
because the information about the superselection sectors is contained in the
violation of Haag duality i.e. in the nontrivial inclusion \cite{KLM}
\begin{align}
\mathcal{A}(E)  &  \subset\mathcal{A}(E^{\prime})^{\prime}\\
E  &  =I_{1}\cup I_{2},\,I_{i}\,\,disjoint\nonumber\\
E^{\prime}  &  =S^{1}\backslash E\nonumber
\end{align}
The content of this double interval inclusion may be be canonically
reprocessed into an endomorphism. The latter in turn is isomorphic to the
so-called Longo-Rehren (LR) endomorphism which for rational (finite number of
DHR sectors) theories has the form
\begin{equation}
\rho_{LR}=\sum_{i}\rho_{i}\otimes\rho_{i}^{opp}on\,\,\mathcal{A}%
\otimes\mathcal{A}^{opp}%
\end{equation}
This mathematical structure corresponds to the physical picture that although
in the vacuum representation the global charge vanishes, the two interval
situation allows for a charge/conjugate charge (or particle/antiparticle)
split in which all existing local superselection charges participate in a
democratic fashion \cite{Haag}. This is yet another manifestation of the
doctrin that all physical information resides in the vacuum representation of
observable nets which has no analog in QM.

It is interesting to note that simple particle physics ideas which date back
to the beginnings of AQFT, as the working hypothesis that the vacuum state on
nets of algebras contains all of particle physics and only needs the right
mathematical tools in order to explicitly reveal its complete content, are now
being vindicated by beautiful operator algebraic methods.

\subsection{Example: Modular Construction of Interaction-Free Nets}

In this section I will briefly sketch how one obtains the interaction-free
local net operator algebras directly from the Wigner particle theory without
passing through pointlike fields.

For the purpose of explanatory simplicity we start from the complex wave
function (momentum) space of the $(m,s=0)$ representation for a neutral
(selfconjugate) scalar particle
\begin{align}
&  H_{Wig}=\left\{  \varphi(p)|\int\left|  \varphi(p)\right|  ^{2}\frac
{d^{3}p}{2\sqrt{p^{2}+m^{2}}}<\infty\right\} \\
&  \left(  u(\Lambda,a)\varphi\right)  (p)=e^{ipa}\varphi(\Lambda
^{-1}p)\nonumber
\end{align}
The first step consists in defining a real subspace which describes
wedge-localized wave functions. For the construction of the standard $t$-$x$
wedge $W_{st}=(x>\left|  t\right|  ,y,z$ arbitrary) we use the x-t Lorentz
boost $\Lambda_{x-t}(\chi)$ and the $t$-$x$ reflection $r:$ ($x,t)\rightarrow
(-x$,$-t)$ which according to well-known theorems is represented antiunitarily
in the Wigner theory\footnote{In case of charged particles the Wigner theory
needs a particle/antiparticle doubling.}. One then starts from the unitary
boost group $u(\Lambda(\chi)$ and forms (by the standard functional calculus)
the unbounded ``analytic continuation'' in the rapidity $\chi$. Using a
notation which harmonizes with that used in the later Tomita-Takesaki modular
theory, we define the following operators in $H_{Wig}$
\begin{align}
\frak{s}  &  =\frak{j}\delta^{\frac{1}{2}}\label{pol}\\
\frak{j}  &  =u(r)\nonumber\\
\delta^{it}  &  =u(\Lambda(-2\pi t))\nonumber
\end{align}
where $u(\Lambda(\chi)$ and $u(r)$ are the unitary/antiunitary representations
of these geometric transformations in the (doubled, if required by
antiparticles of opposite charge) Wigner theory. Note that $u(r)$ is apart
from a $\pi$-rotation around the x-axis the one-particle version of the TCP operator.

Since the antiunitary $t$-$x$ reflection commutes with the $t$-$x$ boost
$\delta^{it}$, it inverts the unbounded $\delta$ i.e. $j\delta=\delta^{-1}j$
which is formally the analytically continued boost at the imaginary value
$t=-i.$ As a result of this commutation relation the unbounded antilinear
operator $\frak{s}$ is involutive on its domain of definition i.e.
$\frak{s}^{2}\subset1$ so that it may be used to define a real subspace
(closed in the real sense i.e. its complexification is not closed)
\begin{equation}
H_{R}(W)=\left\{  \varphi\in H_{Wig}|\,\frak{s}\varphi=\varphi\right\}
\end{equation}
These unusual properties, which are not met anywhere else in QT, encodes
geometric localization properties within abstract operator domains \cite{Sch1}
\cite{BGL}. They also preempt the relativistic locality properties of QFT
which Wigner looked for in his representation approach but without finding the
correct one (he found instead the Newton-Wigner localization \cite{Haag} which
is not covariant\footnote{The inappropriateness of the the Born probability
interpretation for the definition of covariant spacetime localization is
related to several problems (viz. the Klein paradox) one faces if one imposes
QM concepts onto relativistic wave equations. In some way the radical
difference between the local algebras $\mathcal{A}(\mathcal{O})$ (hyperfinite
type III$_{1}$ von Neumann algebras) and quantum mechanical algebras is
already preempted in the Wigner theory.}). The localization in the opposite
wedge i.e. the $H_{R}(W^{opp})$ subspace turns out to correspond to the
symplectic (or real orthogonal) complement of $H_{R}(W)$ in $H$ i.e.
$\operatorname{Im}(\psi,H_{R}(W))=0\curvearrowright\psi\in H_{R}(W^{opp}).$
One furthermore finds the following properties for the subspaces called
``standardness''
\begin{align}
&  H_{R}(W)+iH_{R}(W)\,\,is\,\,dense\,\,in\,\,H\\
&  H_{R}(W)\cap iH_{R}(W)=\left\{  0\right\} \nonumber
\end{align}
The subspaces have instead the following covariance properties%
\begin{equation}
u(a,\Lambda)H_{R}(W)=H_{R}(\Lambda W+a)
\end{equation}
The last line expresses the covariance of this family of wedge-localized real
subspaces and follows from the covariance of the operator $\frak{s}$. Having
arrived at the wedge localization spaces, one may construct localization
spaces for smaller spacetime regions by forming intersections over all wedges
which contain this region $\mathcal{O}$
\begin{equation}
H_{R}(\mathcal{O})=\bigcap_{W\supset\mathcal{O}}H_{R}(W) \label{int}%
\end{equation}
These spaces are again standard and covariant. They have their own
``pre-modular'' (the true Tomita modular operators appear below) object
$\frak{s}_{\mathcal{O}}$ $\frak{\ }$and the radial and angular part
$\delta_{\mathcal{O}}$ and $j_{\mathcal{O}}$ in their polar decomposition
(\ref{pol}), but this time their action cannot be described in terms of
spacetime diffeomorphisms since for massive particles the action is not
implemented by a geometric transformation in Minkowski space. To be more
precise, the action of $\delta_{\mathcal{O}}^{it}$ is only local in the sense
that $H_{R}(\mathcal{O})$ and its symplectic complement $H_{R}(\mathcal{O}%
)^{\prime}=H_{R}(\mathcal{O}^{\prime})$ are transformed onto themselves
(whereas $j$ interchanges the original subspace with its symplectic
complement), but for massive Wigner particles there is no geometric modular
transformation (in the massless case there is a modular diffeomorphism of the
compactified Minkowski space). Nevertheless the modular transformations
$\delta_{\mathcal{O}}^{it}$ for $\mathcal{O}$ running through all double cones
and wedges (which are double cones ``at infinity'') generate the action of an
infinite dimensional Lie group. Except for the finite parametric Poincar\'{e}
group (or conformal group in the case of zero mass particles) the action is
``fuzzy'' i.e. not implementable by a diffeomorphism on Minkowski spacetime.
The emergence of these \textit{fuzzy acting Lie groups is a pure quantum
phenomenon}; there is no analog for the classical mechanics of a particle.
They describe hidden symmetries \cite{SW1}\cite{S-W3}) which the Lagrangian
formalism does not expose.

Note also that the modular formalism characterizes the localization of
subspaces, but (in agreement with particle localization measurements through
counters) is not able to distinguish individual elements in that subspace.
There is a good physical reason for that, because as soon as one tries to do
that, one is forced to leave the unique Wigner $(m,s)$ representation
framework and pick a particular covariant wave functions by selecting one
specific intertwiner among the \textit{infinite set} of $u$ and $v$
intertwiners which link the unique Wigner $(m,s)$ representation to the
countably infinite covariant possibilities \cite{Sch1}. In this way one would
then pass to the framework of covariant fields explained and presented in the
first volume of Weinberg's book on QFT\cite{Wei}. The description of an
individual wave function in $H_{R}(W)$ or $H_{R}(\mathcal{O})$ in the standard
setting depends on the choice of covariant intertwiners\footnote{The ambiguity
in the intertwiners covers only the linear part of choices of field
coordinatizations. The full ambiguity is related to the Borchers class of
relatively local fields which in the free case consists of all Wick-polynomial
composites.}. A selection by e.g. invoking Euler equations and the existence
of a Lagrangian formalism may be convenient for doing particular perturbative
computations or as a mnemotechnical device for classifying polynomial
interaction densities, but is not demanded as an intrinsic attribute of local
quantum physics.

The way to avoid the highly nonunique covariant fields is to pass from real
subspaces \textit{directly} to von Neumann subalgebras of the algebra of all
operators in Fock space $B(H_{Fock}$). This step is well-known. For integral
spin $s$ one defines with the help of the Weyl (or CAR in case of Fermions)
functor $Weyl(\cdot)$ the local von Neumann algebras \cite{Sch1}\cite{BGL}
generated from the Weyl operators as
\begin{equation}
\mathcal{A}(W):=alg\left\{  Weyl(f)|f\in H_{R}(W)\right\}  \label{Weyl}%
\end{equation}
a process which is sometimes misleadingly called ``second quantization''.
These Weyl generators have the formal appearance
\begin{align}
Weyl(f)  &  =e^{ia(f)}\\
a(f)  &  =\sum_{s_{3}=-s}^{s}\int(a^{\ast}(p,s_{3})f_{s_{3}}(p)+h.c.)\frac
{d^{3}p}{2\omega}\nonumber
\end{align}
i.e. unlike the covariant fields they are independent of the nonunique $u$-$v$
intertwiners which appear in the definition of $(m,s)$ non-unique covariant
fields (\ref{field}) and depend solely on the unique Wigner data. An analogue
statement holds for the halfinteger spin case for which the CAR functor maps
the Wigner wave function into the fermionic generators of von Neumann
subalgebras. The particle statistics turns out to be already preempted by the
pre-modular theory on Wigner space\footnote{This preempting of multiparticle
statistcs and spacelike field commutations in the structure of one-particle
wave functions (via premodular properties) is one of several surprising
phenomena which indicate that relativistic wave function are closer to LQP
then they are to Schr\"{o}dinger wave functions. Another well-known indication
is the so called Klein paradox which occurs if one couples relativistic wave
functions to external fields.} \cite{Sch1} (see also additional remarks
further down). The close connection of the Wigner particle structure via
modular theory with localization makes it easier to understand why in the
standard framework of particle physics it never has been possible to find a
nonlocal alternative associated with an elementary length. Recent attempts
based on noncommutative geometry certainly are outside the Wigner particle
framework and their main problem is to maintain consistency with observed
particle physics and its underlying principles.

The local net $\mathcal{A}(\mathcal{O})$ may be obtained in two ways, either
one first constructs the spaces $H_{R}(\mathcal{O})$ via (\ref{int}) and then
applies the Weyl functor, or one first constructs the net of wedge algebras
(\ref{Weyl}) and then intersects the algebras according to
\begin{equation}
\mathcal{A(O)}=\bigcap_{W\supset\mathcal{O}}A(W)
\end{equation}

The functorial mapping $\Gamma$ between the orthocomplemented lattice of real
Wigner subspaces and subalgebras of $B(H_{Fock})$ maps the above pre-modular
operators into those of the Tomita-Takesaki modular theory
\begin{equation}
J,\Delta,S\frak{=\Gamma(j,\delta,s)}%
\end{equation}
(for the fermionic CAR-algebras there is an additional modification by a
``twist'' operator). Whereas the ``pre-modular'' operators denoted by small
letters act on the Wigner space, the modular operators $J,\Delta$ have an $Ad$
action ($AdUA\equiv UAU^{\ast}$) on von Neumann algebras in Fock space which
makes them objects of the Tomita-Takesaki modular theory
\begin{align}
&  SA\Omega=A^{\ast}\Omega,\,S=J\Delta^{\frac{1}{2}}\\
&  Ad\Delta^{it}\mathcal{A}=\mathcal{A}\\
&  AdJ\mathcal{A}=\mathcal{A}^{\prime}\nonumber
\end{align}
The operator $S$ is that of Tomita i.e. the unbounded densely defined normal
operator which relates the dense set $A\Omega$ to the dense set $A^{\ast
}\Omega$ for $A\in\mathcal{A}$ and gives $J$ and $\Delta^{\frac{1}{2}}$ by
polar decomposition. The nontrivial miraculous properties of this
decomposition are the existence of an automorphism $\sigma_{\omega
}(t)=Ad\Delta^{it}$ which propagates operators within $\mathcal{A}$ and only
depends on the state $\omega$ (and not on the implementing vector $\Omega)$
and a that of an antiunitary involution $J$ which maps $\mathcal{A}$ onto its
commutant $\mathcal{A}^{\prime}.$ The theorem of Tomita assures that these
objects exist in general if $\Omega$ is a cyclic and separating vector with
respect to $\mathcal{A}.$ An important thermal aspect of the Tomita-Takesaki
modular theory is the validity of the Kubo-Martin-Schwinger (KMS) boundary
condition \cite{Haag}
\begin{equation}
\omega(\sigma_{t-i}(A)B)=\omega(B\sigma_{t}(A)),\,\,A,B\in\mathcal{A}
\label{KMS}%
\end{equation}
i.e. the existence of an analytic function $F(z)\equiv\omega(\sigma_{z}(A)B)$
holomorphic in the strip $-1<Imz<0$ and continuous on the boundary with
$F(t-i)=\omega(B\sigma_{t}(A)).$ The fact that the modular theory applied to
the wedge algebra has a geometric aspect (with $J$ equal to the TCP operator
times a spatial rotation and $\Delta^{it}=U(\Lambda_{W}(2\pi t))$) is not
limited to the interaction-free theory \cite{Haag}. These formulas are
identical to the standard thermal KMS property of a temperature state $\omega$
in the thermodynamic limit if one formally sets the inverse temperature
$\beta=\frac{1}{kT}$ equal to $\beta=-1.\,$This thermal aspect is related to
the Unruh-Hawking effect of quantum matter enclosed behind event/causal horizons.

Our special case at hand, in which the algebras and the modular objects are
constructed functorially from the Wigner theory, suggest that the modular
structure for wedge algebras may always have a geometrical significance with a
fundamental physical interpretation in any QFT. This is indeed true, and
within the Wightman framework this was established by Bisognano and Wichmann
\cite{Haag}.

If we had taken the conventional route via interwiners and local fields as in
\cite{Wei}, then we would have been forced to use the Borchers construction of
equivalence classes\footnote{The class of local covariant free fields
belonging to the same (m,s)-Wigner representation is a linear subclass of the
full equivalence class of relative local fields associated with a free field
which comprises all Wick-polynomials. Each cyclic field in that class
generates the same net of algebras. In the analogy with coordinates in
differential geometry this subclass corresponds to linear coordinate
transformations.} \cite{St-Wi} in order to see that the different free fields
associated with the $(m,s)$ representation with the same momentum space
creation and annihilation operators in Fock space are just different
generators of the same coherent families of local algebras i.e. yield the same
net. This would be analogous to working with particular coordinates in
differential geometry and then proving at the end that the important objects
of interests as the physical S-matrix are independent of the interpolating
fields (i.e. independent of the ``field-coordinatizations'').

The above method can be extended to all $(m,s)$ positive energy Wigner
representations. The boost transformation for $s\neq0$ has a nontrivial matrix
part whose analytic continuation for the construction of $\Delta$ requires
some care. It is very interesting to note that the spin-statistics connection
can be already seen on the level of the pre-modular structure of the Wigner
representation before one arrives at the operator algebras in Fock space.

It is interesting to note that not all positive energy Wigner representations
will lead to compactly localized algebras with pointlike generating fields.
The two notable exceptions are:

\begin{enumerate}
\item  Wigner's famous ``continuous spin'' zero mass representations in which
the two-dimensional euclidean fixed point group of a lightlike vector
$p=(1,0,0,1)$ is faithfully represented (which, different from the helicity of
the photon-neutrino family, requires an infinite dimensional Hilbert space).
The spaces $H_{R}(\mathcal{O})$ are trivial for compact $\mathcal{O}$ i.e. the
intersection of the nontrivial wedge spaces (\ref{int}) only contains the zero vector.

\item  The Wigner representation theory for massive particles in d=2+1 admits
any spin value (``any''-ons). For $s\neq(half)integer$ the spaces
$H_{R}(\mathcal{O})$ are trivial if $\mathcal{O}$ is compact and nontrivial if
$\mathcal{O}$ is a spacelike cone. For $s=(half)integer$ the double cone
spaces $H_{R}(\mathcal{O})$ are nontrivial as in higher dimensions.
\end{enumerate}

The general pre-modular theory for positive energy representations allows to
prove \cite{BGL} the standardness and nontriviality of $H_{R}(W)$ and
$H_{R}(W_{1}\cap W_{2})\equiv H_{R}(W_{1})\cap H_{R}(W_{2})$ for two
orthogonal $W_{i}^{\prime}s,$ but the nontriviality of any smaller noncompact
or compact region depends on the nature of the stability group of a physical
(positive energy) momentum. The optimal noncompact localization properties of
the famous Wigner continuous spin positive energy representations have not
been investigated. Whether one can relate physically acceptable objects with
these irreducible Wigner representations depends very much on the answer to
the best possible localization properties.

It is easy to see that for any case s$\neq$integer there is a mismatch between
the geometrically opposite and the symplectic opposite i.e.
\begin{align}
&  H_{R}(W^{\prime})\neq H_{R}(W)^{^{\prime}}\\
&  W^{\prime}=W^{opp}=Rot(\pi)W\nonumber\\
&  H_{R}(W)^{^{\prime}}=TH_{R}(W^{\prime})
\end{align}
One needs an additional ``twist'' $T$ in order to transform one into the
other. The distinction between the geometric and the symplectic opposite in
$H_{Wig}$ i.e. the appearance of $T$ \ is also the reason why the Weyl functor
is only appropriate for integer spin. For halfinteger spin for which $T$ turns
out to be multiplication by $i$ the geometric complement suggest to look at
the complement in the sense of the real bilinear form $Imi(f,g)\simeq
Re(f,g).$ Without going into details we mention that this modification leads
entails the necessity to use the CAR functor for fermions in case of
halfinteger spin. The Fock space version of the multiplication with $i$ turns
out to be the twist operator appearing in the DHR work on Fermions
\cite{Haag}. But whereas for s=halfinteger this twist does not force the
compact localization spaces to be trivial and only changes the multiparticle
symmetrized tensor products into the antisymmetric ones, the twist for anyonic
spin has quite different more dramatic consequences for the localization and
the multiparticle structure. As we have already seen the localization cannot
be better than semiinfinite string-like and as far as the multiparticle
structure is concerned one can show that it cannot be described by a tensor
product at all if one wants sharper than wedge localizations. This follows
from a No-Go theorem by \cite{Mund} who proved that spacelike cone localized
anyonic fields which have nonvanishing matrix elements between the vacuum and
the anyonic spin one-particle state and fulfill braid group commutation
relations (which they are required to do by the general spin-statistics
theorem) cannot be fields which applied to the vacuum create a pure
one-particle state vector without the admixture of vacuum polarization
components. In the terminology of the last section one may say there are no
spacelike cone localized anyonic PFG's. Such one particle creating operators
only exist for the larger wedge localization, a fact which strongly suggest to
use the Wigner description only for the construction of the wedge algebras of
``free'' anyons and use those to descend to smaller localizations by the
method of intersections. Normally the presence of these particle-antiparticle
``clouds'' in addition to one-particle components are thought of as a
characteristic property of the interaction, but here they are caused by the
braid group statistics even in the absence of genuine interactions. The
distinction of \textit{free versus interacting} based on Lagrangian
quantization is clearly not very appropriate in such a situation.

It is of paramount importance to explicitly construct these free anyonic
fields for a given spin-statistics structure. Even though they are like
Boson/Fermion fields uniquely determined by their Wigner particle structure,
some of the conceptual problems in their explicit construction are still open.
Their physical importance results from the fact that besides braid group
statistics and the related quantum symmetries there is nothing else which
distinguishes LQP in low dimensions from Fermions/Bosons in d%
%TCIMACRO{\TEXTsymbol{>}}%
%BeginExpansion
$>$%
%EndExpansion
1+2. The nonrelativistic limit does not eradicate the braid group statistics
inasmuch as the Fermi/Bose alternative and the spin-statistics connection is
not lost in this limit. Hence d=1+2 braid group statistics particles cannot be
descibed by QM even if one is only interested in their nonrelativistic
behavior. The complementary statement (which sounds more provocative) would be
to say that QM of Bosons/Fermions owes its physical relevance to the
fortuitous fact that there are relativistic fields (namely ordinary free
fields) which create one-particle state vectors without any vacuum
polarization admixture.

So if the new phenomena of high $T_{c}$-superconductivity and the fractional
quantum Hall effect are characteristic for low (two spatial) dimensions they
should be related to the braid group spin-statistics structure and hence
outside the range of the standard Lagrangian quantization approach. The
appearance of amplification factors from statistical dimension for plektons
(of potential use for high $T_{c}$) and rational statistical phases (of
potential use in the fractional Hall effect) are very encouraging, but there
is still a long way to go before the quantum field theory of anyons/plektons
and their electromagnetic couplings to external fields is understood on the
level of say the understanding of the electromagnetic properties of Dirac
Fermions. The present understanding of the fractional Hall effect has been
obtained via the ``edge current'' approximation in which the fractional
statistics effect enters via the simpler statistics structure of chiral theories.

In the presence of interactions, \ the structure of the wedge algebras is not
only determined by the Wigner theory but the S-matrix also enters in the
characterization of wedge-localized state vectors. There exist however
wedge-localized operators which, if only applied once to the vacuum, create a
one-particle state vector; whereas for any smaller localization region this
would not be compatible with the presence of interactions unless there are in
addition vacuum polarization clouds. In certain interacting cases in low
spacetime dimensions the ``polarization free generators'' (PFG's) have nice
(temperedness) analytic properties which keep them close to free systems; in
fact their Fourier transforms obey a Faddeev-Zamolodchikov algebra. In the
last section we will explain this situation in some more detail.

In passing we briefly remind the reader of the standard way of combining the
Wigner particle picture with Einstein causality through the introduction of
pointlike covariant ``field coordinatizations''.

The covariant field construction is synonymous with the introduction of
intertwiners between the unique Wigner $(m,s)$ representation and the
multitude of Lorentz-covariant momentum-dependent spinorial (dotted and
undotted) tensors, which under the homogenous L-group transform with the
irreducible $D^{\left[  A,B\right]  }(\Lambda)$ matrices.%

\begin{equation}
u(p)D^{(s)}(R(\Lambda,p))=D^{[A,B]}(\Lambda)u(\Lambda^{-1}p) \label{1}%
\end{equation}
The only restriction imposed by this intertwining is:
\begin{equation}
\mid A-B\mid\leq s\leq A+B \label{2}%
\end{equation}
This leaves many $A,B$ (half integer) \textit{choices} for a given $s$. Here
the $u(p)$ intertwiner is a rectangular matrix consisting of $2s+1$ column
vectors $u(p,s_{3}),\;s_{3}=-s,...,+s$ of length $(2A+1)(2B+1)$. Its explicit
construction using Clebsch-Gordan methods can be found in Weinberg's book
\cite{Wei}. Analogously there exist antiparticle (opposite charge) $v(p)$
intertwiners: $D^{(s)\ast}(R(\Lambda,p)\longrightarrow D^{[A,B]}(\Lambda)$.
The covariant field is then of the form:
\begin{align}
\psi^{\lbrack A,B]}(x)  &  =\frac{1}{(2\pi)^{3/2}}\int\{e^{-ipx}\sum_{s_{3}%
}u(p_{1},s_{3})a(p_{1},s_{3})+\label{field}\\
&  +e^{ipx}\sum_{s_{s}}v(p_{1},s_{3})b^{\ast}(p_{1},s_{3})\}\frac{d^{3}%
p}{2\omega}\nonumber
\end{align}
where $a^{\#}$ and $b^{\#}$ are the creation/annihilation operators for
particles/antiparticles, i.e. the n-fold application of the
particle/antiparticle creation operators generate the symmetrized (for integer
spin) or antisymmetrized (for half-integer spin) tensor product subspaces of
Fock space.

Since the range of the $A$ and $B$ (undotted/dotted) spinors is arbitrary
apart from the fact that they must fulfil the inequality (\ref{2}) with
respect to the given physical spin s\footnote{For the massless case the
helicity inequalities with respect to the spinorial indices are more
restrictive, but one Wigner representation still admits a countably infinite
number of covariant representations.}, the number of covariant fields is
countably infinite. Fortunately it turns out that this loss of uniqueness does
not cause any harm in particle physics. If one defines the polynomial $^{\ast
}$-algebras $\mathcal{P}(\mathcal{O})$ as the operator algebras generated from
the smeared field with Schwartz test functions of support $supp$
$f\in\mathcal{O}$ \cite{St-Wi}
\[
P(\mathcal{O})=^{\ast}-alg\left\{  \psi(f)\,|\,suppf\subset\mathcal{O}%
\right\}
\]
one realizes that these localized algebras do not depend on the representative
covariant field chosen from the $(m,s)$ class. In fact all the different
covariant fields which originate from the $(m,s)$ representation share the
same creation/annihilation operators. This gave rise to the linear part of the
Borchers equivalence classes of relatively local fields. The full Borchers
class \cite{St-Wi} generalized the family of Wick polynomials to the realm of
interactions and gave a structural explanation of the insensitivity of the
S-operator. Although the local operator algebras cannot be directly obtained
from the fields, the polynomial algebras of the latter are (under some weak
domain assumptions) affiliated to the von Neumann algebras $\mathcal{A(O)}$

An important property of free fields which fulfill an equation of motion is
the validity of the quantum version of the Cauchy initial value problem. The
algebraic counterpart is the causal shadow property (see beginning of this
section) which for simple connected spacetime regions $\mathcal{O}$ reads
\begin{equation}
\mathcal{A(O)=A(O}^{\prime\prime}\mathcal{)}%
\end{equation}
where $\mathcal{O}^{\prime}$ denotes the causal complement and the causal
complement of the causal complement is the causal completion (or causal
shadow) of $\mathcal{O}.$ As stated previously the causal completion of a
piece of timeslice or a piece of spacelike hypersurface is the double cone
subtended by those regions. In order to derive this property one does not have
to invoke the Cauchy initial value problem of pointlike fields; it is a
functorial consequence of an analog property of localized real subspaces of
$H_{W}$%
\begin{equation}
H_{R}(\mathcal{O})=H_{R}(\mathcal{O}^{\prime\prime})
\end{equation}
If a higher dimensional theory which fulfills the causal shadow property is
restricted to a lower dimensional manifold containing the time direction (this
is sometimes called a (mem)brane), then one obtains a physically unacceptable
theory in which, as one moves upward in time, new degrees of freedom enter
sideways. On the other hand if one tries to extend a free theory in a brane to
the ambient space the resulting theory is only causally consistent if the
objects are independent of the transversal directions in the ambient space. If
the degrees of freedom in the brane are pointlike fields, the degrees of
freedom in the ambient fields are the same, they just look like spacelike
strands going into the transverse direction\footnote{These transversal strands
are not the ``dynamical strings'' of string theory because the latter have
more degrees of freedom than fields.}. So the extension into an ambient world
are not described by standard field degrees of freedom (for a more see the
last section).

Another problem which even in the Wigner setting of noninteracting particles
has not yet been solved is the pre-modular theory for disconnected or
topologically nontrivial regions e.g. in the simplest case for disjoint double
intervals of the massless s=$\frac{1}{2}$ model on the circle. This could be
the first inroad into the terra incognita of nongeometric ``quantum
symmetries'' of purely modular origin without a classical counterpart.

\subsection{Appendix A: Coherence Relations involving Exchange Operators}

With the help of the definition (\ref{ep}) of the exchange operator
$\varepsilon(\rho_{1},\rho_{2})\in(\rho_{2}\rho_{1}|\rho_{1}\rho_{2})$ in
terms of charge transporters one can derive a set of consistency relations
which are most easily remembered in form of their graphical representations.
Irreducible endomorphisms are represented by vertical lines, the later acting
ones to the right of the former acting. Intertwiners are represented by
graphs, an intertwiner $T\in(\rho_{1}...\rho_{n}|\rho_{1}^{\prime}...\rho
_{m}^{\prime})$ has m lines which enter from below and n lines which leave
above. The multiplication $S\circ T$ is represented by juxtaposing the S-graph
on top of the T-graphs (only defined for matching source lines of S with the
range lines of T, however note that if necessary left lines may be added for
matching without changing the operator). The graph of $T^{\ast}$ is the
upside-down mirror image of that of $T.$ The flip $\varepsilon(\rho_{1}%
,\rho_{2})$ is represented by $\rho_{1}$-line which passes from right down to
left above overneath a $\rho_{2}$-line which in turn starts from from left
down to right above underneath the $\rho_{1}$ line (indicated by a breaking of
the $\rho_{2}$ line around the point of crossing). The graphical
representation of an action of $\rho$ on an intertwiner $T$ i.e. $\rho(T)$ is
a $\rho$-line on the right of the $T$-graph. Since the basic $T_{e}$-vertices
with $\rho^{\prime}s$ from the reference set $\nabla_{0}$ form a complete set,
any intertwiner can in principle be written as a linear combination of
products of $T_{e}^{\prime}s$ and their Hermitian adjoints$.$

After these graphical rules have been justified one can immediately check that
the composite exchange operators obey the following formulas (\cite{FRSI}%
\cite{FRSII})
\begin{align}
\varepsilon(\rho_{3},\rho_{1}\rho_{2})  &  =\rho_{1}(\varepsilon(\rho_{3}%
,\rho_{2}))\varepsilon(\rho_{3},\rho_{1})\\
\varepsilon(\rho_{1}\rho_{2},\rho_{3})  &  =\varepsilon(\rho_{1},\rho_{3}%
)\rho_{1}(\varepsilon(\rho_{2},\rho_{3}))\nonumber
\end{align}
whereas the $\varepsilon$ together fulfills the following coherence relations
with intertwiners $T\in(\rho_{2}|\rho_{1})$%
\begin{align}
\rho_{3}(T)\varepsilon(\rho_{1},\rho_{3})  &  =\varepsilon(\rho_{2},\rho
_{3})T\\
\rho_{3}(T)\varepsilon(\rho_{3},\rho_{1})^{\ast}  &  =\varepsilon(\rho
_{3},\rho_{2})^{\ast}T\nonumber
\end{align}
The proof is straightforward and uses in addition to the graphical rules of
the $T_{e}$ intertwiners and the action of the $\rho$ the representation
(\ref{ep}) in terms of a trivial crossing for the exchange of the spectator
endomorphisms which again allows for a graphical representation. It is much
simpler to remember these intertwiner relations in terms of their graphs.

There are some special cases which, because of their importance will be
separately mentioned. One is the exchange-fusion (pentagon) relation which is
the last formula for $T=T_{e}$ i.e. one of the basic intertwiners. For
$T=R\in(\bar{\rho}\rho|id)$ and $\varepsilon(\rho_{1},\rho_{2})=1$ if one of
the $\rho_{i}^{\prime}s$ is $id,$ one gets
\begin{equation}
\rho_{3}(R)=\varepsilon(\bar{\rho}\rho,\rho_{3})R=\varepsilon(\rho_{3}%
,\bar{\rho}\rho)^{\ast}R
\end{equation}
Finally there is the famous Artin relation (adapted to colored braids)
\begin{equation}
\rho_{3}(\varepsilon(\rho_{1},\rho_{2}))\varepsilon(\rho_{1},\rho_{3})\rho
_{1}(\varepsilon(\rho_{2},\rho_{3}))=\varepsilon(\rho_{2},\rho_{3})\rho
_{2}(\varepsilon(\rho_{1},\rho_{3}))\varepsilon(\rho_{1},\rho_{2})
\end{equation}
The usual braid group and the usual Artin relation results from specialization
to one color $\rho_{i}\equiv\rho,\,i=1,2,3$ in which case the Artin generators
of the braid $B_{n}$ group on n+1 strands are $g_{k}=\rho^{k-1}(\varepsilon)$
$k=1...n$ with $\varepsilon\equiv\varepsilon(\rho,\rho)$ which fulfill the
Artin braid relations%
\begin{equation}
g_{k}g_{k+1}g_{k}=g_{k+1}g_{k}g_{k+1} \label{Ar}%
\end{equation}
Note that this construction from a local net of observable algebras represents
the Artin generators as composites of charge transporters and endomorphisms
acting on them, so that the Artin relations are a consequence of the more
basic relations between charge transporters. This is possible because the
braid structure is embedded in a the ambient net of algebras which has a very
rich algebraic structure. In particular the braid group structure comes
equipped with a natural representation structure in terms of Markov traces on
the $B_{\infty}$ group algebra, which is the subject of the next appendix.

\begin{remark}
The braid group $B_{n}$ (and its special case the permutation group $S_{n})$
has a natural inclusive structure $B_{n}\subset B_{n+1}$ which permits to take
the inductive limit B$_{\infty}.$ This property is related to its importance
for particle statistics. Particle physics fulfills the so-called cluster
property: the physics of n particles is contained in that of n+1 particles)
and results from the latter by shifting one particle to spatial infinity, thus
converting it into a ``spectator''. Particle statistics is the discrete
structure which remains after one removes the localization aspect and the
relic of the cluster property is reflected in the inclusive aspect of the
statistics group and in the Markov property of the Markov trace on B$_{\infty
}$.
\end{remark}

Often the matrix representors of the braid group relations (\ref{Ar}) are
called Yang-Baxter relations but this is neither physically nor mathematically
correct; physically, because the more complicated true Yang-Baxter relation
belong to the concept of scattering theory and not to particle/field
statistics and mathematically because whereas the representation theory of
braids and knots is a well established area of the V. Jones subfactor theory,
the Yang-Baxter relation have yet no firm position in mathematics (despite
serious attempts to get to that structure by ``Baxterization'' of braid group representations).

\subsection{Appendix B: Classification of admissable B$_{\infty}$ Representations}

The charge-carrying fields, which in the LQP setting are operators in the
field bundle, form an exchange algebra in which R-matrices which represent the
infinite braid group $B_{\infty}$ appear. The admissable physical
representations define a so called Markov trace on the braid group, a concept
which was introduced by V. Jones but already had been used for the special
case of the permutation group S$_{\infty}$ in the famous 1971 work of
Doplicher, Haag and Roberts \cite{Haag}. Since this very physical method has
remained largely unknown\footnote{Particle physicists who are very familiar
with group theory use a deformation theory known as the ``quantum group''
method. Although its final results are compatible with the structure of
quantum theory, the intermediate steps are not (no Hilbert space and operator
algebras, appearance of null-ideals). The present method is ``quantum''
throughout.} outside a small circle of specialists, its renewed presentation
in this appendix may be helpful

In this classification approach one starts with fusing and decomposing braided
endomorphisms. The simplest case is a basic irreducible endomorphism $\rho$
whose iteration leads to a ``two channel'' irreducible decomposition
\cite{FRSI}
\begin{align}
\rho^{2}  &  \simeq id\oplus\rho_{1}\\
i.e.\left[  \rho^{2}\right]   &  =\left[  id\right]  \oplus\left[  \rho
_{1}\right] \nonumber
\end{align}
where $id$ denotes the identity endomorphism. This is the famous case leading
to the Jones-Temperley-Lieb algebra, whereas the more general two-channel
case
\begin{equation}
\rho^{2}\simeq\rho_{1}\oplus\rho_{2} \label{He}%
\end{equation}
gives rise to the Hecke algebra. Finally the special 3-channel fusion
\cite{Re1}
\begin{equation}
\rho^{2}\simeq id\oplus\rho_{1}\oplus\rho_{2}%
\end{equation}
yield the so-called Birman-Wenzl algebra.

Each single case together with the Markov trace yields of a wealth of braid
group representations. The first case comprises all the selfconjugate minimal
models and is asymptotically associated (see below) with $SU(2)$ which is a
pseudo self-conjugate group, whereas the second is similarly associated with
$SU(n)$ for $n>2$. Finally the third one has an assoiation with SO(n). There
are of course also isolated exceptional fusion laws which do not produce
families and whose basic fusion law cannot be viewed as arising from looking
at higher composites of the previous families. In all such cases one finds a
``quantization'' from the positivity of the Markov-trace \cite{GHJ}; in the
first case this is the famous Jones quantization. All cases have realizations
in chiral QFT as exchange algebra (or reduced field bundle) operators
associated with the current or W observable algebras.

The classification of the admissable braid group representation associated to
the above fusion laws (and the associated knot- and 3-manifold- invariants) is
a purely combinatorial problem of which a simpler permutation group version
(for which only (\ref{He}) occurs ) was already solved in 1971 by DHR
\cite{Haag}. The method requires to study tracial states on the mentioned
abstract $C^{\ast}$-algebras and the resulting concrete von Neumann algebras
are factors of type II$_{1}.$ These operator algebras which are too ``small''
in order to be able to carry continuous translations and allowing localization
are often referred to as ``topological field theories''. In the present
approach these combinatorial data are part of the superselection structure. If
combined with the nature of the charge-carrying fields i.e. the information
whether they form multiplets as in the case of \ current algebras or whether
there are no such group theoretic multiplicities the have the same R-matrices
and the same statistical dimensions (quantum dimensions) but their statistical
phases and therefore their anomalous dimensions may be different. The
numerical R-matrices determined from the Markov trace formalism fix the
structure of the exchange algebras.

The DHR-Jones-Wenzl technique constructs the tracial states via iterated
application of the left inverse of endomorphisms (or by the iteration of the
related V. Jones basic construction in subfactor theory). Under the assumption
of irreducibility of $\rho$ (always assumed in the rest of this section) the
previously introduced left inverse $\phi$ maps the commutant of $\rho
^{2}(\mathcal{A})$ in $\mathcal{A}$ into the complex numbers:
\begin{equation}
\phi(A)=\varphi(A)\underline{1},\quad A\in\rho^{2}(\mathcal{A})^{\prime}%
\end{equation}
and by iteration a faithful tracial state $\varphi$ on $\cup_{n}\rho
^{n}(\mathcal{A})^{\prime}$ with:
\begin{align*}
\phi^{n}(A)  &  =\varphi(A)\underline{1},\quad A\in\rho^{n+1}(\mathcal{A}%
)^{\prime}\quad\\
\varphi(AB)  &  =\varphi(BA),\quad\varphi(\underline{1})=1
\end{align*}
Restricted to the $\mathbf{C}RB_{n}$ algebra generated by the ribbon
braid-group which is a subalgebra of $\rho^{n}(\mathcal{A})^{\prime}$ the
$\varphi$ becomes a tracial state, which can be naturally extended
$(B_{n}\subset B_{n+1})$ to $\mathbf{C}RB_{\infty}$ in the above manner and
fulfills the ``Markov-property'':
\begin{equation}
\varphi(a\sigma_{n+1})=\lambda_{\rho}\varphi(a),\quad a\in\mathbf{C}RB_{n}%
\end{equation}
The terminology is that of V. Jones and refers to the famous Russian
probabilist of the last century as well as to his son, who among other things
constructed knot invariants from suitable functionals on the braid
group.\textit{\ }The ``ribbon'' aspect refers to an additional generator
$\tau_{i}$ which represents the vertical $2\pi$ rotation of the cylinder braid
group ($\simeq$ projective representation of $B_{n})$ \cite{FRSI}\cite{FRSII}$.$

It is interesting to note in passing that the Markov-property is the
combinatorial relict of the physicist's cluster property which relates the
n-point correlation function in local QFT to the n-1 point correlation (or in
QM the physics of n particles to that of n-1 by converting one of the
particles into a spectator by removing it to infinity. This Russian
``matrushka'' structure of inclusive relations requires to deal with the
inductive limit $B_{\infty}$ of the $B_{n}$ braid groups. This picture is
similar to that of cluster properties which was already used in our attempts
to describe the QM statistics in the first section. The existence of a Markov
trace on the braid group of (low dimensional) multi-particle statistics is the
imprint of the cluster property on particle statistics. As such it is more
basic than the notion of internal symmetry. It precedes the latter and
according to the DR theory it may be viewed as the other side of the same coin
on which one side is the old (compact group-) or new (quantum-) symmetry. With
these remarks the notion of internal symmetry which historically started with
Heisenberg's isospin in nuclear physics becomes significantly demystified.

Let us now return to the above 2-channel situation \cite{FRSI}. Clearly the
exchange operator $\varepsilon_{\rho}$ has maximally two different
eigenprojectors since otherwise there would be more than two irreducible
components of $\rho^{2}.$ On the other hand $\varepsilon_{\rho}$ cannot be a
multiple of the identity because $\rho^{2}$ is not irreducible. Therefore
$\varepsilon_{\rho}$ has exactly two different eigenvalues $\lambda
_{1},\lambda_{2}$ i.e.
\begin{equation}
(\varepsilon_{\rho}-\lambda_{1}\underline{1})(\varepsilon_{\rho}-\lambda
_{2}\underline{1})=0
\end{equation}%
\begin{equation}
\Longleftrightarrow\varepsilon_{\rho}=\lambda_{1}E_{1}+\lambda_{2}%
E_{2}\,\,,\quad E_{i}=\left(  \lambda_{i}-\lambda_{j}\right)  ^{-1}\left(
\varepsilon_{\rho}-\lambda_{j}\right)  ,\quad i\neq j
\end{equation}
which after the trivial re-normalization of the unitaries $g_{k}:=-\lambda
_{2}^{-1}\rho^{k-1}(\varepsilon_{\rho})$ yields the generators of the Hecke
algebra:
\begin{align}
g_{k}g_{k+1}g_{k}  &  =g_{k+1}g_{k}g_{k+1}\\
g_{k}g_{l}  &  =g_{l}g_{k}\,\,,\quad\left|  j-k\right|  \geq2\nonumber\\
g_{k}^{2}  &  =(t-1)g_{k}+t\,\,,\quad t=-\frac{\lambda_{1}}{\lambda_{2}}%
\neq-1\nonumber
\end{align}
The physical cluster property in the algebraic form of the existence of a
tracial Markov state leads to a very interesting ``quantization'' \footnote{In
these notes we use this concept always in the original meaning of Planck as a
spectral discretization, and not in its form as a deformation.} which was
first pointed out by V. Jones \cite{GHJ}. Consider the sequence of
projectors:
\begin{equation}
E_{i}^{(n)}:=E_{i}\wedge\rho(E_{i})\wedge...\wedge\rho^{n-2}(E_{i})\,\,,\quad
i=1,2
\end{equation}
and the symbol $\wedge$ denotes the projection on the intersection of the
corresponding subspaces. The notation is reminiscent of the totally
antisymmetric spaces in the case of Fermions. The above relation $g_{1}%
g_{2}g_{1}=g_{2}g_{1}g_{2}$ and $g_{1}g_{n}=g_{n}g_{1},$ $n\geq2$ in terms of
the $E_{i}$ reads:
\begin{align}
E_{i}\rho(E_{i})E_{i}-\tau E_{i}  &  =\rho(E_{i})E_{i}\rho(E_{i})-\tau
\rho(E_{i})\,\,,\quad\tau=\frac{t}{\left(  1+t\right)  ^{2}}\label{TL}\\
E_{i}\rho^{n}(E_{i})  &  =\rho^{n}(E_{i})E_{i}\,\,,\quad n\geq2\nonumber
\end{align}
The derivation of these equations from the Hecke algebra structure is
straightforward. The following recursion relation \cite{Wenzl} of which a
special case already appeared in the DHR work \cite{Haag} is however tricky
and will be given in the sequel

\begin{proposition}
The projectors $E_{i}^{(n)}$ fulfill the following recursion relation
($t=e^{2\pi i\alpha},\,-\frac{\pi}{2}<\alpha<\frac{\pi}{2}):$
\begin{align}
E_{i}^{(n+1)}  &  =\rho(E_{i}^{(n)})-\frac{2\cos\alpha\sin n\alpha}%
{\sin(n+1)\alpha}\rho(E_{i}^{(n)})E_{j}\rho(E_{i}^{(n)})\,\,,\quad i\neq
j,\quad n+1<q\label{rec}\\
E_{i}^{(q)}  &  =\rho(E_{i}^{(q-1)})\quad,\quad q=\inf\left\{  n\in
\mathbf{N,\,}n\left|  \alpha\right|  \geq\pi\right\}  \,\,\,if\,\,\alpha
\neq0,\,\,q=\infty\,\,if\,\,\alpha=0\nonumber
\end{align}
The DHR recursion for the permutation group $S_{\infty}$ is obtained for the
special case t=0 i.e. $\alpha=0.$ In this case the numerical factor in front
of product of three operators is $\frac{n}{n+1}.$
\end{proposition}

The proof is by induction. For n=1 the relation reduces to the completeness
relation between the two spectral projectors of $\varepsilon_{\rho}%
:\,E_{i}=1-E_{j},\,i\neq j.$ For the induction we introduce the abbreviation
$F=E_{j}\rho(E_{i}^{(n)})=\rho(E_{i}^{(n)})E_{j}$ and compute $F^{2}.$ We
replace the first factor $\rho(E_{i}^{(n)})$ according to the induction
hypothesis by:
\begin{equation}
\rho(E_{i}^{(n)})=\rho^{2}(E_{i}^{(n-1)})-\frac{2\cos\alpha\sin(n-1)\alpha
}{\sin n\alpha}\rho^{2}(E_{i}^{(n-1)})\rho(E_{j})\rho^{2}(E_{i}^{(n-1)})
\end{equation}
We use that the projector $\rho^{2}(E_{i}^{(n-1)})$ commutes with the algebra
$\rho^{2}(\mathcal{A})^{\prime}$ (and therefore with $E_{j}\in\rho
^{(2)}(\mathcal{A})^{\prime}$), and that its range contains that of
$\rho(E_{i}^{(n)})$ i.e. $\rho^{2}(E_{i}^{(n-1)})\rho(E_{i}^{(n)})=\rho
(E_{i}^{(n)}).$ Hence we find:
\begin{equation}
F^{2}=E_{j}\rho(E_{i}^{(n)})-\frac{2\cos\alpha\sin(n-1)\alpha}{\sin n\alpha
}\rho^{2}(E_{i}^{(n-1)})E_{j}\rho(E_{j})E_{j}\rho(E_{i}^{(n)})
\end{equation}
Application of (\ref{TL}) with $\tau=\frac{1}{2\cos\alpha}$ to the right-hand
side yields:
\begin{equation}
F^{2}=E_{j}\rho(E_{i}^{(n)})-\frac{\sin(n-1)\alpha}{2\cos\alpha\sin\alpha}%
\rho^{2}(E_{i}^{(n-1)})E_{j}\rho(E_{i}^{(n)})=\frac{\sin(n+1)\alpha}%
{2\cos\alpha\sin n\alpha}F \label{A}%
\end{equation}
where we used again the above range property and a trigonometric identity. For
$n=q-1$ the positivity of the numerical factor fails and by $F^{2}%
E_{j}=(FF^{\ast})^{2}$ and $FE_{j}=FF^{\ast}$ the operator F must vanish and
hence $E_{j}$ is orthogonal to $\rho(E_{j}^{(q-1)})$ which is the second
relation in (\ref{rec}). For $n<q-1$ the right-hand side of the first relation
in (\ref{rec}) with the help of (\ref{A}) turns out to be a projector which
vanishes after multiplication from the right with $\rho^{k}(E_{j}%
),k=1,...,n-2$ as well as with $E_{j}.$ The remaining argument uses the fact
that this projector is the largest with this orthogonality property and
therefore equal to $E_{i}^{(n+1)}$ by definition of $E_{i}^{(n+1)}$ Q.E.D.

The recursion relation (\ref{rec}) leads to the desired quantization after
application of the left inverse $\phi:$%
\begin{align}
\phi(E_{i}^{(n+1)})  &  =E_{i}^{(n)}\left(  1-\frac{2\cos\alpha\sin n\alpha
}{\sin(n+1)\alpha}\eta_{j}\right)  ,\quad i\neq j\label{pos}\\
\eta_{j}  &  =\phi(E_{j}),\,\,0\leq\eta_{j}\leq1,\,\,\eta_{1}+\eta_{2}%
=1\quad\nonumber
\end{align}
>From this formula one immediately recovers the permutation group DHR
quantization in the limit $\alpha\rightarrow0.$ In that case positivity of the
bracket restricts $\eta_{j}$ to the values $\frac{1}{2}(1\pm\frac{1}%
{d}),\,\,d\in\mathbf{N}\cup0$ and the resulting permutation group
representation is associated to the $SU(d)$-group$.$ For $\alpha\neq0$ one
first notes that from the second equation (\ref{rec}) one obtains (application
of $\phi$):
\begin{equation}
\eta_{j}E_{i}^{(q-1)}=\phi(E_{j}\rho(E_{i}^{(q-1)}))=\phi(E_{j}E_{i}%
^{(q)})=0,\quad i\neq j
\end{equation}
where the vanishing results from the orthogonality of the projectors. Since
$\eta_{1}+\eta_{2}=1$ we must have $E_{i}^{(q-1)}=0$ for i=1,2, q$\geq4,$
because $E_{i}^{(q-1)}\neq0$ would imply $\eta_{j}=0$ and $E_{j}^{(q-1)}=0.$
This in turn leads to $E_{j}\equiv E_{j}^{(2)}=0$ which contradicts the
assumption that $\varepsilon_{\rho}$ possesses two different eigenvalues. This
is obvious for $q=3$ and follows for $q>3$ from the positivity of $\phi$
(\ref{pos}) for n=2:
\begin{equation}
\phi(E_{j}^{(3)})=-\frac{\sin\alpha}{\sin3\alpha}E_{j}^{(2)}\quad
\curvearrowright E_{j}^{(2)}=0\quad\curvearrowright E_{i}^{(q-1)}%
=0,\,\,i=1,2,\,q\geq4
\end{equation}
Using (\ref{pos}) iteratively in order to descend in n starting from $n=q-2,$
positivity demands that there exists an $k_{i}\in\mathbf{N,\,}2\mathbf{\leq
}k_{i}\leq q-2,$ with:
\begin{equation}
\eta_{i}=\frac{\sin(k_{i}+1)\alpha}{2\cos\alpha\sin k_{i}\alpha}%
,\,\,i=1,2\,\,\,\,\,\curvearrowright\sin(k_{1}+k_{2})\alpha=0
\end{equation}
where the relation results from summation over $i.$ Since the only solutions
are $\alpha=\pm\frac{\pi}{q},\,k_{1}=d,\,k_{2}=q-d,\,d\in N,\,2\leq d\leq
q-2,$ one finds for the statistics parameters of the plektonic 2-channel
family the value:
\begin{equation}
\lambda_{\rho}=\sum_{i=1}^{2}\lambda_{i}\eta_{i}=-\lambda_{2}\left[  \left(
t+1\right)  \eta_{1}-1\right]  =-\lambda_{2}e^{\pm\pi i(d+1)/q}\frac{\sin
\pi/q}{\sin d\pi/q}%
\end{equation}
a formula which allows for a nice graphical representation. We have
established the following theorem:

\begin{theorem}
Let $\rho$ be an irreducible localized endomorphism such that $\rho^{2}$ has
exactly two irreducible subrepresentations. Then \cite{FRSI}:
\end{theorem}

\begin{itemize}
\item $\varepsilon_{\rho}$ has two different eigenvalues $\lambda
_{1},\,\lambda_{2}$ with ratio
\begin{equation}
\frac{\lambda_{1}}{\lambda_{2}}=-e^{\pm2\pi i/q},\quad q\in\mathbf{N}%
\cup\{\infty\},\,q\geq4
\end{equation}

\item  The modulus of the statistics parameter $\lambda_{\rho}=\phi
(\varepsilon_{\rho})$ has the possible values
\begin{equation}
\left|  \lambda_{\rho}\right|  =\left\{
\begin{array}
[c]{l}%
\frac{\sin\pi/q}{\sin d\pi/q},\,\,q<\infty\\
\frac{1}{d},0\,\,\,\,\,\,\,\,\,\,\,\,\,q=\infty
\end{array}
\right.  ,d\in N,\,\,\,2\leq d\leq q-2
\end{equation}

\item  The representation $\varepsilon_{\rho}^{(n)}$ of the braid group
$B_{n}$ which is generated by $\rho^{(k-1)}(\varepsilon_{\rho}),\,k=1,...,n-1$
in the vacuum Hilbert space is an infinite multiple of the Ocneanu-Wenzl
representation tensored with a one dimensional (abelian) representation. The
projectors $E_{2}^{(m)}$ and $E_{1}^{(m)}$ are ``cutoff'' (vanish) for
$d<m\leq n$ and $q-d<m\leq n$ respectively

\item  The iterated left inverse $\varphi=\phi^{n}$ defines a Markov trace
$tr$ on $B_{n}:$%
\begin{equation}
tr(b)=\varphi\circ\varepsilon_{\rho}(b)
\end{equation}
\end{itemize}

The ``elementary'' representation which is characterized by two numbers $d$
and $q$ gives rise to a host of composite representation which appear if one
fuses the $\rho,\rho_{1},\rho_{2}$ and reduces and then iterates this process
with the new irreducible $\rho^{\prime}$ etc. We will not present the
associated composite braid formalism. With the same method one can determine
the statistical phases up to an anyonic (abelian) phase. In order to have a
unique determination, one needs (as in the original DHR work) an information
on the lowest power of $\rho$ which contains the identity endomorphism (the
vacuum representation) for the first time. A special case of this is $\rho
^{2}\supset id$ i.e. the selfconjugate Jones-Temperley-Lieb fusion. Here\ we
will not present these computations of phases.

The problem of 3-channel braid group statistics \cite{Re1} has also been
solved with the projector method in case that one of the resulting channels is
an automorphism $\tau$:
\begin{equation}
\rho^{2}=\rho_{1}\oplus\rho_{2}\oplus\tau
\end{equation}
In that case $\varepsilon_{\rho}$ has 3 eigenvalues $\mu_{i}$ which we assume
to be different:
\begin{equation}
(\varepsilon_{\rho}-\mu_{1})(\varepsilon_{\rho}-\mu_{2})(\varepsilon_{\rho
}-\mu_{3})=0
\end{equation}
The relation to the statistics phases $\omega_{\rho},$ $\omega_{i}$ is the
following: $\mu_{i}^{2}=\frac{\omega_{i}}{\omega^{2}}.$ In addition to the
previous operators $G_{i}=\rho^{i-1}(\varepsilon_{\rho})=(G_{i}^{-1})^{\ast}$
we define projectors:
\begin{equation}
E_{i}=\rho^{i-1}(TT^{\ast})
\end{equation}
where $T\in(\rho^{2}\left|  \tau)\right.  $is an isometry and hence $E_{i}$
are the projector onto the eigenvalue $\lambda_{3}=\lambda_{\tau}$ of $G_{i}.$
In fact one finds the following relations between the $G_{i}$ and $E_{i}:$%
\begin{align}
&  E_{i}=\frac{\mu_{3}}{(\mu_{3}-\mu_{1})(\mu_{3}-\mu_{2})}(G_{i}-(\mu_{1}%
+\mu_{2})+\mu_{1}\mu_{2}G_{i}^{-1})\\
&  E_{i}G_{i}=\mu_{3}E_{i}\nonumber
\end{align}
This together with the trilinear relations between the $G_{i}^{\prime}s$ and
$E_{i}^{\prime}s$ as well as the commutativity of neighbors with distance
$\geq2$ gives (upon a renormalization) the operators $g_{i}$ and $e_{i}$ which
fulfill the defining relation of the Birman-Wenzl algebra which again depends
on two parameters. The Markov tracial state classification again leads to a
quantization of these parameters except for a continuous one-parameter
solution with statistical dimension $d=2$ which is realized in conformal QFT
as sectors on the fixed point algebra of the $U(1)$ current algebra (which has
a continuous one-parameter solution) under the action of the charge
conjugation transformation (often called ``orbifolds'' by analogy to
constructions in differential geometry).

Finally one may ask the question to what extend these families and their
descendents and some known isolated exceptional cases exhaust the
possibilities of plektonic exchange structures. Although there are some
arguments in favor, the only rigorous mathematical statement is that of Rehren
who proved that for exchange dimension $d<\sqrt{6\text{ }}$that this is indeed
the case \cite{Re2}.

\section{Conformally invariant Local Quantum Physics}

There are two situations for which the algebraic methods go significantly
beyond kinematics and reveal constructive aspects of dynamical properties,
i.e. in the present context properties which are important in the explicit
construction of interacting models. One such situation arises if the
superselection sectors are related to time- or light-like distances rather
then the DHR superselection structure which is associated with spacelike
causality. This is the case for higher dimensional (timelike) and chiral
(lightlike) conformal theories and will be subject of this section. The other
situation is that of modular localization. It turns out that in theories with
massive particles and an asymptotically complete scattering interpretation,
the modular theory for the wedge localized algebras is directly related to the
scattering S-operator; this will be explained in the next section.

The hope is that conformal theories stay sufficiently close to free theories
so that they remain mathematically controllable but without becoming
completely free. There is a mathematical theorem which supports this idea. It
states that the interpolating field associated with a zero mass particle is
necessarily a free field with canonical scale dimension \cite{PLB}. The only
way to evade this undesired free situation (totally ``protected'' in the
parlance of recent perturbative supersymmetric gauge theory) is to have at
least one field with anomalous dimension in the theory.

The observables of a conformal quantum field theory (CQFT) obey in addition to
the spacelike commutativity (Einstein causality) also timelike commutativity
(Huygens causality); the interaction is limited to lightlike distances. In
fact Huygens causality is mathematically equivalent with a net on the
compactification $\bar{M}$ of Minkowski space $M$. This at first sight seems
to force us back into the realm of free zero mass theories and its
integer-valued (for Bosons to which we will mostly limit ourselves) spectrum
of scale dimensions but fortunately there is one saving grace namely the fact
that charge-carrying fields\footnote{The fact that conformal observables on
$\bar{M}$ have an integer-valued scale spectrum does by itself not imply the
absence of interactions since there is no reason why the observable
correlations are identical to those of Wick polynomials of free fields or why
such observables imply the existence of a zero mass particle state.}
associated with such Huygens algebras live on the very rich covering space
$\widetilde{M}$ which comes with a new global causality concept. With other
words the existence of any field which does not live on $\bar{M}$ but rather
requires $\widetilde{M},$ is the indicator of a conformal interaction.

The projections of these globalized-charge carrying operators turn out to be
nonlocal, but they have physically and mathematically completely controllable
noncausality which is due to their natural origin within the setting of LQP.
Whereas in massive QFT the timelike region is the arena of interaction and
remains the unknown dynamical ``black box'', conformal theories permit for the
first time to expose this dynamical region for a systematic study resulting in
timelike exchange algebra relations \cite{anomalous} and their classification.
There is good reason to believe that this additional structure will lead to
the first explicit construction of 4-dimensional QFTs and hence may
successfully solve the more than 70 year old existence problem for QFTs in
physical spacetime.

In the following let us recall the geometric aspects of $\bar{M}$ and
$\widetilde{M}$ before we will adapt the superselection analysis of the
previous section to the realm of conformal theories.

The simplest type of conformal QFT is obtained by realizing that zero mass
Wigner representation of the Poincar\'{e} group with positive energy (and
discrete helicity) allow for an extension the conformal symmetry group
$SO(4,2)/Z_{2}$ without enlargement of the Hilbert space. Besides scale
transformations, this larger symmetry also incorporates the fractional
transformations (proper conformal transformations in 4-dim. vector notation)
\begin{equation}
x^{\prime}=\frac{x-bx^{2}}{1-2bx+b^{2}x^{2}} \label{sc}%
\end{equation}
It is often convenient to view this formula as the action of the translation
group $T(b)$ conjugated with a (hyperbolic) inversion $I$
\begin{align}
I  &  :x\rightarrow\frac{-x}{x^{2}}\\
x^{\prime}  &  =IT(b)Ix
\end{align}
which does not belong to the above conformal group, although it is unitarily
represented and hence a Wigner symmetry in the Wigner representation space.
For fixed $x$ and small $b$ the formula (\ref{sc}) is well defined, but
globally it mixes finite spacetime points with infinity and hence requires a
more precise definition; in particular in view of the positivity
energy-momentum spectral properties in its action on quantum fields. Hence as
preparatory step for the quantum field theory concepts one has to achieve a
geometric compactification. This starts most conveniently from a linear
representation of the conformal group $SO(d,2)$ in d+2-dimensional auxiliary
space $\mathbb{R}^{(d,2)}$ (i.e. without direct field theoretic significance)
with two negative (time-like) signatures
\begin{equation}
G=\left(
\begin{array}
[c]{ccc}%
g_{\mu\nu} &  & \\
& -1 & \\
&  & +1
\end{array}
\right)
\end{equation}
and restricts this representation to the (d+1)-dimensional forward light cone
\begin{equation}
LC^{(d,2)}=\{\xi=(\mathbf{\xi},\xi_{4},\xi_{5});\mathbf{\xi}^{2}+\xi_{d}%
^{2}-\xi_{d+1}^{2}=0\}
\end{equation}
where $\mathbf{\xi}^{2}=\xi_{0}^{2}-\vec{\xi}^{2}$ denotes the d-dimensional
Minkowski length square. The compactified Minkowski space $\bar{M}_{d}$ is
obtained by adopting a projective point of view (stereographic projection)
\begin{equation}
\bar{M}_{d}=\left\{  x=\frac{\mathbf{\xi}}{\xi_{d}+\xi_{d+1}};\xi\in
LC^{(d,2)}\right\}
\end{equation}
It is then easy to verify, that the linear transformations which keep the last
two components invariant, consist of the Lorentz group and that those
transformations which only transform the last two coordinates yield the
scaling formula
\begin{equation}
\xi_{d}\pm\xi_{d+1}\rightarrow e^{\pm s}(\xi_{d}\pm\xi_{d+1})
\end{equation}
leading to $x\rightarrow\lambda x,\lambda=e^{s}\,.$ The remaining
transformations, namely the translations and the fractional proper conformal
transformations, are obtained by composing rotations in the $\mathbf{\xi}_{i}%
$-$\xi_{d}$ and boosts in the $\mathbf{\xi}_{i}$-$\xi_{d+1}$ planes.

A convenient description of Minkowski spacetime $M$ in terms of this d+2
dimensional auxiliary formalism is obtained in terms of a ``conformal time''
$\tau$%
\begin{align}
M_{d}  &  =(sin\tau,\mathbf{e,}cos\tau),\,\,e\in S^{d-1}\\
t  &  =\frac{sin\tau}{e^{d}+cos\tau},\,\,\vec{x}=\frac{\vec{e}}{e^{d}+cos\tau
}\label{M}\\
e^{d}+cos\tau &  >0,\,\,-\pi<\tau<+\pi\nonumber
\end{align}
so that the Minkowski spacetime is a piece of the d-dimensional wall of a
cylinder in d+1 dimensional spacetime which becomes tiled with the closure of
infinitely many Minkowski worlds. If one cuts the wall on the backside in the
$\tau$-direction, this carved out piece representing d-dimensional
compactified Minkowski spacetime has the form of a d-dimensional double cone
symmetrically around $\tau=0,\mathbf{e=(0},e^{d}=1\mathbf{)}$ without its
boundary\footnote{The graphical representations are apart from the
compactification (which involves identifications between past and future
points at time/light-infinity) the famous Penrose pictures of $M.$}%
$\mathbf{.}$ The above directional compactification leads to an identification
of boundary points at ``infinity'' and give e.g. for d=1+1 the compactified
manifold the topology of a torus. The points which have been added at the
infinity to $M$ namely $\bar{M}\backslash M$ are best described in terms of
the d-1 dimensional submanifold of points which are lightlike with respect to
the (past infinity) apex at $m_{-\infty}=(0,0,0,0,1,\tau=-\pi)$. The cylinder
walls form the universal covering $\widetilde{M_{d}}=S^{d-1}\times\mathbb{R} $
which is tiled in both $\tau$-directions by infinitely many Minkowski
spacetimes (``heavens and hells'') \cite{LM}. If the only interest would be
the description of the compactification $\bar{M},$ then one may as well stay
with the original x-coordinates and write the d+2 $\xi$-coordinates (following
Dirac and Weyl) as%

\begin{align}
&  \xi^{\mu}=x^{\mu},\,\,\mu=0,1,2,3\\
&  \xi^{4}=\frac{1}{2}(1+x^{2})\nonumber\\
&  \xi^{5}=\frac{1}{2}(1-x^{2})\nonumber\\
&  i.e.\,\,\,\left(  \xi-\xi^{\prime}\right)  ^{2}=\left(  x-x^{\prime
}\right)  ^{2}\nonumber
\end{align}
Since $\xi$ is only defined up to a scale factor, we conclude that only
lightlike differences retain an objective meaning in $\bar{M}.$

An example of a physical theory on $\bar{M}$ is provided by non-interacting
photons. The impossibility of a distinction between space- and time- like
finds its mathematical formulation in the optical Huygens principle which says
that the lightlike separation is the only one where the physical fields
propagate and hence where intuitively speaking an interaction can happen. In
the terminology of LQP this means that the commutant of an observable algebra
localized in a double cone consists apparently of a (Einstein causal)
connected spacelike- as well as two disconnected (Huygens causality) timelike-
pieces. But taking the compactification into consideration one realizes that
all three pieces are connected and the space/time-like distinction is
meaningless on $\bar{M}.$ In terms of Wightman correlation functions this is
equivalent to the rationality of the analytically continued Wightman functions
of observable fields which includes an analytic extension into timelike Jost
points \cite{To1}\cite{anomalous}.

Therefore in order to make contact with particle physics aspects, the use of
either the covering $\widetilde{M}$ or of more general fields (see next
section) on $\bar{M}$ is important since only in this way one can implement
the pivotal property of causality together with the associated localization
concepts. As first observed by I. Segal \cite{Se} and later elaborated and
brought into the by now standard form in field theory by L\"{u}scher and Mack
\cite{LM}, a global form of causality can be based on the sign of the invariant%

\begin{subequations}
\begin{align}
&  \left(  \xi(\mathbf{e},\tau)-\xi(\mathbf{e}^{\prime},\tau^{\prime})\right)
^{2}\gtrless0,\,\,hence\\
&  \left|  \tau-\tau^{\prime}\right|  \gtrless2\left|  Arcsin\left(
\frac{\mathbf{e}-\mathbf{e}^{\prime}}{4}\right)  ^{\frac{1}{2}}\right|
=\left|  Arccos\left(  \mathbf{e\cdot e}^{\prime}\right)  \right| \nonumber
\end{align}
where the $<$ inequality characterizes global spacelike distances and $>$
corresponds to positive and negative global timelike separations. Whereas the
globally spacelike region of a point is compact, the timelike region is not.
The concept of global causality solves the so called Einstein causality
paradox of CQFT \cite{HSS} which consisted in the observation that there are
massless QFT (example: the massless Thirring model) for which the unitary
implementation of (\ref{sc}), which for sufficiently large $b$-parameters
transforms spacelike into timelike distances (passing through lightlike
infinity), would create a causality clash since the anomalous dimension fields
are only Einstein but not Huygens local. The notion of global causality
(\ref{cau}) in the sense of the covering space avoids the intermediate
trespassing through lightlike separations and uses unitaries which implement
the covering group acting on the covering space instead of (\ref{sc}).

For a particle physicist the use of covering space with its many heavens and
hells above and below is not so attractive because the experimental hardware
is not conformal covariant. Therefore it is helpful to know that there is a
way of re-phrasing the physical content of globally causal conformal fields
(which violate the Huygens principle and instead exhibit the phenomenon of
``reverberation'' \cite{HSS} inside the forward light cone) in the setting of
the ordinary Minkowski world $M$ of particle physics without running into the
trap of the causality paradox of the previous section. In this way the use of
the above $\xi$- parametrization would loose some of its importance and the
changed description may be considered as an alternative to the L\"{u}%
scher-Mack approach on covering space.

This was achieved a long time ago in a joint paper involving the present
authors \cite{S-S}. The main point of this work was to point out that the
global causality structure could be encoded into a global decomposition theory
of fields (conformal block decomposition) with respect to the center of the
conformal covering. Local fields, although they behave apparently irreducibly
under infinitesimal conformal transformations, transform in general reducibly
under the action of the global center of the covering $Z(\widetilde
{SO(d,2)}).$ This central reduction was the motivation in for the global
decomposition theory of conformal fields in \cite{S-S}. In the present setting
it reads:
\end{subequations}
\begin{align}
F(x)  &  =\sum_{\alpha,\beta}F_{\alpha,\beta}(x),\,\,F_{\alpha,\beta}(x)\equiv
P_{\alpha}F(x)P_{\beta}\label{com}\\
Z  &  =\sum_{\alpha}e^{2\pi i\theta_{\alpha}}P_{\alpha}\nonumber
\end{align}
These component fields behave analogous to trivializing sections in a fibre
bundle; the only memory of their origin from an operator on covering space is
their quasiperiodicity
\begin{align}
ZF_{\alpha,\beta}(x)Z^{\ast}  &  =e^{2\pi i(\theta_{\alpha}-\theta_{\beta}%
)}F_{\alpha,\beta}(x)\label{tr}\\
U(b)F(x)_{\alpha,\beta}U^{-1}(b)  &  =\frac{1}{\left[  \sigma_{+}(b,x)\right]
^{\delta_{F}-\zeta}\left[  \sigma_{-}(b,x)\right]  ^{\zeta}}F(x)_{\alpha
,\beta}\nonumber\\
\zeta &  =\frac{1}{2}(\delta_{F}+\theta_{\beta}-\theta_{\alpha})\nonumber\\
U(\lambda)F(x)_{\alpha,\beta}U^{-1}(\lambda)  &  =\lambda^{\delta_{F}%
}F(\lambda x)_{\alpha,\beta}\nonumber
\end{align}
where the second line is the transformation law of special conformal
transformation of the components of an operator $F$ with scale dimension
$\delta_{F}$ sandwiched between superselected subspaces $H_{\alpha}$ and
$H_{\beta}$ associated with central phase factors $e^{2\pi i\theta_{\alpha}}$
and $e^{2\pi i\theta_{\beta}}$ and the last line is the scale transformation.
Using the explicit form of the conformal 3-point function it is easy to see
that phases are uniquely given in terms of the scaling dimensions $\delta$
which occur in the conformal model \cite{S-S}.
\begin{equation}
e^{2\pi i\theta}\in\left\{
\begin{array}
[c]{c}%
\{e^{2\pi i\delta}|\,\delta\in scaling\,spectrum\}\,\,\,\,Bosons\\
\{e^{2\pi i(\delta+\frac{1}{2})}|\,\delta\in scaling\,spectrum\}\,\,\,Fermions
\end{array}
\right.
\end{equation}
A central projector projects onto the subspace of all vectors which have the
same scaling phase i.e. onto the so called conformal block associated with the
center, so the labelling refers to (in case of Bosons) the anomalous
dimensions $mod(1).$

The prize one has to pay for this return to the realm of particle physics on
$M$ in terms of component fields (\ref{com}) is that these projected fields
are not pointlike Wightman fields and hence there is no chance to associate
them with a Lagrangian or Euclidean action; the timelike decomposition theory
transcends the standard QFT approach though not its underlying principles.
Unlike ordinary pointlike fields the component fields depend on a source and
range projector and if applied to a vector, the source projector has to match
the Hilbert space i.e. $F_{\alpha,\beta}$ annihilates the vacuum if $P_{\beta
}$ is not the projector onto the vacuum sector. This is very different from
the behavior of the original $F$ which, in case it was localized in a region
with a nontrivial spacelike complement, cannot annihilate the vacuum. This
kind of projected fields are well known from the exchange algebra formalism of
chiral QFT \cite{R-S} and they appear in a rudimentary form already in
\cite{S-S}.

The crucial question now is: what is the timelike/spacelike structure of the
double-indexed component fields? Whereas it is easy to see that the
$F_{\alpha,\beta}$ are genuinely nonlocal fields without any spacelike
commutation relation, consistency considerations using analytic properties of
F-correlation functions suggest the following ($\pm$) timelike relations%

\begin{align}
F_{\alpha,\beta}(x)G_{\beta,\gamma}(y)  &  =\sum_{\beta^{\prime}}%
R_{\beta,\beta^{\prime}}^{(\alpha,\gamma)}G_{\alpha,\beta^{\prime}}%
(y)F_{\beta^{\prime},\gamma}(x),\,\,x>y\label{ex}\\
R  &  \rightarrow R^{-1}\,\,for\text{\thinspace\thinspace}x<y\nonumber
\end{align}
i.e. a commutation relation with R-matrices which generate a representation of
the infinite braid group $B_{\infty}$. The consisteny problems of the
simultaneous validity of these timelike braid group structure with the
spacelike permutation group (Bosons/Fermions) commutation relations have been
analysed within the analytic framework of vacuum correlation functions and
their analytic continuations \cite{anomalous}. This suggests \cite{private}
the presence of the following group $G_{\infty}$ which generalizes the braid
group on infinitely many strands $B_{\infty}$%
\begin{align}
b_{i}t_{j}  &  =t_{j}b_{i},\,\,\left|  i-j\right|  \geq2\\
b_{i}t_{j}t_{i}  &  =t_{j}t_{i}b_{j},\,\left|  i-j\right|  =1\nonumber\\
b_{i}b_{j}t_{i}  &  =t_{j}b_{i}b_{j},\,\,\left|  i-j\right|  =1\,\,\nonumber
\end{align}
Here the b's are the generators of the braid group, i.e. they fulfill the
Artin relations (\ref{Ar}) among themselves, whereas the t's are the
transpositions which generate the permutation group. The above ``mixed''
relations are the consistency relations between the timelike braiding and the
spacelike permuting. There is yet no systematic study of the representation
theory of $G_{\infty}$ apart from the determination of a particular family of
abelian representations \cite{private}. Neither does there presently exist a
derivation of the mixed group $G_{\infty}$ within the DHR superselection
formalism. \ 

\subsection{Chiral Conformal QFT}

For d=1+1 one encounters a very special situation which leads to a significant
simplification of the above formalism. Already for the classical wave equation
the 2-dimensional situation is very different from the higher dimensional one.
Whereas in higher dimension the characteristic initial value problem is
uniquely defined by giving data on one lightfront only, for d=1+1 one needs
the characteristic data on the right and left lightray in order to have a
unique specification throughout spacetime\footnote{For the propagation in the
massive case the data on one light ray suffice.}. This leads to the well-known
doubling of degrees of freedom: the general wave solution in massless d=1+1
consists of right- and left- moving chiral contributions.

The starting point of the chiral factorization of d=1+1 massless conformal QFT
is the observation that the conformal group $PSO(2,2)$ factorizes into two
$PSL(2)\simeq PSU(1,1).$ One then naturally expects that the subtheories which
commute with the left/right Moebius group $PSL(2)=SL(2)/\mathbb{Z}_{2}$ are
the two chiral components into which the theory tensor-factorizes, a fact
which one can rigorously prove in the LOP setting \cite{Rehren}
\begin{align}
H  &  =H_{+}\otimes H_{-}\\
\mathcal{A}(M^{(1,1)})  &  =\mathcal{A}(\mathbb{R})\otimes\mathcal{A}%
(\mathbb{R})\nonumber\\
\mathcal{A}(\bar{M}^{(1,1)})  &  =\mathcal{A}(S^{1})\otimes\mathcal{A}(S^{1})
\end{align}
where the last line is the factorization of the compactification on which the
$PSU(1,1)$ acts independently on each factor. The Moebius group $PSL(2,R)$ is
generated by the following transformations (translations, dilations, proper
conformal transformations)
\begin{align}
x  &  \rightarrow x+a\\
x  &  \rightarrow\lambda x\nonumber\\
x  &  \rightarrow\frac{x}{1-bx}\nonumber
\end{align}
which form a finite dimensional subgroup of the diffeomorphism group of the
one-point compactified real line $\mathbb{\text{\r{R}}\simeq}S^{1}.$ In most
of the literature one finds the following formula for the action of the
(global) diffeomorphism $x\rightarrow f(x)$ on the fields
\begin{equation}
A(x)\rightarrow\left(  f^{\prime}(x)\right)  ^{d_{A}}A(f(x))\, \label{law}%
\end{equation}
This formula is incorrect for fields with anomalous dimension $d_{A}.$ In fact
these fields live on the covering space and cannot obey a transformation law
as (\ref{law}) in which the x-dependent prefactor is not operator- but only
numerical-valued. Numerical valuedness is only possible for the transformation
of the component fields $A_{\alpha,\beta}$ in the central decomposition
(\ref{ex}). But even in that case the correct law depends on the source and
range projectors through a frequency split e.g. under proper conformal
transformations is (see also \cite{S-S})
\begin{align}
A_{\alpha,\beta}(x)  &  \rightarrow\left(  \frac{1}{\left(  1-bx\right)  ^{2}%
}\right)  _{+}^{d_{A}-\xi}\left(  \frac{1}{(1-bx)^{2}}\right)  _{-}^{\xi
}A_{\alpha,\beta}(\frac{x}{1-bx})\label{tra}\\
\xi &  =\frac{1}{2}(d_{A}+\theta_{\beta}-\theta_{\alpha})\nonumber
\end{align}
This frequency split maintains the spectrum property\footnote{In one
mathematical treatise \cite{Schot} this inconsisteny of (\ref{law}) was
noticed but in order to maintain harmony with the physics literature its
clarification was not pursued.}. Only inside analytically continued
correlation functions of products of observable fields (integer scale
dimensions) the transformation law appears as if it would be coming from
(\ref{law})
\begin{align*}
w(z_{1},...,z_{n})  &  \rightarrow\prod_{i}\left(  f^{\prime}(z_{i})\right)
^{d_{A}}w(f(z_{1}),...,f(z_{n}))\\
w(z_{1},...,z_{n})  &  =anal.cont.\,of\,\,\left\langle 0\left|  A(x_{1}%
)...A(x_{n})\right|  0\right\rangle
\end{align*}
Diffeomorphisms beyond the Moebius group change the vacuum into another state
and destroys the analytic properties of the vacuum expectations by generating
singularities within the natural (BWH) analyticity domain of Wightman
functions \cite{St-Wi}. There is however no mathematical concept by which one
can avoid the more complicated transformation laws of operators under those
conformal transformations which change infinity. Fields which depend
holomorphically on complex coordinates $z,$ as they appear sometimes in the
literature on chiral QFT and in string theory, are meaningless since analytic
properties are not part of the operator algebra but enter through states and
manifest themselves in state vectors and vacuum correlation
functions\footnote{Whereas it is true that fields applied to the vacuum define
normalizable state vectors if the imaginary parts of their analytical
continued spacetime arguments are ordered, this ordering prescription prevents
an encoding into universal operator domains which (as in the Wightman
approach) are preserved under successive applications.}; in thermal
representations of the same operator algebra they are completely different.
Therefore the terminology of physicists of calling (local) chiral fields
``holomorphic'' is somewhat unfortunate. As exemplified in formula
(\ref{tra}), the transformation properties of anomalous dimension component
fields $A_{\alpha\beta}$ has subtle phases in case the diffeomorphisms change
the infinity of Minkowski spacetime.

Unlike to the higher dimensional conformal theories where there is a
consistency problem between space- and time-like causality of observables, the
separations in the chiral theories are lightlike and ``chiral causality''
simply means disjointness of the lightlike intervals. In this case the
previous decomposition theory of fields on the covering (\ref{ex}) only
involves positive/negative lightlike distances and the compatibility problem
between space- and time-like algebraic structure is absent. For chiral
theories one knows many models in which a nontrivial braid group structure
does occur.

The chiral case is also more easily susceptible to an algebraic analysis in
terms of DHR localized endomorphisms on $\mathcal{A}(S^{1}),$ since one only
has to pay attention to one kind of Haag duality. All the content of the
previous chapter is applicable, one just has to replace the double cones by
intervals on $S^{1}$. The observable algebra is best described as a
pre-cosheaf which is a map
\begin{equation}
I\rightarrow\mathcal{A}(I),\,I\subset\subset S^{1}%
\end{equation}
which is isotonic, local (i.e. operators commute for disjoint localization)and
fulfills the positive energy condition as in section 2. Haag duality in the
compactified description follows automatically from conformal invariance and
locality. The ordering structure on a light ray does not only permit the more
general braid group structure instead of the usual permutation group, but one
can even proof that a chiral model with only (anti)commuting fields is
associated to a free field theory in Fock space.

There are two important questions which go beyond the content of second chapter:

1) How does one systematically construct chiral theories from the
superselection data?

2) How does one use these data for the construction of d=1+1 conformal field theories?

The attentive reader will notice that the first question implies a change of
strategy. Whereas for the structural analysis of LQP it was advantageous to
start with the observable algebra in order to classify the admissable
superselection structures and in particular the admissible (braid group)
statistics, for the actual construction of models it may turn out to be easier
to look directly for the spacetime carriers of the superselection charges
without constructing first the observable algebras. This is precisely the
message from the interaction free case since the free fields themselves are
more easily constructed (see the second section) than their associated local
observables which usually are given in terms of local composites. For chiral
theories in particular this would mean to look for an alternative method to
the construction of anomalous dimensional fields than that via representation
theory of current algebras.

Whereas the first problem is still at the beginning of its understanding (see
remarks in next section), the extension of chiral theories to two-dimensional
local theories is a well-studied subject \cite{Rehren}. It is part of a
general extension problem of $\mathcal{A}\subset\mathcal{B,}$ in this case of
the extensions of endomorphisms on $\mathcal{A}$ to endomorphisms on
$\mathcal{B}$.

\subsection{Charge Transport around S$^{1}$}

Up to now we were not completely precise about in what mathematical way a
global algebra with its localizable representations and endomorphisms should
be related to the net of local algebras. This cavalier attitude did not cause
any problem as long as the local algebras were indexed by double cones (more
general compact regions) in Minkowski spacetime. In this case the system of
local algebras is really a bona fide net in the strict mathematical sense
since two double cones are always contained in a sufficiently large third
double cone i.e. the local system is directed towards infinity. But then there
is a natural globalization namely the $C^{\ast}$-algebra of the inductive
limit $\mathcal{A}_{qua}$ in the operator norm topology introduced in section
2 (\ref{gl}). This $C^{\ast}$-algebra still remembers its local origin: its
elements can be approximated uniformly by local elements. If we would have
taken the von Neumann closure, the limit would have been too big for an
interesting representation theory. It is then easy to show that a
representation of the net of local algebras (see below) is uniquely associated
with a representation (in the usual sense) of $\mathcal{A}_{qua},$ a fact
which we have already used in section 2. In conformal theories however this
procedure would be very clumsy and artificial since e.g. in the chiral case
infinity in the compact $S^{1}$ description is a special point like any other
point which is not left invariant by the conformal transformations. In that
case the globalization is most efficiently done in terms of a universal
$C^{\ast}$-algebra $\mathcal{A}_{univ}(S^{1})$ which is different (it turns
out to be bigger) from the ``non-compact'' DHR quasilocal algebra
$\mathcal{A}_{qua}(\mathbb{R}).$

In order to understand its construction, we note that the net $\left\{
\mathcal{A}(I)\right\}  _{I\subset S^{1}}$ is not directed; in fact because of
this it is not a net in the mathematical use of that word, but rather a
``pre-cosheaf''. The globalization of a pre-cosheaf is done relative to a
class of distinguished representations and leads to the so-called universal
C$^{\ast}$-algebra $\mathcal{A}_{univ}$ as follows

\begin{definition}
$\mathcal{A}_{univ}$ is the $C^{\ast}$algebra which is uniquely determined by
the system (precosheaf\footnote{The terminology results from the fact that it
is dual to the pre-sheaf of partial normal states associated with the
$\mathcal{O}^{\prime}s.$ Often physicists by abuse of laguage continue to use
``nets'' even if the system of $\mathcal{O}^{\prime}s$ is not directed as in
case of compact spacetimes.}) of local algebras $(A(I))_{I\in\mathcal{T}},$
$\mathcal{T}=$ family of proper intervals $I\subset S^{1}$ and the following
universality condition:

(i)\thinspace\thinspace\thinspace\thinspace there are unital embeddings
$i^{I}:\mathcal{A}(I)\rightarrow\mathcal{A}_{univ}$ s. t..
\begin{equation}
i^{J}\mid_{\mathcal{A}(I)}=i^{I}\text{ \thinspace\thinspace}if\,\,\text{
}I\subset J,\,\,I,J\in\mathcal{T}%
\end{equation}
and $\mathcal{A}_{univ}$ is generated by the algebras $i^{I}(\mathcal{A}%
(I)),\,\,I\in\mathcal{T};$

(ii)\thinspace\thinspace\thinspace\thinspace for every coherent family of
representations $\pi^{I}:\mathcal{A}(I)\rightarrow\mathcal{B}(H_{\pi})$ there
is a unique representation $\pi$ of $\mathcal{A}_{univ}$ in $H_{\pi}$ s. t..
\begin{equation}
\pi\circ i^{I}=\pi^{I}%
\end{equation}
\end{definition}

The universal algebra inherits the action of the M\"{o}bius group as well as
the notion of positive energy representation through the embedding.

The universal algebra has more (global) elements than the quasilocal algebra
of the DHR theory, in fact one obtains $\mathcal{A}_{qua}(\mathbb{R})$ by
puncturing $\mathcal{A}_{univ}(S^{1})$ at ``infinity'': $\mathcal{A}%
_{qua}\equiv\mathcal{A}\subset\mathcal{A}_{univ}$ with the consequence that
the vacuum representation $\pi_{0}$ ceases to be faithful ($\mathcal{A}%
_{univ}$ includes annihilators of the vacuum) and the global superselection
charge operators which are outer for $\mathcal{A}_{qua}$ become inner for
$\mathcal{A}_{univ}$. From this observation emerges the algebra of Verlinde
\cite{Verlinde} (explained below) which originally was obtained by geometric
(rather than local quantum physics) arguments in the limited context of chiral
conformal theories. The removal of a point $\xi$ from $S^{1}$ (this removal
recreates the infinity of $\mathcal{A}_{qua}$) forces $\mathcal{A}_{univ}$ to
shrink to $\mathcal{A}.$

Most of this new features can be seen by studying global intertwiners in
$\mathcal{A}_{univ}.$ Let $I,J$ $\in\mathcal{T}$ and $\xi,\zeta\in I^{\prime
}\cap J^{\prime}$ (i.e. two points removed from the complements) and choose
$\rho$ and $\sigma$ s. t.. loc$\rho,$ loc$\sigma\subset I$ and $\hat{\rho}%
\in\left[  \rho\right]  $ with loc$\hat{\rho}\subset J.$ Then the statistics
operators $\varepsilon(\rho,\sigma)$ and $\varepsilon(\sigma,\rho
)\in\mathcal{A}(I)\subset A_{\xi}\cap A_{\zeta}$ are the same (i.e. they don't
need a label $\xi$ or $\zeta)$ independently of whether we use the quasilocal
algebra $\mathcal{A}_{\xi}$ or $\mathcal{A}_{\zeta}$ for their definition. By
Haag duality a charge transporter $V:$ $\pi_{0}\rho\rightarrow\pi_{0}\hat
{\rho}$ lies both in $\pi_{0}(\mathcal{A}_{\xi})$ and $\pi_{0}(\mathcal{A}%
_{\zeta}).$ However its pre-images with respect to the embedding are different
iff the monodromy operator is nontrivial i.e. iff the braidgroup
representation does not reduce to that of the permutation group. In fact:
\begin{align}
V_{\rho}  &  \equiv V_{+}^{\ast}V_{-}\text{ with}\,\,\,V_{+}\in\mathcal{A}%
_{\xi},\quad V_{-}\in\mathcal{A}_{\zeta}\label{V}\\
V_{\rho}  &  \in\left(  \rho,\rho\right) \nonumber
\end{align}
is a global selfintertwiner, which is easily shown to be independent of the
choice of $V$ and $\hat{\rho}$ (i.e. $\hat{\rho}$ is a ``spectator'' in the
aforementioned sense)$.$ The representation of the statistics operators in
terms of the charge transporters $\varepsilon(\rho,\sigma)=\sigma(V_{+}%
)^{\ast}V_{+},\,\,\varepsilon(\sigma,\rho)^{\ast}=\sigma(V_{-})^{\ast}V_{-}$
leads to:
\begin{equation}
\sigma(V_{\rho})=\varepsilon(\rho,\sigma)V_{\rho}\varepsilon(\sigma
,\rho)\,\,\curvearrowright\pi_{0}\sigma(V_{\rho})=\pi_{0}\left[
\varepsilon(\rho,\sigma)\varepsilon(\sigma,\rho)\right]  \label{around}%
\end{equation}
The first identity is very different from the relation between $\varepsilon
^{\prime}s$ due to local intertwiners. The global intertwiner $V_{\rho}$ is
trivial in the vacuum representation, thus showing its lack of faithfulness
with respect to $\mathcal{A}_{univ}$. The global aspect of $V_{\rho}$ is only
activated in charged representations where it coalesces with monodromy
operators. From its definition it is clear that it represents a charge
transport once around the circle$\footnote{Note that in $A_{univ}$ which
corresponds to a compact quantum world it is not possible to ``dump'' unwanted
charges to ``infinity'', as in the case of $A_{quasi}$, but instead one
encounters ``polarization'' effects upon charge transportation once around
i.e. the round transport in the presence of charged endomorphism is different
from that in the vacuum endomorphism.}.$ In the analytically continued vacuum
expectation values the algebraic monodromy of charge transport aquires the
monodromy around a branch cut in complex function, but in the vein of previous
remarks, the multivaluedness of branchings in analytical continuations has no
direct place in the operator algebra theory.

The left hand side of the first equation in (\ref{around}) expresses a
transport ``around'' in the presence of another charge $\sigma,$ i.e. a kind
of charge polarization. Let us look at the invariant version of $V_{\rho}$
namely the global ``Casimir'' operators $W_{\rho}=R_{\rho}^{\ast}V_{\rho
}R_{\rho}:id\rightarrow id.$ This operator lies in the center $\mathcal{A}%
_{univ}\cap\mathcal{A}_{univ}{}^{\prime}$ and depend only on the class
(=sector) $\left[  \rho\right]  $ of $\rho.$ By explicit
computation\cite{FRSII} one shows that after the numerical renormalization
$C_{\rho}:=d_{\rho}W_{\rho}$ one encounters the fusion algebra:
\begin{align}
(i)\,\,\,\,C_{\sigma\rho}  &  =C_{\sigma}\cdot C_{\rho}\\
(ii)\,\,\,\,C_{\rho}^{\ast}  &  =C_{\bar{\rho}}\nonumber\\
(iii)\,\,\,\,C_{\rho}  &  =\sum_{\alpha}N^{\alpha}C_{\alpha}%
\,\,\,\,if\,\,\,\rho\simeq\oplus_{\alpha}N^{\alpha}\rho_{\alpha}\nonumber
\end{align}
Verlinde's modular algebra emerges upon forming matrices with row index equal
to the label of the central charge and the column index to that of the sector
in which it is measured:
\begin{equation}
S_{\rho\sigma}:=\left|  \sum_{\gamma}d_{\gamma}^{2}\right|  ^{-\frac{1}{2}%
}d_{\rho}d_{\sigma}\cdot\pi_{0}\sigma(W_{\rho})
\end{equation}
In case of nondegeneracy of sectors, which expressed in terms of statistical
dimensions and phases means $\left|  \sum_{\rho}\kappa_{\rho}d_{\rho}%
^{2}\right|  ^{2}=\sum_{\rho}d_{\rho}^{2},$ the above matrix $S$ is equal to
Verlinde's matrix $S$ \cite{Verlinde}which together with the diagonal matrix
$T=\kappa^{-1}Diag(\kappa_{\rho}),$ with$\,\,\kappa^{3}=(\sum_{\rho}%
\kappa_{\rho}d_{\rho}^{2})/\left|  \sum_{\rho}\kappa_{\rho}d_{\rho}%
^{2}\right|  $ satisfies the defining equations of the generators of the genus
1 mapping class group which is $SL(2,\mathbb{Z})$
\begin{align}
SS^{\dagger}  &  =1=TT^{\dagger},\quad TSTST=S\\
S^{2}  &  =C,\quad C_{\rho\sigma}\equiv\delta_{\bar{\rho}\sigma}\nonumber\\
TC  &  =CT\nonumber
\end{align}
It is remarkable that these properties are common to chiral conformal theories
and to d=2+1 plektonic models even though the localization properties of the
charge-carrying fields are quite different i.e. the S-T structure is not
limited to conformal theories as the original Verlinde argument (which uses
geometry properties ascribed to correlation functions instead of charge
transporters of LQP) may suggest. One also obtains the general validity of the
phase relation:
\begin{equation}
\frac{\kappa}{\left|  \kappa\right|  }=e^{-2\pi ic/8}%
\end{equation}
where $c$ is a parameter which in the chiral conformal setting is known to
measure the strength of the two-point function of the energy-momentum tensor
(the Virasoro central term). This identification may be derived by studying
the (modular) transformation properties of the Gibbs partition functions for
the compact Hamiltonian $L_{0}$ of the conformal rotations under thermal
duality transformations $\beta\rightarrow1/\beta$. For d=2+1 plektons, no
similar physical interpretation is known. It is interesting to confront these
results with the structure of superselection rules of group algebras in the
appendix of the first section.

\begin{lemma}
The matrix $S$ is similar to the character matrix in the appendix section 1.2.
However in distinction to nonabelian finite groups (which also yield a
\textit{finite} set of charge sectors of the fixed point observable algebra)
the present nonabelian sectors produce a symmetric ``character'' matrix $S$
which signals a perfect self-duality between charge measurers $\left\{
Q\right\}  $ and charge creators $\left\{  \rho\right\}  $ i.e. as if the
representation dimensions and the size of the conjugacy classes coalesce$.$
The group theoretic case does not lead to the S-T modular group structure.
Furthermore the algebra $\mathcal{Q}$ generated by the central charges and the
action of the endomorphisms on those charges\footnote{This action leads out of
the center and generates a global subalgebra of $\mathcal{A}_{univ}.$} do not
contain the DR group theoretic structure since it only involves endomorphisms
with nontrivial monodromy.
\end{lemma}

This strongly suggests to try to understand the new ``quantum symmetry''
property in terms of the structural properties of the algebra $\mathcal{Q}.$
As a generalization of the Verlinde matrix $S$ one finds for the $Q^{\prime}s$
in the presence of more than one polarization charges the entries of the
higher genus mapping class group matrices. The reason is that in addition to
the above process whose schematic description
\begin{equation}
vacuum\overset{split}{\longrightarrow}\rho\bar{\rho}\overset{global\,\,\rho
}{\underset{selfintertw.}{\longrightarrow}}\rho\bar{\rho}\overset
{fusion}{\longrightarrow}vacuum
\end{equation}
led to the global intertwiner $W_{\rho}=R_{\rho}^{\ast}V_{\rho}R_{\rho},$
there are the more involved global intertwiners associated with processes in
which the global selfintertwining occurs after a split of a \textit{non-vacuum
charge} $\sigma$ and a later fusion to $\mu$ which appear in a $\rho\bar{\rho
}$ reduction:
\begin{equation}
\sigma\overset{split}{\longrightarrow}\alpha\beta\overset{global\,\alpha
}{\underset{intertw.}{\longrightarrow}}\alpha\beta\overset{fusion}%
{\longrightarrow}\mu,\quad\sigma,\mu\subset\rho\bar{\rho}\quad
\end{equation}
with the global intertwiner $V_{\alpha}\in\left(  \alpha,\alpha\right)  $
being used in $T_{e(\sigma)}^{\ast}V_{\alpha}T_{e(\mu)}$ where $T_{e(\mu)}$ is
the $\alpha\beta\rightarrow\mu$ fusion intertwiner and the Hermitian adjoints
represent the corresponding splitting intertwiner. As in the vacuum case, the
selfintertwiners $V_{\alpha}$ become only activated after the application of
another endomorphism say $\eta,$ i.e. in the presence of another charge $\eta$
(hence the name ``polarization'' mechanism). It can be shown that the
following $T_{e(\sigma)}^{\ast}V_{\alpha}T_{e(\mu)}$ operators are the
building blocks of the mapping class group matrices which have
multicharge-``measurer'' ($Q$-) and multicharge-creator ($\rho$-) column and
row multiindices and are formed from repeated use of operators of the form
\begin{equation}
\phi_{\lambda}((T_{g(\eta)}\eta(T_{e(\mu)}^{\ast}V_{\alpha}T_{e(\sigma
)})T_{f(\eta)}^{\ast})):\eta\rightarrow\eta
\end{equation}
Here $T_{f(\eta)}$ and $T_{g(\eta)}$ are the intertwiners corresponding to the
charge edges $f(\eta):\lambda\sigma\rightarrow\eta$ and $g(\eta):\lambda
\mu\rightarrow\eta,$ whereas $\phi_{\lambda}$ is the left inverse of the
endomorphism $\lambda.$ Besides the global intertwiners $V,$ we only used the
local splitting intertwiners and their Hermitian adjoints which represent the
fusion intertwiners. The main question is: why do we organize the numerical
data of the global charge-measurer and charge-creator algebra $\mathcal{Q}$ as
entries in a multiindex matrix? This is ultimately a question about physical
interpretation and the use of this algebra $\mathcal{Q}$ in LQP. The
difficulty here is, that although in the present stage of development of LQP
one understands the combinatorial properties of superselected charges
including their braid group statistics, there is yet no understanding of the
d=2+1 spacetime carriers of these properties which would be needed for
applications to fractional quantum Hall- or High $T_{c}$-phenomena. One would
expect the above matrix $S$ and its multicharge mapping class group
generalization to show up in scattering of ``plektons''. The formalism and its
interpretation for charged fields with braid group statistics is expected to
be quite different from standard Lagrangian physics and attempts to treat
plektons within the standard setting by manipulating Chern-Simons Lagrangians
have remained inconclusive. In the operator algebra setting this
\textit{natural non-commutativity} (i.e. without changing the classical
spacetime indexing of nets into something noncommutative) caused by braid
group statistics is more visible and suggests a constructive approach based on
the (Tomita) modular wedge localization (see next section) which is presently
under way, but there is still a long way to go.

Finally some additional remarks on the higher dimensional conformal case
treated at the beginning of this section are called for. In that case we have
stayed away from the formulation in terms of endomorphism of $\mathcal{A}%
_{univ}(\bar{M})$ and the ensuing charge transport around the compact
Dirac-Weyl world $\bar{M}.\,$\ \ The reason is not that we have doubts about
its validity, but rather that for the consistency with the spacelike
Boson/Fermion DHR statistics we would have to understand how the $\bar{M}$
timelike Haag duality is related to the spacelike Haag dualization on $M.$
Since dualizations in the pointlike field setting do not change the pointlike
fields themselves but only the way in which algebras indexed by spacetime
regions are generated by these fields, we found it safer to use the pointlike
framework in the hope a future more rigorous treatment using endomorphisms and
showing consistency of timelike/spacelike aspects will confirm our findings.

\section{Constructive use of Modular Theory}

In order to formulate the modular localization concept in the case of
interacting particles, one must take note of the fact that the scattering
matrix\footnote{Mathematicians who are not familiar with the physically
pivotal scattering theory which relates interacting to free theories may look
up \cite{Haag} or take (\ref{mo}) as a definition of S. The modular
construction is independent of the scattering interpretation of this
operator.} $S$ of local QFT is the product of the interacting TCP-operator
$\Theta$ (mentioned in the second section) with the free (incoming) TCP
operator $\Theta_{0}$ and (since the rotation by which the Tomita reflection
$J$ differs from $\Theta$ is interaction-independent as all connected
Poincar\'{e} transformations are interaction-independent) we have:
\begin{equation}
S=\Theta\cdot\Theta_{0},\quad S=J\cdot J_{0} \label{mo}%
\end{equation}
and as a result we obtain for the Tomita involution $\check{S}$ (to avoid
confusion with the S-matrix we now write $\check{S}$ for the Tomita involution
which was called $S$ in the previous section)
\begin{equation}
\check{S}=J\Delta^{\frac{1}{2}}=SJ_{0}\Delta^{\frac{1}{2}}=S\check{S}_{0}
\label{rel}%
\end{equation}
Again we may use covariance in order to obtain $\check{S}(W)$ and the
localization domain of $\check{S}(W)$ as $\mathcal{D}(\check{S}%
(W))=\mathcal{H}_{R}(W)+i\mathcal{H}_{R}(W)$ i.e. in terms of a net of closed
real subspaces $\mathcal{H}_{R}(W)\in\mathcal{H}_{Fock}$ of the incoming Fock
space. However now the construction of an associated von Neumann algebra is
not clear since an ``interacting'' functor from subspaces of the Fock space to
von Neumann algebras is not known. In fact whereas the existence of a functor
from the net of real localized \textit{Wigner subspaces} $H_{R}(W)\subset
H_{Wig}$ to a net of von Neumann algebras is equivalent to
\begin{equation}
H_{R}(W_{1}\cap W_{2})=H_{R}(W_{1})\cap H_{R}(W_{2})
\end{equation}
The equality can be shown to become an inequality $\subset$ for the above
localized subspaces $\mathcal{H}_{R}$ of \textit{Fock space}.

As in the free case, the modular wedge localization does not use the full
Einstein causality but only the so-called ``weak locality'', which is just a
reformulation of the TCP invariance \cite{St-Wi}. Weakly relatively local
fields form an equivalence class which is much bigger than the local Borchers
class, but they are still associated to the same $S$-matrix (or rather to the
same TCP operator \cite{St-Wi}). Actually the $S$ in local quantum physics has
two different interpretations: $S$ with the standard scattering interpretation
in terms of large time limits of suitably defined operators obtained from
localized (compactly as in the sense of DHR or along spacelike cones
\cite{Haag}) operators, and on the other hand $S$ in its role to provide
modular localization in interacting theories as in the above formulas. There
is \textit{no parallel} \textit{outside local quantum physics} to this
two-fold role of S as a scattering operator and simultaneously as a relative
modular invariant between an interacting- with its associated free- system.
Whereas most concepts and properties which have been used in standard QFT as
e.g. time ordering and interaction picture formalism are shared by
nonrelativistic theories, modular localization is a new characteristic
structural element in LQP which is closely related to the vacuum polarization property.

The so-called \textit{inverse problem} in QFT is the question whether an
admissable $S$ (i.e. one which fulfills the general S-matrix properties as
unitarity and the analytic crossing symmerty) has a unique associated QFT.
Since the S-matrix has no unique attachment to a particular field coordinate
but is rather affiliated with a local equivalence of field coordinatizations,
the natural arena for this typ of question is the LQP algebraic setting.
Indeed this modular localization setting allows to show that if an admissable
S-operator has any associated LQP theory at all, it must be unique.

\subsection{Polarization-Free Generators}

The special significance of the wedge-localization in particle physics is due
to the fact that it is the smallest localization region for which there exists
operators $G$ such that their one-fold application to the vacuum $G\Omega$ is
a one-particle state vector without the admixture of particle-antiparticle
vacuum polarization clouds\footnote{Here we assume the usual LSZ setting of
scattering theory i.e. the existence of particle states. As previously
mentioned this excludes interacting conformal theories.}. We call such
operators ``polarization-free generators'' (PFG) \cite{Sch1} \cite{SW2}%
\cite{BBS}. Since they are necessarily unbounded, we present a more precise definition.

\begin{definition}
A closed operator G is called a polarization-free generator (PFG) if (i) it is
affiliated with a wedge algebra $\mathcal{A}(W)$, (ii) has the vacuum $\Omega$
in its domain and in that of its adjoint $G^{\ast}$ and (iii) $G\Omega$ and
$G^{\ast}\Omega$ are in the mass $m$ one-particle subspace $\mathcal{H}%
^{(1)}=E_{m}\mathcal{H}$ ($E_{m}$=one-particle projector).
\end{definition}

The existence of these operators is a consequence of the following general
theorem of modular theory:

\begin{theorem}
Let $\Phi$ be any vector in the domain of the Tomita modular operator
$\check{S}$ associated with the modular theory of ($\mathcal{A}$,$\Omega$).
Then there exists a closed operator $F$ which is affiliated with $\mathcal{A}
$ and together with its adjoint $F^{\ast}$ contains $\Omega$ in its domain and
satisfies
\[
F\Omega=\Phi,\,\,F^{\ast}\Omega=S^{\ast}\Phi
\]
\end{theorem}

\begin{proof}
The previous theorem is then a consequence for $\mathcal{A}=\mathcal{A}(W)$
and the fact that there exists a dense set of one particle states in the
domain of $\Delta_{W}^{\frac{1}{2}}$=$U(\Lambda_{W}(\chi=i\pi))$ (i.e. the
analytically continued L-boost) which is identical to the domain of $\check
{S}_{W}.$

Although the PFG's for wedge regions always exist, their use for the
construction of the wedge algebras from the wedge localized subspace is
presently limited to their ``temperedness''.

\begin{definition}
A polarization-free generator G is called tempered if there exists a dense
subspace $\mathcal{D}$ (domain of temperedness) of its domain which is stable
under translations such that for any $\Psi\in\mathcal{D}$ the function
$x\rightarrow GU(x)\Omega$ is strongly continuous and polynomial bounded in
norm for large x, and the same holds also true for $G^{\ast}.$
\end{definition}
\end{proof}

It turns out that tempered PFG's generate wedge algebras which stay close to
interaction-free theories, in fact for $d\geqslant1+2$ the S-matrix is trivial
i.e. equal to the identity. In d=1+1 it is easy to exhibit large classes of
nontrivial examples by modifying commutation relations in momentum space.
Using the rapidity parametrization for on-shell momentum $p=m(cosh\theta
,sinh\theta),$ the commutation relation for free creation and annihilation
operators reads ($\pm$ (anti)/commutator)
\begin{align}
\left[  a(\theta),a(\theta^{\prime})\right]  _{\pm}  &  =0\\
\left[  a(\theta),\alpha^{\ast}(\theta^{\prime})\right]  _{\pm}  &
=\delta(\theta-\theta^{\prime})\nonumber
\end{align}

The iterative application of the creation operator defines a basis in Fock
space. We start with the Fock space of free massive Bosons or Fermions. In
order to save notation we will explain the main ideas first in the context of
selfconjugate (neutral) scalar Bosons. Using the Bose statistics we use for
our definitions the ``natural'' rapidity-ordered notation for n-particle state
vectors
\begin{equation}
a^{\ast}(\theta_{1})a^{\ast}(\theta_{2})...a^{\ast}(\theta_{n})\Omega
,\,\,\,\theta_{1}>\theta_{2}>...>\theta_{n}%
\end{equation}
and define new creation operators $Z^{\ast}(\theta)$ as follows: in case of
$\theta_{i}>\theta>\theta_{i+1}$ and with the previous convention we set
\begin{align}
&  Z^{\ast}(\theta)a^{\ast}(\theta_{1})...a^{\ast}(\theta_{i})...a^{\ast
}(\theta_{n})\Omega=\\
&  S(\theta-\theta_{1})...S(\theta-\theta_{i})a^{\ast}(\theta_{1})...a^{\ast
}(\theta_{i})a^{\ast}(\theta)...a^{\ast}(\theta_{n})\Omega\nonumber
\end{align}
where $S(\theta)$ represents $\theta$-dependent functions of modulus one. With
$Z(\theta)$ as the formal adjoint one finds the following two-particle
commutation relations
\begin{align}
&  Z^{\ast}(\theta)Z^{\ast}(\theta^{\prime})=S(\theta-\theta^{\prime})Z^{\ast
}(\theta^{\prime})Z^{\ast}(\theta)\label{ab}\\
&  Z(\theta)Z^{\ast}(\theta^{\prime})=S(\theta^{\prime}-\theta)Z^{\ast}%
(\theta^{\prime})Z(\theta)+\delta(\theta-\theta^{\prime})\nonumber
\end{align}
where the formal $Z$ adjoint of $Z^{\ast}$ is defined in the standard way. The
$\ast$-algebra property requires $S(\theta)^{\ast}=S(\theta)^{-1}=S(-\theta).$
This structure leads in particular to
\begin{align}
Z^{\ast}(\theta_{1})...Z^{\ast}(\theta_{n})\Omega &  =a^{\ast}(\theta
_{1})...a^{\ast}(\theta_{n})\Omega\\
Z^{\ast}(\theta_{n})...Z^{\ast}(\theta_{1})\Omega &  =\prod_{i>j}S(\theta
_{i}-\theta_{j})a^{\ast}(\theta_{1})...a^{\ast}(\theta_{n})\Omega\nonumber
\end{align}
for the \textit{natural/opposite} order, all other orders giving partial
products of S's. Note that for momentum space rapidities it is not necessary
to worry about their coalescence since only the $L^{2}$ measure-theoretical
sense (and no continuity) is relevant here. In fact the mathematical control
of these operators i.e. the norm inequalities involving the number operator
\textbf{N} hold as for the standard creation/annihilation operators
\[
\left\|  \mathbf{N}^{-\frac{1}{2}}\int Z^{\ast}(\theta)f(\theta)d\theta
\right\|  \leq\left(  f,f\right)  ^{\frac{1}{2}}%
\]
Let us now imitate the free field construction and ask about the localization
properties of these F-fields
\begin{equation}
F(x)=\frac{1}{\sqrt{2\pi}}\int(e^{-ipx}Z(\theta)+h.c.)
\end{equation}
This field cannot be (pointlike) local if $S$ depends on $\theta$ since the
on-shell property together with locality leads to the free field formula. In
fact it will turn out (see next section) that the smeared operators $F(f)=\int
F(x)f(x)d^{2}x$ with
\begin{equation}
suppf\in W_{0}=\left\{  x;\,\,x^{1}>\left|  x^{0}\right|  \right\}
\end{equation}
have their localizations in the standard wedge $W$. But contrary to smeared
pointlike localized fields, the wedge localization cannot be improved by
improvements of the test function support inside $W.$ Instead the only way to
come to a local net of compactly localized algebras (and, if needed, to their
possibly existing pointlike field generators) is by intersecting oppositely
localized wedge algebras (see below). This improvement of localization by
algebraic means instead of by sharpening the localization of test functions
(quantum- versus classical- localization) is the most characteristic
distinction from the standard formalism. \textit{It takes care of
noncommutative features of LQP without violating its principles unlike the use
of noncommutative geometry (the introduction of noncommutative spacetime) in
particle physics}.

It follows from modular theory that the wedge localization properties of the
above PFG's $F(f)$ are most conveniently established via the KMS properties of
their correlation functions.

\begin{theorem}
The KMS-thermal property of the pair ($A(W),\Omega$) is equivalent to the
crossing symmetry of the S-coefficient in (\ref{ab})
\end{theorem}

For a sketch of the proof we consider the KMS property of the affiliated PFG's
$F(f).$ For their 4-point function the claim is%

\begin{align}
\left(  \Omega,F(f_{1^{^{\prime}}})F(f_{2^{^{\prime}}})F(f_{2})F(f_{1}%
)\Omega\right)   &  \equiv\left\langle F(f_{1^{^{\prime}}})F(f_{2^{^{\prime}}%
})F(f_{2})F(f_{1})\right\rangle _{therm}\\
&  \overset{KMS}{=}\left\langle F(f_{2^{^{\prime}}})F(f_{2})F(f_{1}%
)F(f_{1^{^{\prime}}}^{-2\pi i})\right\rangle _{therm}\nonumber\\
&  \Leftrightarrow S(\theta)=S(i\pi-\theta)
\end{align}
Here we only used the cyclic KMS property (the imaginary $2\pi$-shift in the
second line corresponds to the modular holomorphy (\ref{KMS}) re-expressed in
terms of the Lorentz boost parameter) for the four-point function. The
relation is established by Fourier transformation and contour shift
$\theta\rightarrow\theta-i\pi.$ One computes%

\begin{align}
&  F(\hat{f}_{2})F(\hat{f}_{1})\Omega=\int\int f_{2}(\theta_{2}-i\pi
)f_{1}(\theta_{1}-i\pi)Z^{\ast}(\theta_{1})Z^{\ast}(\theta_{2})\Omega+c.t.\\
&  =\int\int f_{2}(\theta_{2}-i\pi)f_{1}(\theta_{1}-i\pi)\{\chi_{12}a^{\ast
}(\theta_{1})a^{\ast}(\theta_{2})\Omega+\nonumber\\
&  +\chi_{21}S(\theta_{2}-\theta_{1})a^{\ast}(\theta_{2})a^{\ast}(\theta
_{1})\Omega\}+c\Omega\nonumber
\end{align}
where the $\chi$ are the characteristic function for the differently permuted
$\theta$-orders. The analogous formula for the bra-vector is used to define
the four-point function as an inner product. If S has no pole in the physical
strip, the KMS property is obviously equivalent to the crossing symmetry of
$S(\theta).$ If $S(\theta)$ has a (necessarily crossing symmetric) pole in the
in the physical strip, the contour shift will produce an unwanted terms which
wrecks the KMS relation. The only way out is to modify the previous relation
by adding a compensating bound state contribution to the scattering term
\begin{align}
&  F(\hat{f}_{2})F(\hat{f}_{1})\Omega=(F(\hat{f}_{2})F(\hat{f}_{1}%
)\Omega)_{scat}\label{2F}\\
&  +\int d\theta f_{1}(\theta_{1}+i\theta_{b})f_{2}(\theta_{2}-i\theta
_{b})\left|  \theta,b\right\rangle \left\langle \theta,b\left|  Z^{\ast
}(\theta-i\theta_{b})Z^{\ast}(\theta+i\theta_{b})\right|  \Omega\right\rangle
\nonumber
\end{align}
The second contribution then compensates the pole contribution from the
contour shift. In general the shift will produce an uncompensated term from a
crossed pole whose position is obtained by reflecting in the imaginary axis
around $i\frac{\pi}{2}.$ which creates the analogous crossed bound state
contribution. In our simplified selfconjugate model it is the same term as
above. In the presence of one or several poles one has to look at poles in
higher point functions.

Despite the different conceptual setting one obtains the same formulas as
those for the S-matrix bootstrap of factorizing models and hence one is
entitled to make use of the bootstrap technology in this modular program. In
fact the involution $J$ for the present model turns out to be (the S without
the $\theta$-dependent argument denotes the S-operator in Fock space)
\begin{align}
J  &  =SJ_{0}\\
S^{\ast}a^{\ast}(\theta_{1})...a^{\ast}(\theta_{n})\Omega &  =\prod
_{i>j}S(\theta_{i}-\theta_{j})a^{\ast}(\theta_{1})...a^{\ast}(\theta
_{n})\Omega\nonumber
\end{align}
so even without invoking scattering theory we see that the operator S fulfills
the modular definition of the S-matrix.

Using a suitable formalism it is easy to see that the PFG generators can be
generalized to particle/antiparticle multiplets. In this case the coefficient
functions S are matrix valued and the associativity of the $Z$-algebra is
nothing else than the Yang-Baxter relation. Our notation using the letter $Z$
is intended to indicate that the modular wedge localization (for those cases
with tempered PFG's \cite{BBS}) leads to a derivation and spacetime
interpretation of the Zamolodchikov-Faddeev algebra \cite{Zam}. The
creation/annihilation generators of this algebra are simply the
positive/negative energy contributions to the Fouriertransform of tempered
PFG's and the crossing symmetry of the structure coefficients $S$ is nothing
else then the modular characterization of the wedge localization of the PFG's.
This is a significant conceptual step which does not only equip the formally
useful ZF algebra with a much needed physical interpretation, but also
vindicates the old dream of the S-matrix bootstrap approach concerning the
avoidance of ultraviolet problems. Neither the S-matrix nor the local algebras
know about the short distance properties of the individual field
coordinatizations. Different from the standard approach, the formfactors of
fields are determined \textit{before} their short distance properties can
endanger their existence. The only presently known truely intrinsic
ultraviolet behavior which one can associate with e.g. double cone algebras
regardless of what generating pointlike field coordinate may have been used is
the entropy which can be assigned to the ``split inclusion'' \cite{Haag} which
in physical terms consists of a double cone with a ``collar'' around it in
form of a slightly larger double cone. This localization entropy is expected
to have a divergence in terms of the inverse collar size and a Bekenstein area
dependence as a result of the vacuum fluctuations near the surface of such a
relativistic box \cite{Sch1}.

Our wedge-localized PFG's do not only link the ZF algebra do the general
principles of local quantum physics, but they also reduce the danger that the
mathematically amazing results of the bootstrap-formfactor program for
factorizing models\footnote{For a short list of papers which have motivated
and influenced my work on the PFG generators or are close to its underlying
philosophy see \cite{fac} where the reader also finds a more detailed list of
the many model contributions to the bootstrap-formfactor program.} may end up
to become a sectarian issue dissociated from the rest of particle physics.

After having constructed the generators which are affiliated with the wedge
algebra $\mathcal{A}(W),$ one tries to construct the generators for double
cone intersections%

\begin{align}
\mathcal{A}(\mathcal{O}_{a})  &  :=\mathcal{A}(W_{a})^{\prime}\cap
\mathcal{A}(W)\\
\mathcal{O}_{a}  &  =W_{a}^{opp}\cap W\nonumber
\end{align}
where $W_{a}$ is the a-translated wedge and the double cone $\mathcal{O}_{a}$
is defined in the last line. Making for an operators $A\in\mathcal{A}%
(\mathcal{O}_{a})$ the formal Ansatz as a series in the $Z^{\prime}s,$%

\begin{equation}
A=\sum\frac{1}{n!}\int_{C}...\int_{C}a_{n}(\theta_{1},...\theta_{n}%
):Z(\theta_{1})...Z(\theta_{n}):\in\mathcal{A}(W) \label{series}%
\end{equation}
the relation which characterizes its affiliation with $\mathcal{A}%
(\mathcal{O}_{a})$ has a simple form in terms of PFG's
\begin{equation}
\left[  A,F_{a}(\hat{f})\right]  =0
\end{equation}
where $F_{a}(\hat{f})$ is the previously introduces PFG $F(\hat{f})$
translated by $a.$ On can show \cite{Sch1} that this relation leads to the
``kinematical pole relation'' which is one of Smirnov's \cite{fac} formfactor
axioms\footnote{In the literature the kinematical pole relation is written for
pointlike fields which formally corresponds to $a=0.$ The finite size leads to
a better (Pailey-Wiener) large momentum behavior.}. It relates the residuum of
certain poles of $a_{n}$ to the coefficient $a_{n-2}.$ These meromorphic
coefficient functions are related to the matrix elements of $A$ between
particle states which are called formfactors by the physicist. In the
mathematical sense the collection of such matrix elements or a formal series
as (\ref{series}) define only a sesquilinear form. The control of associated
operators has not yet been achieved and therefore the existence problem of the
factorizing models, which in the LQP setting is the nontriviality
\begin{equation}
\mathcal{A}(\mathcal{O}_{a})\neq\left\{  \mathbb{C}\mathbf{1}\right\}
\end{equation}
presently remains an open mathematical problem. But even though the
construction of factorizing models remains incomplete, there is an important
message concerning the ultraviolet problem in QFT. It is well known that in
the standard approach the short distance behavior sets severe limits; it is
impossible in $d=1+3$ to associate a meaningful perturbation theory with
interacting Lagrangian fields unless their short distance dimensions stay
close to their canonical values (i.e. $1$ for Bosons and $\frac{3}{2}$ for
Fermions) since this would lead to nonrenormalizable situations. As a result
of this restriction the number of different renormalizable coupling types is
finite. The only fields which are allowed to have large values of short
distance dimensions are the composites of the Lagrangian fields. The wedge
localization approach, which avoids field coordinatizations at the outset,
does not suffer from those Lagrangian short distance problems. Whereas the
construction of pointlike fields in the causal perturbation theory is
controlled by their short distance behavior, the existence of nontrivial
theories in the modular localization approach is determined by the
nontriviality of double cone algebras obtained by intersecting wedge algebras.
Although a more detailed investigation may reveal a relation between these two
structure, that the latter requirement appears less restrictive. Having
constructed the double cone algebras, one may of course ``coordinatize'' them
by pointlike fields; as long as one avoids such field coordinatizations in the
process of construction no ultraviolet limitations have been introduced by the
method of construction. In the Lagrangian approach we cannot even be sure that
(apart from certain ``superrenormalizable'' models) the perturbatively
renormalizable (i.e. ultraviolet-wise acceptable) models have rigorously
existing models behind them.

\subsection{Modular Inclusions, Holography, Chiral Scanning and Transplantation}

A modular inclusion in the general mathematical sense is an inclusion of two
von Neumann algebras (in our case they are assumed to be factors)
$\mathcal{N}\subset\mathcal{M}$ with a common cyclic and separating vector
$\Omega$ and such that modular group $\Delta_{\mathcal{M}}^{it}$ for $t<0$
transforms $\mathcal{N}$ into itself (compression of $\mathcal{N}$) i.e.
\begin{align}
&  Ad\Delta_{\mathcal{M}}^{it}\mathcal{N}\subset\mathcal{N}\\
t &  \lessgtr0,\pm halfsided\,\,modular
\end{align}
(when we simply say modular, we mean $t<0$) We assume that $\cup_{t}%
Ad\Delta_{\mathcal{M}}^{it}\mathcal{N}$ is dense in $\mathcal{M}$ (or that
$\cap_{t}\Delta_{\mathcal{M}}^{it}\mathcal{N=}\mathbb{C\cdot}1)$. This modular
inclusion situation may be viewed as a generalization of a situation studied
by Takesaki \cite{Sunder} in which the modular group of $\mathcal{M}$ leaves
$\mathcal{N}$ invariant. This then leads to the modular objects $\Delta
^{it},J$ of $\mathcal{N}$ being restrictions of those of $\mathcal{M}$ as well
as the existence of a conditional expectation $E:\mathcal{M}\rightarrow
\mathcal{N}.$ Whereas for inclusions of abelian algebras conditional
expectations always exist (physical application: the Kadanoff-Wilson
renormalization group ``decimation'' or ``integrating out'' degrees of freedom
in Euclidean QFT), the existence of noncommutative conditional expectations is
tight to the shared modular group of the two algebras.

The above modular inclusion situation has in particular has the consequence
that the two modular groups $\Delta_{\mathcal{M}}^{it}$ and $\Delta
_{\mathcal{N}}^{it}$ generate a two parametric group of (translations,
dilations) in which the translations have positive energy \cite{Wies}. Let us
now look at the relative commutant\ (see appendix of \cite{S-W3}). Let
$(\mathcal{N\subset M},\Omega)$ be modular with nontrivial relative commutant.
Then consider the subspace generated by relative commutant $H_{red}%
\equiv\overline{(\mathcal{N}^{\prime}\cap\mathcal{M)}\Omega}\subset H.$ The
modular unitary group of $\mathcal{M}$ leaves this subspace invariant since
$\Delta_{\mathcal{M}}^{it},t>0$ maps $\mathcal{N}^{\prime}\cap\mathcal{M}$
into itself by the inclusion being modular. Now consider the orthogonal
complement of $H_{red}$ in $H.$ This orthogonal complement is mapped into
itself by $\Delta_{\mathcal{M}}^{it}$ for positive $t$ since for $\psi$ be in
that subspace, then
\begin{equation}
\left\langle \psi,\Delta_{\mathcal{M}}^{it}(\mathcal{N}^{\prime}%
\cap\mathcal{M})\Omega\right\rangle =0\,\,for\,\,t>0.
\end{equation}
Analyticity in $t$ then gives the vanishing for all $t,$ i.e. invariance of
$H_{red}.$

Due to Takesaki's theorem \cite{Sunder}, we can restrict $\mathcal{M}$ to
$H_{red}$ using a conditional expectation to this subspace defined in terms of
the projector $P$ onto $H_{red}$. Then
\begin{align}
E(\mathcal{N}^{\prime}\cap\mathcal{M)} &  \subset\mathcal{M}|_{\overline
{(\mathcal{N}^{\prime}\cap\mathcal{M)}\Omega}}=E(\mathcal{M})\\
E(\cdot) &  =P\cdot P
\end{align}
is a modular inclusion on the subspace $H_{red}.$ $\mathcal{N}$ also restricts
to that subspace, and this restriction $E(\mathcal{N})$ is obviously in the
relative commutant of $E(\mathcal{N}^{\prime}\cap\mathcal{M)\subset
}E(\mathcal{M)}$. Moreover using arguments as above it is easy to see that the
restriction is cyclic with respect to $\,\Omega$ on this subspace. Therefore
we arrive at a reduced modular ``standard inclusion''
\begin{equation}
(E(\mathcal{N)}\subset E(\mathcal{M}),\Omega)
\end{equation}
Standard modular inclusions are known to be isomorphic to chiral conformal
field theories \cite{GLW} i.e. they lead to the canonical construction of a
net $\left\{  \mathcal{A}(I)\right\}  _{I\in\mathcal{K}}$ indexed by intervals
on the circle with the Moebius group PL(2,R) acting in correct manner
(including positive energy).

This theorem and its extension to modular intersections leads to a wealth of
physical applications in QFT, in particular in connection with ``hidden
symmetries'' symmetries which are of purely modular origin and have no
interpretation in terms of quantized Noether currents \cite{SW1}\cite{S-W3}.

The modular inclusion techniques unravel new structures which are not visible
in terms of standard field coordinatizations. In order to provide a simple
example, let us briefly return to d=1+1 massive theories. It is clear that in
this case we should use the two modular inclusions which are obtained by
sliding the (right hand) wedge into itself along the upper/lower light ray
horizon. Hence we chose $\mathcal{M=A(}W\mathcal{)}$ and $\mathcal{N}%
=\mathcal{A}(W_{a_{+}})$ or $\mathcal{N}=\mathcal{A}(W_{a_{-}})$ where
$W_{a\pm}$ denote the two upper/lower light-like translated wedges $W_{a\pm
}\subset W.$ As explained in section 2 following \cite{GLRV} (where cyclicity
is shown for massive free fields), this case leads to a modular inclusion as
above with the additional cyclicity $H_{red}=H.$ In the case of the upper
horizon of $W$ we therefore have (omitting for simplicity the $\pm$
subscripts)
\begin{align}
&  \mathcal{A}(I(0,a))\equiv A(W_{a})^{\prime}\cap\mathcal{A}(W)\\
&  \overline{\mathcal{A}(I(0,a)\Omega}=H\nonumber
\end{align}
where the notation indicates that the localization of $\mathcal{A}(I(0,a))$ is
thought of as the piece of the upper light ray interval between the origin and
the endpoint $a$.

>From the standardness of the inclusion one obtains according to the previous
discussion an associated conformal net on the line, with the following formula
for the chiral conformal algebra on the half line
\begin{equation}
\mathcal{A}(R_{>})\equiv\bigcup_{t\geq0}Ad\Delta_{W}^{it}\left(
\mathcal{A}(I(0,a))\right)  \subseteq\mathcal{A}(W)\text{,}%
\end{equation}
with the equality sign
\begin{equation}
\mathcal{A}(R_{>})=\mathcal{A}(W) \label{ch}%
\end{equation}
following from the cyclicity property%
\begin{equation}
\overline{\mathcal{A}(R_{>})\Omega}=\overline{\mathcal{A}(W)\Omega} \label{cy}%
\end{equation}
for the characteristic data on one light ray together with the before
mentioned theorem of Takesaki (which gives $P=1$ in this case)$.$ An entirely
analogous argument applies to the lower horizon of $W.$

The argument is word for word the same in higher spacetime dimensions, since
the appearance of transversal components do not modify the chain of reasoning.
The cyclicity property (\ref{cy}) can be shown for for all free fields
\cite{GLRV} except for massless d=1+1 fields which need both upper and lower
characteristic data. One of course does not expect such modular properties to
be effected by interactions. The following formal intuitive argument of the
kind used by physicists suggests that the ``characteristic shadow property''
follows generally (apart from the mentioned d=1+1 massless exception) from the
standard causal shadow property of QFT. Let us start from the special
situation $\mathcal{A}(W)=\mathcal{A}(R_{>}^{(\alpha)})$ where $R_{>}%
^{(\alpha)}$ is a spacelike positive halfline with inclination $\alpha$ with
respect to the x-axis. This is a consequence of the standard causal shadow
property (the identity of $\mathcal{A}(\mathcal{O})=\mathcal{A(O}%
^{\prime\prime})$ where $\mathcal{O}^{\prime\prime}$ is the causal completion
of the convex spacelike region $\mathcal{O}$) in any spacetime dimension.The
idea is that if this relation remains continuously valid for $R_{>}^{(\alpha
)}$ approaching the light ray ($\alpha=45%
%TCIMACRO{\unit{\UNICODE{0xb0}}}%
%BeginExpansion
\operatorname{{{}^\circ}}%
%EndExpansion
)$ which then leads to the desired equality. We will call the property
(\ref{ch}) the ``characteristic shadow property''.

Physicists are accustomed to relate massless theories with PSL(2,$\mathbb{R}%
$)-invariant LQP. Although any d=1+1 massless theory is under quite general
circumstances conformally invariant and factorizes into massless chiral
theories, not every PSL(2,$\mathbb{R}$)-invariant chiral theory describes a
massless situation. For the above construction of a chiral theory via modular
inclusion this is obvious, since although the lightray momentum operators
(generators of the lightray \ translations $U_{\pm}(a)$) $P_{\pm}$ have a
gapless nonnegative continuous spectrum going down to zero, the physical mass
spectrum of the original two-dimensional theory is given by $\mathbb{M}%
^{2}=P_{+}P_{-}$ and starts with the zero value belonging to the vacuum being
followed by a discrete one-particle state in the gap between zero and the
start of the continuum.

The above holographic projection which via modular inclusion associates a
chiral conformal theory with the originally two-dimensional massive theory did
not change the algebras in the net but only their spacetime affiliation (and
hence their physical interpretation). This becomes quite obvious if one asks
how the lightlike $U_{-}(a)$ translation acts on the holographic projection
i.e. on the chiral $\mathcal{A}_{+}(\mathbb{R})$ net. Using the natural
indexing of $\mathcal{A}_{+}(\mathbb{R})$ in terms of intervals on
$\mathbb{R}$, we see that $U_{-}(a)$ acts in a totally fuzzy way as an
automorphism since it is not a member of the $PSL(2,\mathbb{R})$ Moebius group
of $\mathcal{A}_{+}(\mathbb{R})$. With other words the holographic image is
not a chiral conformal theory in the usual sense of zero mass physics with an
associated characteristic energy-momentum tensor and an ensuing
Virasoro-algebra (on which the opposite light cone translation would act
trivially), but rather a $PSL(2,\mathbb{R})$ invariant theory with
\textit{additional automorphisms with ``fuzzy'' actions} in terms of the
chiral net indexing. The holographic projection contains the same informations
as the original two-dimensional net in particular their is no change of the
number of degrees of freedom. The lowering of the spacetime dimension in the
holographic process is accompanied by the conversion of some of the originally
geometric automorphisms into fuzzy ones. These properties are very important
if one uses lightray quantization (or the infinite momentum frame method). The
original local information gets completely reprocessed and the reconstruction
of the two-dimensional local quantum physics from its lightray quantization
description is a nontrivial task.

This idea of holographic encoding also works in higher spacetime dimensions.
In that case the formal analog of the lightray theory $\mathcal{A}%
_{+}(\mathbb{R})$ would be a $\left(  d-1\right)  $-dimensional lightfront
net. One again starts with a modular inclusion of wedges by a lightlike
translation. However this does not yet create a net on the lightfront horizon
of the wedge. It turns out that the missing transversal net structure can be
created by d-2 carefully selected additional L-boosts. The best way of
describing the result is actually not in terms of a d-1 dimensional
holographic lightfront net, but rather as a ``scanning'' of the original
theory in terms of an abstract chiral theory which is brought into d-1
scanning positions in the same Hilbert space \cite{Sch2}.

There is one very peculiar case of this holographic association of a theory
with a lower dimensional one: the famous $AdS_{d+1}$-$CQFT_{d}$ relation (a
correspondence between the so called anti deSitter spacetime manifold in d+1
and conformal field theory in d spacetime dimension). In this case the maximal
symmetry groups namely SO(d,2) are the same and in particular the phenomenon
of certain geometrically acting automorphisms turning fuzzy in the holographic
image is absent. Furthermore the correspondence is not related to lightfront
horizons and modular inclusions; it is the only known ``sporadic'' case of
holography based on shared maximal symmetry which is intimately related to the
d+2 dimensional linear formalism which one uses to conformally compactify the
d-dimensional Minkowski space \cite{AdS}. Although it is an isomorphism
between two QFTs, its existence was first conjectured in the setting of string
theory \cite{Ma-Wi}. The rigorous proof in the setting of LQP \cite{Re3} leads
to a correspondence which deviates in one point from the way the original
conjecture in the field theoretic setting was formulated in some publications.
There the correspondence was thought to be one between two Lagrangian field
theories i.e. two quantum field theories which have a presentation in terms of
covariant pointlike fields. This is however not true, inasmuch as it is not
the case in the previous modular inclusion-based holography (where this is
obvious since the original geometric symmetry gets lost in the holographic
projection). One really needs the field coordinate independent LQP setting of
nets of operator algebras also in the AdS-CQFT relation \cite{Re3}.
Intuitively one of course expects this, since the content of an isomorphism
between theories in different spacetime dimensions cannot be described in
terms of pointlike maps. Using concepts from physics one realizes that a
conformal theory which fullfils in addition to the already mentioned (see
previous section) causality properties the requirement of primitive causality
(the causal shadow property) which is the local quantum physical adaptation of
a dynamical propagation law in time (i.e. of a classical hyperbolic
differential equation), then the algebraic net of the canonically associated
AdS theory cannot be generated by pointlike fields. Rather the best localized
generators which the AdS theory possesses are ``strands'' which intersect in a
pointlike manner the conformal boundary of AdS at infinity.

String theorist would perhaps say at this point that this is what they would
have expected anyhow, but unfortunately these stringlike configurations
extending into the bulk do not increase the number of degrees of freedom which
the dynamical strings of string theory proper would do (this is the reason for
calling tem ``stands''). On the other hand these AdS strands are really
strings in the sense of LQP localization, whereas localization in the
target-space formalism of string theory has remained one of the most
inconclusive and obscure issues.

Opposite to the discussed situation is the one where one starts with a
pointlike field situation in AdS. Then the associated CQFT violates the causal
shadow property: as one moves up into the causal shadow region of a piece of a
(thin) timeslice there are more and more degrees of freedom entering from the
AdS bulk. The d+1 dimensional AdS theory generated by pointlike fields
contains too many degrees of freedom which destroy the equality between the
algebra indexed by a connected spacetime region $\mathcal{O}$ and the algebra
indexed by its causal completion (=causal shadow) $\mathcal{O}^{\prime\prime
}.$ Since the AdS spacetime from the point of view of particle physics is an
auxiliary concept\footnote{It arises if one wants to reorganize the content of
conformal theory in such a way that the so called conformal hamiltonian (which
is the higher dimensional analog of the chiral rotation) becomes the true
hamiltonian (in the sense of the Lagrangian formalism) \cite{anomalous}.}, it
is more reasonable to relax particle physics requirement on the AdS side than
on the conformal side; after all conformal theories are thought of as the
scaling limit of massive physical theories.

This brings up the interesting question whether there are QFTs, which even
after holographic unfolding (reprocessing into higher dimensions by changing
the spacetime net indexing or by giving a spaetime meaning to a fuzzy
subalgebra) do not allow an interpretation in terms of pointlike field
coordinatization, i.e. \textit{which are intrinsically operator-algebraic and
not even formally obtainable by Lagrangian quantization.} The general answer
to this question is not known, but for conformal theories a pointlike field
coordinatization is always possible. For chiral theories proofs have been
published \cite{Joerss} and there is no reason for thinking that these methods
are limited to low dimensions. My personal opinion is that there may exist
massive LQP theories which do not violate known physical principles and which
do not allow a complete description in terms of pointlike fields.

It is interesting to note that there exist relations between QFTs on
spacetimes with the same dimensionality which, although not requiring the use
of modular inclusions, share the aspect of fuzzyness (hidden symmetries) of
certain automorphisms. A nice illustration of such a ``transplantation'' was
recently given in \cite{BMS}. The authors start with a QFT on 4-dim. deSitter
spacetime which is the submanifold $dS=\left\{  x\in\mathbb{R}^{5}|\,x_{0}%
^{2}-x_{1}^{2}-...-x_{4}^{2}=-1\right\}  .$ They then show that the so called
Robertson-Walker spacetime (RW) in a certain parameter range has an isometric
embedding into dS. Although this does not lead to transformation formulas for
pointlike fields, it does allow to transplant the family of algebras of
double-cone shaped regions on dS to corresponding (using the embedding)
regions on RW. As one expects from a map which cannot be expressed in terms of
pointlike fields, part of the geometrical SO(4,1) dS symmetry becomes fuzzy
after the RW transplantation. These hidden symmetries are pure ``quantum'' and
would not be there at all in the (semi)classical RW theory. Apparently the
transplanted theory fulfills all the physical requirements presented in the
second section (adapted to curved spacetime).

It is interesting to analyze some other ideas from string theory as ``branes''
and the Klein-Kaluza dimensional reduction within the algebraic setting of
local quantum physics.

The (mem)brane idea consists in studying a theory which results from
restricting a given theory to a spacetime submanifold containing the time axis
by fixing one spatial coordinate. This is of course perfectly legitimate as an
auxiliary mathematical device. If however one wants to attribute a physical
reality to the restricted as well as to the ambient theory (which includes the
requirement of the causal shadow property), one faces a similar problem as in
the AdS-CQFT discussion above. The causal shadow property prevents to have
pointlike fields in both cases; if the brane theory is generated by pointlike
fields, these pointlike fields develop into transversal strands (which do not
depend on the transversal coordinates) in the ambient theory and these
stringlike configurations do not increase the degree of freedoms, they lack
the dynamical aspects of string theory.

The problem with the Kaluza-Klein limit is more severe, since the attempt to
reduce the number of spatial dimensions by making them compact and then
letting their size go to zero will create uncontrollable vacuum fluctuations.
Not only quantities as those which feature in the Casimir effect will diverge,
but there is not even a good reason to believe that the vacuum state on the
ambient algebra will stay finite on any local operator in the Kaluza-Klein
limit. Operator methods are better suited to make this difficulty visible than
Lagrangian quantization which especially in its functional action form is
formally closer to semiclassical aspects. In the quantization approach it
tends to be overlooked because the Kaluza-Klein reduction is made on the
quantization level before the actual model calculations (and hence the memory
of the ambient spacetime is lost), instead of first computing the correlation
functions of the ambient theory and then taking the K-K limit.

Most of the problems touched upon in this section belong to the ``unfinished
business'' of particle physics, whereas section 2 and part of section 3 (in
particular the DHR theory) consists of material with a well-understood and
firm mathematical and conceptual position. We included some ``unfinished
business''in these notes in order to counteract the widespread impression that
operator algebra methods are limited to formulate already understood aspects
of local quantum physics in a more rigorous fashion.

\subsection{Concluding remarks}

The aim of these lectures is to convince mathematicians that not only are
operator algebra methods useful for their innovative power in problems of
local quantum physics, but also that some of the concepts coming from LQP have
left their mark on mathematics. This can be seen e.g. by the many
contributions coming from LQP which preceeded discoveries in subfactor
theory\footnote{Searching in the http://xxx.lanl.gov/ archives math-ph and
math.OA under the names Doplicher, Evans, Longo, Mueger, Rehren, Roberts,
Xu..., the reader finds many AQFT-induced contributions to operator
algebras.}. Although our main motivation and illustrations of modular
inclusions were related to LQP, there have also been interesting recent
applications of modular inclusions to noncommutative dynamical (Anosov- and
K-) systems \cite{Thir}.

On the physical side the presently most promising ideas in my view are related
to the ongoing development of modular theory. Concepts as modular inclusions
are the first steps towards a seemingly very deep connection between the
relative positioning of copies of one abstract operator algebra in one common
Hilbert space and more geometrical spacetime properties and (finite or
infinite dimensional) Lie groups which are generated by the modular groups of
the various standard pairs (algebra, state vector) which can be associated
with such situations. Apart from the action of Poincar\'{e}/conformal groups
there are infinitely many modular actions which are ``fuzzy'' i.e. their local
action on an algebra associated with a region cannot be encoded into a
diffeomorphism. These actions are totally ``quantum'' i.e. they do not exist
in classical theory and hence remain hidden in quantization procedures.
Hitherto ``fuzziness'' of spacetime aspects was mostly noticed in connection
with ``noncommutative QFT'' (the noncommutativity referring to spacetime), but
here we met this behavior without changing any of the underlying principles as
soon as we use methods which allow us to go beyond the confinements of the
Lagrangian approach. With other words the use of OA-methods in QFT emulates
some of the properties which result from noncommutative QFT.

Another problem which in my view can only be solved by operator algebra
methods is to obtain an intrinsic understanding of \textit{interactions}
independent of field coordinatizations and their short distance behavior. In
fact the physical Leitmotiv underlying these lectures is \textit{how to adapt
Wigner's ideas with the help of Tomita's modular theory to the realm of
interactions}.

As mentioned, the modular methods tend to be more noncommutative than the
standard methods although the noncommutative aspects are resulting from the
same physical principles which underlie the standard approach. An example of
this is the program of constructing the multiparticle spaces and interpolating
fields for d=1+2 Wigner particles with braid group statistics using modular
wedge localization as presented in section 2.2. In the standard setting the
only Lagrangians which have a chance to have a relation with braid group
statistics are those containing Chern-Simons terms, but the whole setup
starting from Euclidean actions is too commutative to produce the real-time
vacuum expectations of products of the spacetime carriers of these plektonic
charges\footnote{In fact one Chern-Simons Lagrangian seems to describe only
plektons without anti-plektons, but the question of discrete invariances as
parity can only be settled in a formalism which has both (similar to
neutrinos) and then it is a matter of how they interact.}. In order to
counteract this there have been attempts to change the classical spacetime to
something noncommutative \cite{Suss}. The operator algebra approach would not
change the classical spacetime structure which enters as the indexing of the
net of algebras in agreement with the physical idea that the physics of d=1+2
plektons is laboratory physics and not quantum gravity. It would rather
attempt to unravel noncommutative aspects of LQP by changing Lagrangian
methods by modular constructions.

On almost all issues considered in these notes the operator algebra framework
offers a more conservative alternative than the more artistic (outside of
perturbation theory) Lagrangian approach. In particular the underlying
philosophy of LQP is not that of a search for better Lagrangians but rather
that of unfolding and sharpening of physical principles. In that respect it is
closer in spirit to condensed matter physics where one deemphasizes individual
Hamiltonians in favor of the notion of universality classes (equivalence
classes which share certain structural aspects, the most prominent being short
distance class).

Since in the history of physics the times of greatest progress were those of
contradictions and not of harmony, I believe that the importance of LQP will increase.

Last not least there are of course also strong inner-mathematical reasons for
studying ideas about noncommutative geometry which are also related to the use
of operator algebras \cite{Connes}. It would be nice if there also would exist
compelling physical reasons for a direct connection between particle physics
and noncommutative geometry. The present problem with such a program is that
the algebraic smoothness which distinguishes noncommutative geometry from
$C^{\ast}$- or von Neumann algebras does not seem to be related to the
smoothness/analyticity of physical correlation functions \cite{Haag} (which is
already accounted for by the von Neumann algebra nets of AQFT). Furthermore
the use of noncommutative spacetime for the control of ultraviolet
divergencies in laboratory (i.e. excluding quantum gravity) particle physics
seems to be far-fetched in view of the fact that more conservative ideas as
the modular localization structure contain the very strong message that the
ultraviolet problem as we know it may well be a fake of singular field
coordinatization which the Lagrangian quantization formalism enforces upon us.

\textbf{Acknowledgement} \textit{I thank Ruy Exel for the invitation to the
2001 summer school in Florianopolis and for the hospitality during this
\ enjoyable visit. I am indebted to Jorge Zubelli for inviting me to give a
series of lectures at the IMPA.}

\end{document}